\documentclass[preprint2, tighten]{aastex62}
\usepackage{amsmath,amstext,longtable,rotating}
\usepackage{apjfonts} 

\newcommand{\mjup}{$M_{\rm Jup}$}

\newcommand{\teff}{$T_{ \rm eff}$}
\newcommand{\logg}{$\log{g}$}

\newcommand{\methane}{CH$_4$}
\newcommand{\water}{H$_2$O}

\newcommand{\ammonia}{NH$_3$}

\newcommand{\microm}{$\mu$m}

\shorttitle{Spitzer Parallax Program}
\shortauthors{Martin et al.}

\begin{document}

\title{Y dwarf Trigonometric Parallaxes from the \it{Spitzer  Space Telescope}}
\author[0000-0002-0618-5128]{Emily C.\ Martin}
\affiliation{Department of Physics and Astronomy, University of California Los Angeles, 430 Portola Plaza, Box 951547, Los Angeles, CA, 90095-1547, USA}
\affiliation{IPAC, MS 100-22, Caltech, 1200 East California Blvd., Pasadena, CA 91125, USA}

\author[0000-0003-4269-260X]{J.\ Davy Kirkpatrick}
\affiliation{IPAC, MS 100-22, Caltech, 1200 East California Blvd., Pasadena, CA 91125, USA}

\author{Charles A.\ Beichman}
\affiliation{IPAC, MS 100-22, Caltech, 1200 East California Blvd., Pasadena, CA 91125, USA}
\affiliation{NASA Exoplanet Science Institute, California Institute of Technology, 770 S. Wilson Ave., Pasadena, CA 91125, USA.}

\author[0000-0002-4424-4766]{Richard L.\ Smart}
\affiliation{Istituto Nazionale di Astrofisica, Osservatorio Astrofisico di Torino, Strada Osservatorio 20, 10025 Pino Torinese, Italy}

\author{Jacqueline K.\ Faherty}
\affiliation{Department of Astrophysics, American Museum of Natural History, New York, NY 10023}

\author{Christopher R.\ Gelino}
\affiliation{IPAC, MS 100-22, Caltech, 1200 East California Blvd., Pasadena, CA 91125, USA}
\affiliation{NASA Exoplanet Science Institute, California Institute of Technology, 770 S. Wilson Ave., Pasadena, CA 91125, USA.}

\author[0000-0001-7780-3352]{Michael C.\ Cushing}
\affiliation{Department of Physics and Astronomy, The University of Toledo, 2801 West Bancroft Street, Toledo, OH 43606, USA.}

\author[0000-0002-6294-5937]{Adam C.\ Schneider}
\affiliation{School of Earth and Space Exploration, Arizona State University, Tempe, AZ, 85282, USA}

\author[0000-0001-5058-1593]{Edward L.\ Wright}
\affiliation{Department of Physics and Astronomy, University of California Los Angeles, 430 Portola Plaza, Box 951547, Los Angeles, CA, 90095-1547, USA}

\author[0000-0001-8014-0270]{Patrick Lowrance}
\affiliation{IPAC, MS 100-22, Caltech, 1200 East California Blvd., Pasadena, CA 91125, USA}

\author{James Ingalls}
\affiliation{IPAC, MS 100-22, Caltech, 1200 East California Blvd., Pasadena, CA 91125, USA}

\author[0000-0002-7595-0970]{C.~G.\ Tinney}
\affiliation{Exoplanetary Science at UNSW, School of Physics, UNSW Sydney, NSW 2052, Australia.}
\affiliation{Australian Center for Astrobiology, UNSW Australia, NSW 2052, Australia.}

\author{Ian S.\ McLean}
\affiliation{Department of Physics and Astronomy, University of California Los Angeles, 430 Portola Plaza, Box 951547, Los Angeles, CA, 90095-1547, USA}

\author[0000-0002-9632-9382]{Sarah E.\ Logsdon}
\affiliation{Department of Physics and Astronomy, University of California Los Angeles, 430 Portola Plaza, Box 951547, Los Angeles, CA, 90095-1547, USA}
\affiliation{NASA Goddard Space Flight Center, 8800 Greenbelt Road, Greenbelt, MD, 20771, USA}

\author{J{\'e}r{\'e}my Lebreton}
\affiliation{IPAC, MS 100-22, Caltech, 1200 East California Blvd., Pasadena, CA 91125, USA}

\correspondingauthor{Emily C. Martin}
\email{emartin@astro.ucla.edu}

\begin{abstract}
Y dwarfs provide a unique opportunity to study free-floating objects with masses $<$30 \mjup \ and atmospheric temperatures approaching those of known Jupiter-like exoplanets. Obtaining distances to these objects is an essential step towards characterizing their absolute physical properties. Using \textit{Spitzer}/IRAC [4.5] images taken over baselines of $\sim$2-7 years, we measure astrometric distances for 22 late-T and early Y dwarfs, including updated parallaxes for 18 objects and new parallax measurements for 4 objects. These parallaxes will make it possible to explore the physical parameter space occupied by the coldest brown dwarfs. We also present the discovery of 6 new late-T dwarfs, updated spectra of two T dwarfs, and the reclassification of a new Y dwarf, WISE J033605.04$-$014351.0, based on \textit{Keck}/NIRSPEC $J$-band spectroscopy. Assuming that effective temperatures are inversely proportional to absolute magnitude, we examine trends in the evolution of the spectral energy distributions of brown dwarfs with decreasing effective temperature. Surprisingly, the Y dwarf class encompasses a large range in absolute magnitude in the near- to mid-infrared photometric bandpasses, demonstrating a larger range of effective temperatures than previously assumed. This sample will be ideal for obtaining mid-infrared spectra with the James Webb Space Telescope because their known distances will make it easier to measure absolute physical properties.
\end{abstract}

\keywords{brown dwarfs --- astrometry:parallaxes --- infrared:astrometry --- $Spitzer$:IRAC --- stars: individual: WISE J033605.04$-$014351.0}

\section{Introduction}
Y dwarfs \citep{cushing2011, kirkpatrick2012} have effective temperatures (\teff) $\lesssim$ 500 K, are extremely faint, and emit the majority of their light in the mid-infrared. The all-sky, space-based \textit{Wide-field Infrared Survey Explorer} mission ($WISE$; \citealt{wright2010}) has specifically designed $W1$ and $W2$ filter bandpasses such that the $W1$ filter covers the strong, fundamental \methane \ bandhead at 3.3 \microm, a known absorber in the atmospheres of cold brown dwarfs, and the $W2$ filter centers on the peak of emission expected at 4.5 \microm. Thus cold brown dwarfs have very red $W1-W2$ colors and can be easily identified.

The first Y dwarfs were confirmed using a combination of ground-based and space-based spectroscopy. With typical $J$- and $H$- band magnitudes $\gtrsim$ 19, these observations are at the limit of the capabilities of the largest ground-based telescopes, and supplemental \textit{Hubble Space Telescope (HST)} observations are often required. However, the faintest Y dwarf candidates, with near-infrared magnitudes $\gtrsim$ 23, are difficult even for \textit{HST}, and will require \textit{James Webb Space Telescope (JWST)} observations to fully characterize their atmospheres. Observations of the brightest Y dwarfs revealed nearly equal flux, sharp emission peaks (in units of $f_{\lambda}$) in the shorter wavelength near-infrared $Y$, $J$, and $H$ bands, and relatively shallower, broader $K$-band fluxes \citep{cushing2011, leggett2016}. \methane \ and \water \ are the major absorbers in the atmospheres of Y dwarfs, carving out large swaths of their spectra in the near- and mid-infrared. Initial atmospheric models \citep{burrows2003} suggested that \ammonia \ would also be present in the atmospheres of Y dwarfs. Observers have yet to find direct spectroscopic evidence of this molecule in the near-infrared \citep{leggett2013, schneider2015}; however, \citet{line2015} and \citet{line2017} find unambiguous detections of \ammonia \ in cold brown dwarf spectra using advanced atmospheric retrieval techniques. Such difficulties in directly observing \ammonia \ absorption features suggests that non-equilibrium chemistry likely plays an important role in mixing the atmosphere faster than it can achieve chemical equilibrium \citep{morley2014}.  

For such cold substellar objects to exist at the current age of the universe, they must inherently have lower masses on average than the M, L, and T dwarf field populations. Based on predictions from evolutionary models (e.g. \citealt{burrows2001, saumon2008}), Y dwarfs occupy the mass range of $\sim 1-30$ \mjup. Y dwarfs represent the very bottom of the stellar/sub-stellar main sequence, as well as the lowest-mass end of the field-mass function, and are thus crucial targets for follow-up to better understand star formation at the lowest masses.

Y dwarfs share similar temperatures, masses, and chemical compositions with gas-giant exoplanets, making them useful testbeds for atmospheric physics of the coldest objects. Atmospheric observations of exoplanets are difficult because of the extreme contrast needed to differentiate the light of the planet from its host star. Single, free-floating brown dwarfs in the field do not suffer from being outshone by a brighter, more massive companion, and thus make excellent laboratories for studying the atmospheres of planetary-mass objects at temperatures ranging from $\sim$ 200--500 K \citep{beichman2014, faherty2016, skemer2016}.

Additionally, because Y dwarfs are so small and faint, most of the known Y dwarfs are located within the nearest $\lesssim$ 15 pc to the Sun. Y dwarfs that are farther than $\sim$ 20 pc are too faint to be observable with $WISE$. The farthest known Y dwarf, WD 0806-661B, at $\sim$ 19 pc, was found as a companion to a white dwarf \citep{luhman2011}, through a common-proper-motion search of the nearest stellar systems. Recent studies (e.g., \citealt{smart2010}, \citealt{winters2017}) have focused on completing the census of low-mass stars in the solar neighborhood. \citet{kirkpatrick2012} presented a preliminary volume-limited survey of the coldest (\teff \ $\lesssim$ 1000 K) substellar objects within the nearest 8 pc, but was only able to place lower limits on the number density of the coldest and lowest-mass brown dwarfs below 600 K. Precise distances of a larger sample of ultracool brown dwarfs will allow us to better characterize the solar neighborhood down to the lowest masses. 

Our current understanding of the star formation process lacks empirical data to place bounds on the lowest mass capable of forming from the collapse and turbulent fragmentation of a massive molecular cloud, if such a bound even exists. The so-called minimum Jeans mass has been examined from a theoretical perspective by several groups (see, e.g. \citealt{low1976, bate2005, padoan2007} and references therein) and shown to vary from $\sim$ 3 \mjup \ to $\sim$ 10 \mjup. \citet{burgasser2004} used simulations of varying birthrates and mass functions along with evolutionary models from \citet{burrows1997} and \citet{baraffe2003} to show the estimated luminosity functions and temperature distributions that could be produced. The local number density of Y dwarfs is shown to be the most critical constraint in determining the minimum Jeans mass. Furthermore, the relatively small number of low-mass brown dwarfs that are companions to nearby stars can be used to infer that gravitational instability is not likely to produce objects below $\sim$ 15 \mjup \ \citep{zuckerman2009}. 

Recent studies have presented trigonometric parallaxes and proper motions for small samples of nearby brown dwarfs. Several of these objects were discovered to be within 3 pc (WISE 1049$−$5319AB, WISE 0855$-$0714 \citealt{luhman2013}; \citealt{luhman2014}) and have dramatically altered our understanding of the solar neighborhood since these systems were found to be the 3rd and 4th closest systems to the Sun. Previous studies of the parallaxes of late-T and Y dwarfs include \citet{dupuy2013} and \citet{leggett2017}, who use data from the \textit{Spitzer Space Telescope} to measure astrometric fits. \citet{beichman2014} uses a combination of \textit{Spitzer} and ground-based astrometry, and \citet{smart2017} and \citet{tinney2014} both utilize ground-based near-infrared observations to measure parallaxes. \citet{luhman2016} published initial parallaxes for three Y dwarfs presented in this paper, using a subset of the data from the $Spitzer$ programs reported here. We provide updated parallaxes for these objects using a longer time baseline.

Our \textit{Spitzer} parallax program (PI: Kirkpatrick) aims to measure distances to all of the nearby late-T and Y dwarfs within 20 pc that are not being covered by ground-based astrometric monitoring. We are astrometrically monitoring 143 objects with $Spitzer$/IRAC channel 2 imaging through 2018 (Cycle 13). In this paper, we present \textit{Spitzer} photometry for 27 objects, including preliminary parallaxes for 19 Y dwarfs and 3 late-T dwarfs in our \textit{Spitzer} parallax program. The \textit{Spitzer} observations cover baselines of $\sim$ 2--7 years. 

We also present spectroscopic confirmation and spectrophotometric distance estimates for several AllWISE late-T and Y dwarf candidates with \textit{Keck}/NIRSPEC $J$ band observations. The AllWISE processing of the $WISE$ database combined all of the photometry from the original $WISE$ mission and selected high-proper motion candidates (see \citealt{kirkpatrick2014} for the initial results from the AllWISE motion survey). The new brown dwarfs presented in this paper were found in the AllWISE processing but were only recently followed-up spectroscopically to confirm their substellar nature.

In \S 2 we present our sample of targets and candidate selection methods. Section 3 describes our ground-based photometric and spectroscopic follow-up. Our $Spitzer$ photometric and astrometric data acquisition and reduction methods are explained in \S 4, and astrometric analysis is detailed in \S 5. We present our results in \S 6, followed by a discussion in \S 7. We summarize our findings in \S 8. 

\begin{center}
\begin{deluxetable*}{lcccccclll}
\tabletypesize{\scriptsize}
\tablecaption{Coordinates, Spectral Types, and Photometry of Target Objects\label{tab:Photometry}}
\tablehead{
\colhead{WISEA} &  
\colhead{Infrared} &
\colhead{Ref} &
\colhead{$J_{\rm MKO}$} &     
\colhead{$H_{\rm MKO}$} & 
\colhead{Ref} &
\colhead{$W1$} &     
\colhead{$W2$} &     
\colhead{[3.6]} &     
\colhead{[4.5]} \\
\colhead{Designation} &  
\colhead{Sp.\ Type} &
\colhead{} &
\colhead{(mag)} &
\colhead{(mag)} &
\colhead{} &
\colhead{(mag)} &
\colhead{(mag)} &
\colhead{(mag)} &
\colhead{(mag)} \\
\colhead{(1)} &                          
\colhead{(2)} &  
\colhead{(3)} &     
\colhead{(4)} &
\colhead{(5)} &                          
\colhead{(6)} &
\colhead{(7)} &
\colhead{(8)} &
\colhead{(9)} &
\colhead{(10)}
}
\startdata
J014656.66+423409.9AB&  T9+Y0\tablenotemark{a} & 2 & 20.69$\pm$0.07 & 21.30$\pm$0.12 & 8& $>$19.137&  15.083$\pm$0.065&  17.360$\pm$0.089& 15.069$\pm$0.022\\
J033605.04$-$014351.0&  Y0\tablenotemark{b} & 1 & $>$21.1         &  $>$20.2         & 1& 18.449$\pm$0.470&  14.557$\pm$0.057&  17.199$\pm$0.076&  14.629$\pm$0.019\\
J035000.31$-$565830.5&  Y1      & 2 & 22.178$\pm$0.073&  22.263$\pm$0.135& 5& $>$18.699       &  14.745$\pm$0.044&  17.832$\pm$0.131&  14.712$\pm$0.019\\
J035934.07$-$540154.8&  Y0      & 2 & 21.566$\pm$0.046&  22.028$\pm$0.112& 5& $>$19.031       &  15.384$\pm$0.054&  17.565$\pm$0.108&  15.357$\pm$0.023\\
J041022.75+150247.9  &  Y0      & 3 & 19.325$\pm$0.024&  19.897$\pm$0.038& 5& $>$18.170       &  14.113$\pm$0.047&  16.578$\pm$0.047&  14.149$\pm$0.018\\
J053516.87$-$750024.6& $\ge$Y1: & 2 & 22.132$\pm$0.071&  23.34$\pm$0.34 &5,9& 17.940$\pm$0.143&  14.904$\pm$0.047&  17.648$\pm$0.112&  15.116$\pm$0.022\\
J055047.86$-$195051.4& T6.5     & 1 & 17.925$\pm$0.021&  ---             & 1& 18.727$\pm$0.437&  15.594$\pm$0.095&  16.536$\pm$0.039&  15.303$\pm$0.021\\
J061557.21+152626.1  & T8.5     & 1 & 18.945$\pm$0.052&  ---             & 1& $>$18.454       &  15.324$\pm$0.117&  17.189$\pm$0.057&  15.199$\pm$0.019\\
J064223.48+042343.1  &  T8      & 1 & 17.677$\pm$0.012&  ---             & 1& $>$18.583	      &  15.418$\pm$0.110&  16.654$\pm$0.039&  15.177$\pm$0.019\\
J064723.24$-$623235.4&  Y1      & 4 & 22.854$\pm$0.066&  23.306$\pm$0.166& 5& $>$19.539       &  15.224$\pm$0.051&  17.825$\pm$0.128&  15.151$\pm$0.021\\
J071322.55$-$291752.0&  Y0      & 2 & 19.98$\pm$0.05  &  20.19$\pm$0.08  & 10& $>$18.776       &  14.462$\pm$0.052&  16.646$\pm$0.052&  14.208$\pm$0.018\\
J073444.03$-$715743.8&  Y0      & 2 & 20.354$\pm$0.029&  21.069$\pm$0.071& 5& 18.749$\pm$0.281&  15.189$\pm$0.050&  17.605$\pm$0.100&  15.271$\pm$0.022\\
J082507.37+280548.2  & Y0.5     & 5 & 22.401$\pm$0.050&  22.965$\pm$0.139& 5& $>$18.444       &  14.578$\pm$0.060&  17.424$\pm$0.097&  14.642$\pm$0.019\\
J105130.02$-$213859.9&  T8.5    & 1 & 18.939$\pm$0.099&  19.190$\pm$0.391& 11& 17.301$\pm$0.141&  14.596$\pm$0.056&  16.467$\pm$0.042&  14.640$\pm$0.019\\
J105553.62$-$165216.5&  T9.5    & 1 & 20.703$\pm$0.212&  $>$20.1         & 1& $>$18.103       &  15.067$\pm$0.078&  17.352$\pm$0.085&  15.011$\pm$0.021\\
J120604.25+840110.5  &  Y0      & 5 & 20.472$\pm$0.030&  21.061$\pm$0.062& 5& $>$18.734       &  15.058$\pm$0.054&  17.258$\pm$0.088&  15.320$\pm$0.022\\
J122036.38+540717.3  & T9.5     & 1 & 20.452$\pm$0.100&  ---             & 1& 19.227$\pm$0.517&  15.757$\pm$0.091&  17.896$\pm$0.101&  15.694$\pm$0.022\\
J131833.96$-$175826.3&  T8    & 1 & 18.433$\pm$0.187\tablenotemark{c} & 17.714$\pm$0.232\tablenotemark{c} & 10& 17.513$\pm$0.160&  14.666$\pm$0.058&  16.789$\pm$0.056&  14.712$\pm$0.019\\
J140518.32+553421.3  & Y0 pec?  & 3 & 21.061$\pm$0.035&  21.501$\pm$0.073& 5& 18.765$\pm$0.396&  14.097$\pm$0.037&  16.850$\pm$0.059&  14.069$\pm$0.017\\
J154151.65$-$225024.9\tablenotemark{d}
                     &  Y1      & 5 & 21.631$\pm$0.064&  22.085$\pm$0.170& 5& 16.736$\pm$0.165&  14.246$\pm$0.063&  16.512$\pm$0.046&  14.227$\pm$0.018\\
J163940.84$-$684739.4& Y0 pec   & 6 & 20.626$\pm$0.023&  20.746$\pm$0.029& 5& 17.266$\pm$0.187&  13.544$\pm$0.059&  16.293$\pm$0.029&  13.679$\pm$0.016\\
J173835.52+273258.8  &  Y0      & 3 & 19.546$\pm$0.023&  20.246$\pm$0.031& 5& 17.710$\pm$0.157&  14.497$\pm$0.043&  16.973$\pm$0.064&  14.475$\pm$0.018\\
J182831.08+265037.6  & $\ge$Y2  & 3 & 23.48$\pm$0.23  &	 22.85$\pm$0.24 &2,9& $>$18.248       &  14.353$\pm$0.045&  16.907$\pm$0.018&  14.321$\pm$0.018\\
J205628.88+145953.6  &  Y0      & 3 & 19.129$\pm$0.022&  19.643$\pm$0.026& 5& 16.480$\pm$0.075&  13.839$\pm$0.037&  16.068$\pm$0.032&  13.905$\pm$0.017\\
J220304.18+461923.4  & T8    & 1 & 18.573$\pm$0.017&  ---             & 1& $>$18.919       &  14.967$\pm$0.069&  16.351$\pm$0.021&  14.643$\pm$0.016\\
J220905.75+271143.6  &  Y0:     & 7 & 22.859$\pm$0.128&  22.389$\pm$0.152& 5& $>$18.831	      &  14.770$\pm$0.055&  17.733$\pm$0.121&  14.735$\pm$0.019\\
J222055.34$-$362817.5&  Y0      & 2 & 20.447$\pm$0.025&  20.858$\pm$0.035& 5& $>$18.772	      &  14.714$\pm$0.056&  17.180$\pm$0.072&  14.742$\pm$0.020\\
\enddata
\tablenotetext{a}{\added{Object is a known binary, so the combined-light magnitudes are not used elsewhere in this paper.}}
\tablenotetext{b}{See Section~\ref{sec:newspt} for discussion on the spectral type of this object.}
\tablenotetext{c}{Photometry is on the 2MASS system, not MKO. These values are not used elsewhere in this paper because\replaced{we think they are faulty.}{the two photometric systems are not comparable.} }
\tablenotetext{d}{This object does not appear in the AllWISE Source Catalog, so $WISE$ data are drawn from the $WISE$ All-Sky Source Catalog instead. See \citet{kirkpatrick2012} for discussion regarding the possible erroneous $W1$ measurement for this object.}
\tablecomments{References to spectral types and $JH$ photometry: (1) This paper, (2) \citealt{kirkpatrick2012}, (3) \citealt{cushing2011}, (4) \citealt{kirkpatrick2014}, (5) \citealt{schneider2015}, (6) \citealt{tinney2012}, (7) \citealt{cushing2014}, (8) \citealt{dupuy2015}, (9)\citealt{leggett2013}, (10) \citealt{leggett2015}, (11) \citealt{mace2013a}.}
\end{deluxetable*}
\end{center}

\section{Sample}
\label{sec:sample}
Objects in this paper were selected from two separate lists. The first was a list of nineteen previously published Y dwarfs (\citealt{cushing2011},  \citealt{kirkpatrick2012}, \citealt{tinney2012}, \citealt{kirkpatrick2014}, \citealt{schneider2015}) that includes one object, WISE J033605.04$-$014351.0 (hereafter WISE 0336$-$0143)\footnote{Source designations are abbreviated as WISE hhmm$\pm$ddmm.  Full designations are given in Table~\ref{tab:Photometry}.}, published earlier as a late-T dwarf (\citealt{mace2013a}) but now identified here as an early Y (See \S~\ref{sec:newspt}). The second was a list of eight objects selected from either the $WISE$ All-Sky Source Catalog or the AllWISE Source Catalog as having colors and magnitudes suggesting a late spectral type ($\geq$ T6). Specifically, these eight objects -- all classified as late-T dwarfs and listed in Table~\ref{tab:Photometry} -- were selected as (1) having $W1-W2$ $>$ 2.7 mag and $W2-W3$ $<$ 3.5 mag, (2) detected with S/N$>$3 in $W2$, and (3) not flagged as a known artifact in $W2$. \deleted{These eight late-T dwarfs were also followed up with both $Spitzer$/IRAC channel 1 (3.6 $\mu$m band; hereafter, [3.6]) and channel 2 (4.5 $\mu$m band; hereafter, [4.5]) and ground-based near-infrared imaging. If the resulting $Spitzer$ [3.6]$-$[4.5] color and $J-$W2 or $H-$W2 color further verified the late type, the object was scheduled for Keck/NIRSPEC spectroscopic follow-up. See Figures 1, 7, 8, and 11 of \cite{kirkpatrick2011} for color trends as a function of spectral type for T and Y dwarfs.}

\deleted{In this paper, we present $Spitzer$/IRAC [3.6] and [4.5] photometry for all 27 objects. Five of these targets were color-selected too late to have sufficient astrometric monitoring, however we were able to confirm their late-T dwarf nature. We present updated (18) and new (4) parallaxes for the remaining 22 late-T and Y dwarfs.}

\section{Photometric and Spectroscopic Follow-up}
\label{sec:obs}

\added{In this paper, we present $Spitzer$/IRAC channel 1 (3.6 $\mu$m band; hereafter, [3.6]) and channel 2 (4.5 $\mu$m band; hereafter, [4.5]) photometry for all 27 objects. Five of these targets were color-selected too late to have sufficient astrometric monitoring, however we were able to confirm their late-T dwarf nature. We present updated (18) and new (4) parallaxes for the remaining 22 late-T and Y dwarfs.}

\added{Of the T-dwarfs in the sample, if the resulting $Spitzer$ [3.6]$-$[4.5] color hinted at its substellar nature, it was selected for ground-based near-infrared photometric follow-up. Then, if the $J-$W2 or $H-$W2 color further verified the late type, the object was scheduled for Keck/NIRSPEC spectroscopic follow-up. See Figures 1, 7, 8, and 11 of \cite{kirkpatrick2011} for color trends as a function of spectral type for T and Y dwarfs.}

\begin{center}
\begin{deluxetable*}{llcccc}
\tabletypesize{\small}
\tablecaption{NIRSPEC Observations\label{tab:NIRSPEC}}
\tablehead{
\colhead{Short Name} &  
\colhead{SpT} &
\colhead{UT Date of Observation} &     
\colhead{Integration Time [s]} &     
\colhead{A0V Calibrator}  &
\colhead{Seeing Conditions}
}
\startdata
WISE0336$-$0143&  Y0\tablenotemark{a}   &2016 Feb 10&  2400& HD 27700&clear\\
WISE0550$-$1950&  T6.5 &2016 Feb 10&  3000& HD 44704&clear\\
WISE0615+1526  &  T8.5 &2016 Feb 01&  1200& HD 43583&clear\\
---            &  ---  &2016 Feb 11&  3000& HD 43583&clear\\
WISE0642+0423  &  T8   &2016 Feb 01&  4200& HD 43583&clear\\
WISE1051$-$2138&  T8.5 &2016 Feb 11&  4200& HD 95642&clear\\
WISE1055$-$1652&  T9.5:&2016 Feb 01&  6600& HD 98884&clear\\
---            &  ---  &2016 Feb 10&  3600& HD 92079&clear\\
WISE1220+5407  &  T9.5 &2016 Feb 01&  1800&  81 UMa&variable seeing\\
---            &  ---  &2016 Feb 11&  3600& HD 99966&clear\\
WISE1318$-$1758&  T8   &2016 Feb 11&  2400&HD 112304&windy\\
WISE2203+4619  &  T8   &2014 Oct 06&  4800&HD 219238&clear\\
\enddata
\tablenotetext{a}{See Section~\ref{sec:newspt} for discussion on the spectral type of this object.}
\end{deluxetable*}
\end{center}

\subsection{Ground-based photometry with Palomar/WIRC}
Near-infrared images of WISE 0336$-$0143, WISE 0550$-$1950, WISE 0615+1526, WISE 0642+0423, WISE 1055$-$1652, WISE 1220+5407, and WISE 2203+4619 were obtained using the Wide-Field Infrared Camera (WIRC; \citealt{wilson2003}) on the 200 inch Hale Telescope at Palomar Observatory on 4 Jan 2012 (WISE 0336$-$0143, WISE 1055$-$1652), 7 Mar 2014 (WISE 2203+4619) and 26 Feb 2016 (WISE 0550$-$1950, WISE 0615+1526, WISE 0642+0423, WISE 1220+5407).  WIRC has a pixel scale of 0$\farcs$2487/pixel providing a total field of view of 8$\farcm$7.  For each object, fifteen 2-minute images were obtained in the $J$ filter (30 minutes total exposure time).  The sky was clear during the observations on all nights.  

Images obtained in 2012 and 2014 were reduced using a suite of IRAF scripts and FORTRAN programs provided by T. Jarrett.  These scripts first linearize and dark subtract the images.  From the list of input images, a sky frame and flat field image are created and subtracted from and divided into (respectively) each input image.  At this stage, WIRC images still contain a significant bias that is not removed by the flat field.  Comparison of Two Micron All Sky Survey (2MASS; \citealt{skrutskie2006}) and WIRC photometric differences across the array shows that this flux bias has a level of $\approx$10\% and the pattern is roughly the same for all filters.  Using these 2MASS-WIRC differences for many fields, we can create a flux bias correction image that can be applied to each of the ``reduced'' images.  

In April 2014, the primary science-grade detector experienced a catastrophic failure and was replaced with a lower quality engineering-grade detector (there are more cosmetic defects, for example).  The previous reduction scripts were fine-tuned for the original detector and produced sub-optimal results with the new chip.  A WIRC reduction package written in IDL by J. Surace was used for the 2016 data as it was able to better handle the non-uniformity in one of the quadrants.  In addition to the quadrant cleaning, the Surace package differed from the Jarrett package in that the reduced data from the former did not exhibit, and thus did not require, a flux bias correction.  The other data reduction steps were essentially the same. The processed frames were mosaicked together using a median and had their astrometry and photometry calibrated using 2MASS stars in the field.\added{\footnote{The mosaicked images are available in a tarball on the AAS Journals website, linked to this paper.}} 

Table~\ref{tab:Photometry} lists the photometry, using Vega system magnitudes. Additional photometry for the remaining targets in this sample was taken from the literature. The majority of the near-infrared photometry listed in Table~\ref{tab:Photometry} is on the MKO system, though some of the Y dwarfs have synthetic photometry measured with $HST$ and corrected to match MKO filter profiles (see \citealt{schneider2015} for further details). We caution the reader that the photometric filter system can significantly change the near-infrared photometry of Y dwarfs. 

\subsection{Ground-based Spectroscopy with Keck/NIRSPEC}
\begin{sidewaysfigure*}
\begin{center}
\vspace{3.5in}
\includegraphics[scale=.8, angle=0]{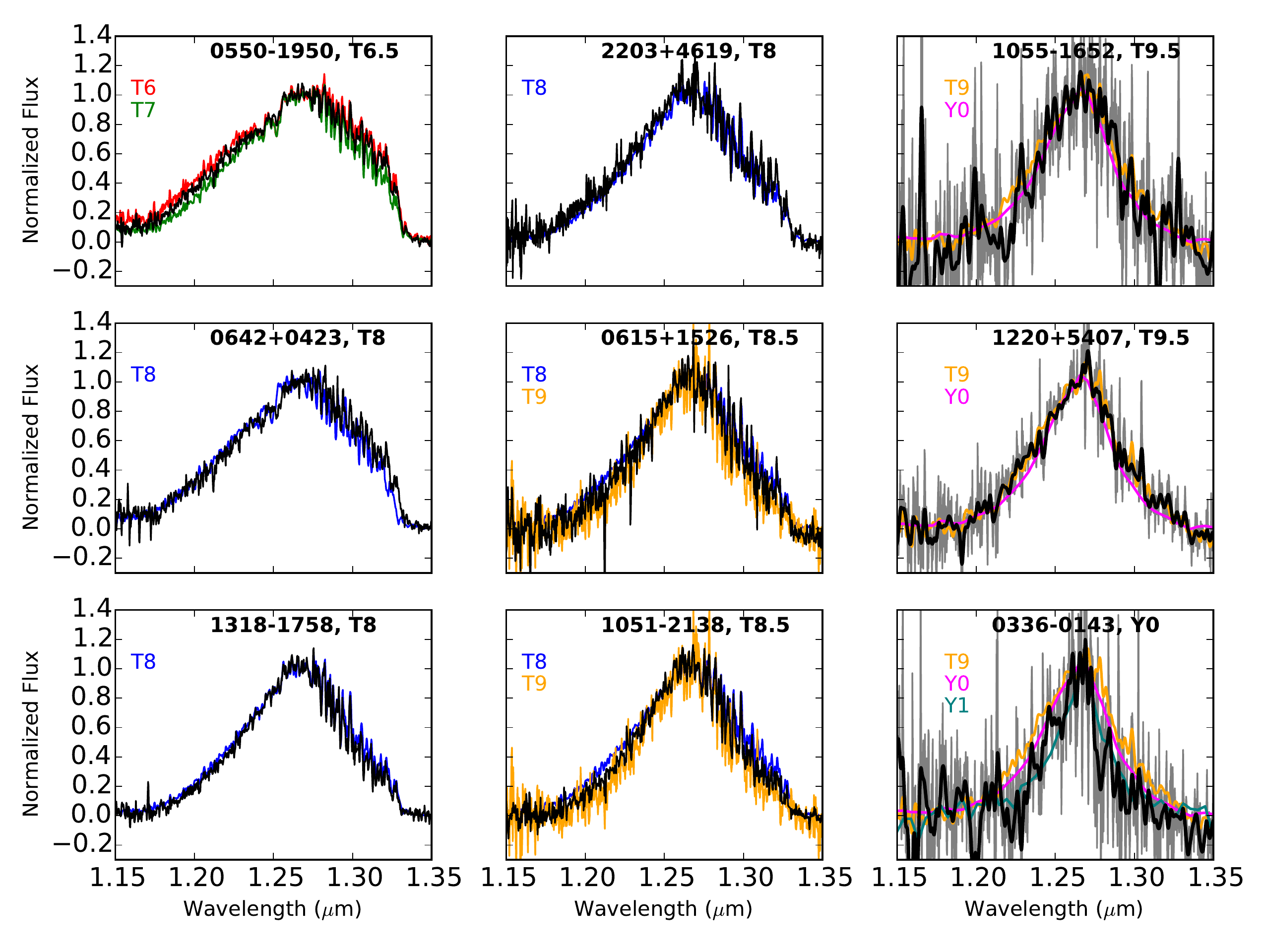}
\caption{NIRSPEC J-band spectra compared to spectral standards. The target spectrum is shown in black and the spectral standards are shown in color, and labeled in each subplot. Spectra for the spectral standards are NIRSPEC observations from \citet{mclean2003}, \citet{kirkpatrick2012}, and \citet{mace2013a}. The three latest-type objects have low SNR so their observed spectra are plotted in gray, and the binned spectra ($R\sim$500, smoothed with a gaussian kernel) are overplotted in black. 
\label{fig:spectra}}
\end{center}
\end{sidewaysfigure*}

Using the NIRSPEC instrument at the W.M. Keck Observatory \citep{mclean1998}, we made $J$-band spectroscopic observations of 4 targets from the original $Spitzer$ Parallax Program with unknown or uncertain spectral types: WISE0336$-$0143, WISE1051$-$2138, WISE1055$-$1652, and WISE 1318$-$1758. We observed an additional five targets that were likely to be late-type T or Y dwarfs based on their $W1-W2$ colors from the AllWISE processing \citep{kirkpatrick2014}. Spectral types and observation information for these targets are listed in Table~\ref{tab:NIRSPEC}. All targets were observed using AB nod pairs along the 0$\farcs$57 (3-pixel) slit, producing a spectral resolution of $R=\lambda / \Delta \lambda \sim $ 1500 per resolution element.

Spectroscopic reductions were made using a modified version of the REDSPEC package\footnote{Available at \url{http://www2.keck.hawaii.edu/inst/nirspec/redspec.html}} , following a similar procedure to \citet{mace2013a}. Frames were spatially and spectrally rectified to remove the instrumental distortion on the image plane of the detector. Frames were then background-subtracted and divided by a flat-field. Spectra from each nod pair were extracted by summing over 9--11 pixels before combining the nods. The extracted spectrum was then divided by an A0V calibrator spectrum to remove telluric features and lastly, corrected for barycentric velocity. Observations made of the same target on separate nights were combined into a single spectrum after being reduced separately. \added{Raw spectra in this paper are available in the Keck Observatory Archive\footnote{\url{https://koa.ipac.caltech.edu}} and reduced spectra are available on the NIRSPEC Brown Dwarf Spectroscopic Survey website\footnote{\url{http://bdssarchive.org}}.}

\subsection{New late-T and Y dwarfs and updated spectral types}
\label{sec:newspt}
Here we present new and updated spectral types for 9 objects in our sample that we observed with NIRSPEC. $J$ band spectra for these objects and the spectral standards used to classify them are shown in Figure~\ref{fig:spectra}.

\begin{figure}
\begin{center}
\includegraphics[scale=.43, angle=0]{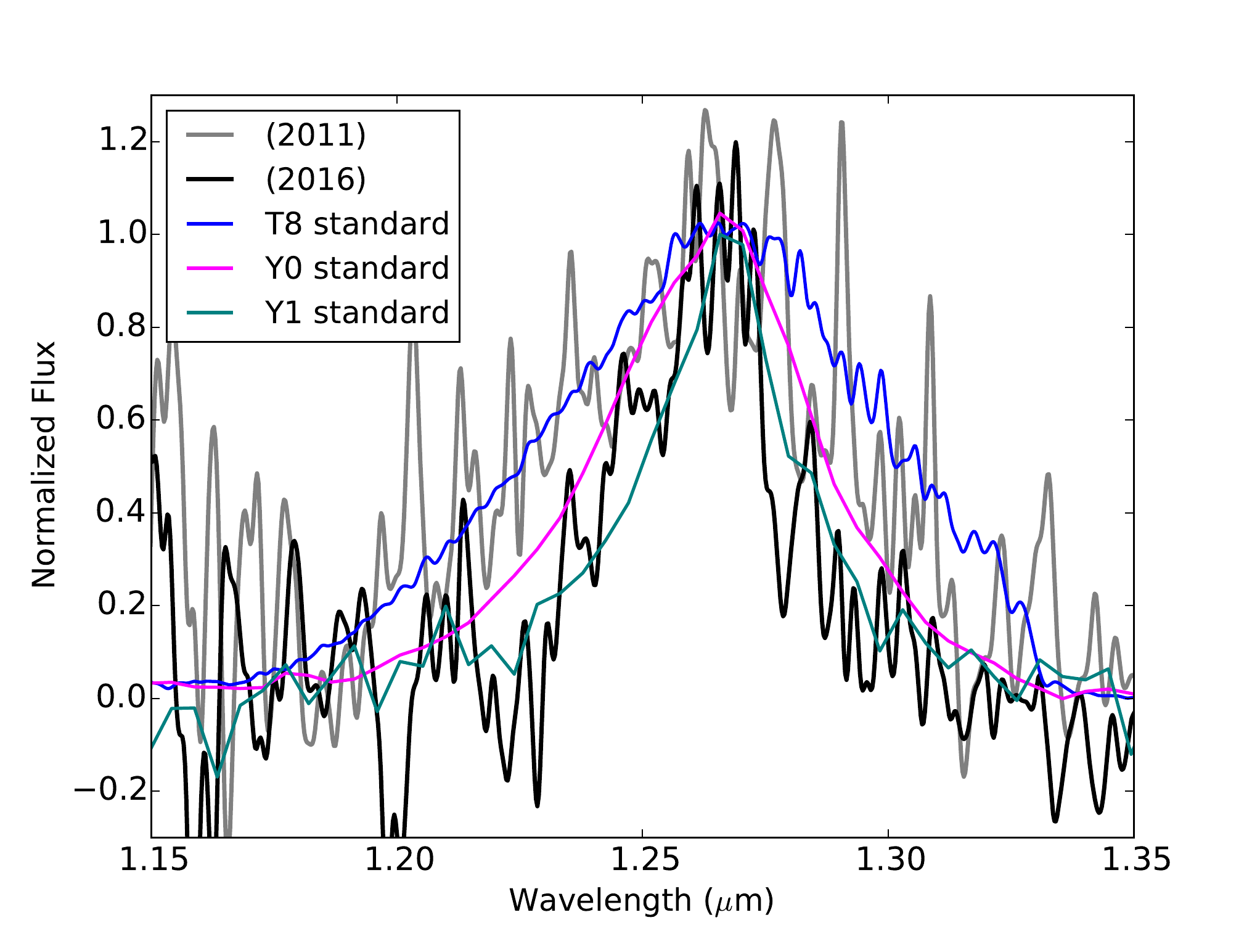}
\caption{Comparison of the (smoothed) NIRSPEC spectra of WISE 0336$-$0143 as observed in 2011 by \citet{mace2013a} (grey) and as observed in 2016 (this paper, black). The T8 (blue), Y0 (magenta), and Y1 (teal) spectral standards are over-plotted for comparison. 
\label{fig:0336_comparison}}
\end{center}
\end{figure}

\textit{WISE 0336$-$0143} was originally classified as T8: by \citet{mace2013a}. In 2016, we sought to re-observe WISE 0336$-$0143 for two reasons. First, the spectrum published in \citet{mace2013a} had a low signal-to-noise (SNR) and we wished to obtain a higher SNR spectrum. Second, we hypothesized based on its [3.6]$-$[4.5] color of 2.57 mag that WISE 0336$-$0143 should be much colder than a T8 to explain its extreme redness. Typical [3.6]$-$[4.5] colors for T8 objects are $\sim$ 1.5--2 mag (see Figure 7 in \citealt{mace2013b}; WISE 0336$-$0143 is the obvious T8 outlier in that plot.) In Figure~\ref{fig:0336_comparison}, we plot the normalized NIRSPEC spectra of the 2011 and 2016 observations. The 2016 observations match much better to a Y dwarf (see also Figure~\ref{fig:spectra}), so we  will henceforth classify this object as a Y0:. We have only been able to obtain limits on the near-infrared photometry for this object. With $J$ > 21, WISE 0336$-$0143 will require additional observations with an 8- or 10-m class ground-based telescope, or observations with \textit{HST} or JWST to further characterize its spectrum. 

\textit{WISE 0550$-$1950}, \textit{WISE 0615+1526}, \textit{WISE 0642+0423}, \textit{WISE 1220+5407}, and \textit{WISE 2203+4619} are new T dwarfs found using the AllWISE color cuts discussed in \S~\ref{sec:sample}. We find spectral types of T6.5, T8.5, T8, T9.5, and T8, respectively, based on comparison of their $J$-band spectra to spectral standards.

\textit{WISE 1051$-$2138} was given a spectral type of T9: in \citet{mace2013a}. Our re-observed spectrum, shown in Figure~\ref{fig:spectra}, indicates that this object should be classified as T8.5.

\textit{WISE 1055$-$1652} was placed on our parallax program without having an observed spectrum to confirm its substellar nature. We present the discovery of this new T9.5: dwarf.

\textit{WISE 1318$-$1758} was classified as a T9: in \citet{mace2013b} based on a noisy Palomar/TripleSpec spectrum and we re-classify it here as a T8. As shown in Figure~\ref{fig:spectra}, the T8 spectral standard is a very good match for WISE 1318$-$1758. 

\section{$Spitzer$ Astrometric Follow-Up}
\label{sec:astrometry}

\added{In order to measure distances for these ultracool dwarfs, we undertook an astrometric campaign using Spitzer IRAC [4.5] images spanning baselines of $\sim$ 2--7 years. We have utilized data from 6 $Spitzer$ programs (Table~\ref{tab:SpitzerObs}) in our analysis. Of these, program 90007 was specifically designed for parallax and proper motion measurements.}

\subsection{Observations}
$Spitzer$ IRAC [4.5] images have a field of view of 5\farcm2 on a side, over 256x256 pixels, producing a pixel scale of 1\farcs2 pix$^{-1}$. The full-width at half-maximum for a centered point response function (PRF) is 1\farcs8, for the warm mission. The raw images have a maximum optical distortion of 1.6 pixels, on the edge of the array. During {\it Spitzer} cryogenic operations, [3.6] was more sensitive than [4.5]. After cryogen depletion, however, the deep image noise\footnote{See "Warm IRAC Characteristics" at \url{http://irsa.ipac.caltech.edu/data/SPITZER/docs/irac/} for a summary of each of the effects discussed here.} was found to be 12\% worse in [3.6] and 10\% better in [4.5], making the channels more comparable in sensitivity for average field stars ([3.6]$-$[4.5] $\sim$ 0 mag) during warm operations (\citealt{carey2010}). The behavior of latent images from bright objects was also found to change during warm operations; whereas latents in [4.5] decay rapidly -- typically within ten minutes -- [3.6] latents decay on timescales of hours. Moreover, the [4.5] intrapixel sensitivity variation (also known as the pixel phase effect) is about half that of [3.6]. Given these points, the fact that the PRF is better sampled in [4.5] than in [3.6], and the fact that our cold brown dwarfs are also much brighter in [4.5] than in [3.6] (1.0 $<$ [3.6]$-$[4.5] $<$ 3.0 mag; Figure 11 of \citealt{kirkpatrick2011}), we chose to do our imaging in [4.5]. All [4.5] {\it Spitzer}/IRAC observations of the targets, the MJD range of usable data for each source, and the number of epochs available in each program, are given in Table~\ref{tab:SpitzerObs}.

\deleted{We have utilized data from 6 $Spitzer$ programs (Table~\ref{tab:SpitzerObs}) in our analysis. Of these, program 90007 was specifically designed for parallax and proper motion measurements and} \added{Our primary program, Program 90007,} used a total integration time of 270s per epoch so that all targets would have SNR$>$100 in [4.5]. To smear out the effects of intrapixel sensitivity variation, which can bias the astrometry in a frame, we chose a 9-point random dither pattern with 30s exposures per dither. Dither sizes vary for this setup, but are on the order of $\sim$ 5--30 \arcsec.\footnote{For more information on dithers, see \url{https://irsa.ipac.caltech.edu/data/SPITZER/docs/irac/calibrationfiles/dither/}} To keep the number of common reference stars between individual exposures high, we chose a dither pattern of medium scale. Timing constraints were imposed so that there was one sample within a few days of maximum parallax factor with (usually) evenly spaced samples throughout the rest of the target's visibility period.

We also used [4.5] data taken as part of earlier programs 551, 70062, and 80109 as well the later program 11059 (PI: Kirkpatrick) to increase the time baseline to help disentangle proper motion from parallax. Program 11059 used the same observing setup as program 90007, described above. All the other programs from which we utilized data (except 551; PI: Mainzer) used a frame time of 30s and a 5-point cycling dither pattern with medium scale, and observations were obtained in both [3.6] and [4.5]. In anticipation of parallax program 90007, we used the same [4.5] setup to re-observe our most promising targets during programs 70062 and 80109 after the original [3.6]+[4.5] Astronomical Observation Request (AOR) was completed. 

Program 551, which targeted only WISE1828+2650, used a frame time of 100s and a 36-point Reuleaux with medium dither in [3.6] and a frame time of 12s and a 12-point Reuleaux pattern of medium dither in [4.5]. Program 10135 (PI: Pinfield), which targeted only WISE2203+4619, used a frame time of 30s and a 16-point spiral dither pattern of medium step in both channels; in this case, two exposures were taken at each dithered position.

\begin{center}
\begin{deluxetable*}{lccc}
\tabletypesize{\small}
\tablecaption{{\it Spitzer} Observations\label{tab:SpitzerObs}}
\tablehead{
\colhead{Short Name} &  
\colhead{{\it Spitzer} Program \# (\# of [4.5] Epochs)} &
\colhead{MJD Range of Observations} &
\colhead{AORs for Photometry}\\
\colhead{(1)} &
\colhead{(2)} &
\colhead{(3)} &
\colhead{(4)}
}
\startdata
WISE0146+4234&   70062(1), 80109(2), 90007(12)&  55656.0 - 56768.1& 41808128\\
WISE0336$-$0143& 70062(1), 80109(1), 90007(12)&  55663.2 - 56777.7& 41462784\\ 
WISE0350$-$5658& 70062(2), 80109(2), 90007(10)&  55457.1 - 56925.1& 40834560\\ 
WISE0359$-$5401& 70062(2), 80109(2), 90007(12)&  55457.2 - 57035.8& 40819712\\ 
WISE0410+1502&   70062(2), 80109(2), 90007(12)&  55490.0 - 56792.5& 40828160\\  
WISE0535$-$7500& 70062(2), 80109(1), 90007(10)&  55486.2 - 56875.6& 41033472\\  
WISE0550$-$1950& 11059(1)                     &  57175.2          & 52669696\\
WISE0615+1526&   11059(1)                     &  57175.1          & 52669952\\
WISE0642+0423&   11059(1)                     &  57175.1          & 52670208\\
WISE0647$-$6232& 70062(2), 80109(1), 90007(10)&  55458.4 - 56887.1& 40829696\\  
WISE0713$-$2917& 80109(1), 90007(12)          &  55928.8 - 56856.5& 44568064\\  
WISE0734$-$7157& 70062(1), 80109(1), 90007(10)&  55670.6 - 56790.7& 41754880\\  
WISE0825+2805&   80109(2), 90007(12)          &  55933.9 - 56849.0& 44221184\\  
WISE1051$-$2138& 70062(1), 90007(11)          &  55633.6 - 56903.4& 41464320\\  
WISE1055$-$1652& 80109(1), 90007(9)           &  56124.9 - 56900.5& 44549632\\  
WISE1206+8401&   70062(2), 80109(1), 90007(12)&  55539.7 - 57049.1& 40823808\\  
WISE1220+5407&   11059(1)                     &  57063.2          & 52671232\\
WISE1318$-$1758& 70062(1), 80109(2), 90007(12)&  55663.5 - 56925.0& 40824832\\  
WISE1405+5534&   70062(1), 80109(2), 90007(10)&  55583.1 - 56902.1& 40836864\\  
WISE1541$-$2250& 70062(1), 80109(2), 90007(12)&  55664.9 - 56812.0& 41788672\\  
WISE1639$-$6847& 90007(12), 11059(1)          & 56431.7 - 57175.3& 52672000\tablenotemark{a}\\  
WISE1738+2732&   70062(2), 80109(2), 90007(12)&  55457.5 - 56864.6& 40828416\\  
WISE1828+2650&   551(1), 70062(1), 80109(2), 90007(12)&  55387.3 - 56878.5& 39526656, 39526912\tablenotemark{b}\\  
WISE2056+1459&   70062(2), 80109(2), 90007(12)&  55540.0 - 57049.0& 40836608\\  
WISE2203+4619&   10135(1)                     &  56922.9          & 50033152\\
WISE2209+2711&   70062(2), 80109(1), 90007(12)&  55561.9 - 56925.3& 40821248\\  
WISE2220$-$3628& 80109(2), 90007(12)          &  55949.1 - 56902.9& 44552448\\ 
\enddata
\tablenotetext{a}{This high motion object was blended with a background star during our original observation in program 80109. We reacquired this observation during program 11059 to make up for the loss of a [4.5] astrometric epoch and the loss of our sole [3.6] photometric data point.}
\tablenotetext{b}{The [3.6] and [4.5] observations of this object in program 551 were broken into separate AORs but were observed concurrently.}
\end{deluxetable*}
\end{center}

\subsection{Astrometric and Photometric Data Reductions}

We used the $Spitzer$ Heritage Archive\footnote{Available at \url{http://irsa.ipac.caltech.edu/}} to download all of the basic calibrated data (BCD) at [4.5] for the programs listed in Table~\ref{tab:SpitzerObs}. Data were reduced using the Mosaicker and Point Source Extractor (MOPEX\footnote{Available at
  \url{http://irsa.ipac.caltech.edu/data/SPITZER/docs/dataanalysistools/tools/mopex/}}) with customized scripts \added{created following instructions in the MOPEX handbook}. These scripts use the individually dithered BCD files to create a coadded image at each epoch (i.e., for each AOR) and to detect and characterize sources on the resulting coadd.

The data and scripts have been modified in two ways to utilize new knowledge gained during the {\it Spitzer} warm mission. First, the headers of the BCD files  available at the $Spitzer$ Heritage Archive have been updated to include a new {\it Spitzer}-produced fifth-order distortion  correction for the IRAC camera, which is an improvement over the third-order correction included previously (\citealt{lowrance2014}). Second, the PRF employed by the code is one created specifically for use on {\it Spitzer} warm data\footnote{ For more information on the PRF maps, see \url{http://irsa.ipac.caltech.edu/data/SPITZER/docs/irac/calibrationfiles/psfprf/} }, sampled onto a 5$\times$5 grid to account for small changes in shape across the array. The MOPEX code performs a simultaneous chi-squared minimization\footnote{For more information, see \url{http://irsa.ipac.caltech.edu/data/SPITZER/docs/dataanalysistools/tools/mopex/mopexusersguide/88/\#\_Toc320000081}} using fits of the PRF to the stack of individual frames to measure the photometry and position of the source in that AOR. It should be noted that the random dithers will help to zero out the astrometric  bias caused by the intrapixel distortion in each individual frame (\citealt{ingalls2012}), so this effect did not have to be specifically addressed in our reduction methodology.

Our [3.6] observations were run identically to the [4.5] data discussed above. We divided the resulting PRF-fit fluxes by the appropriate [3.6] and [4.5] correction factors (1.021 and 1.012, respectively) indicated in Table C.1 of the IRAC Instrument Handbook\footnote{Available at \url{http://irsa.ipac.caltech.edu/data/SPITZER/docs/irac/iracinstrumenthandbook/}} and converted these fluxes to magnitudes using the [3.6] and [4.5] zero points of 280.9$\pm$4.1 Jy and 179.7$\pm$2.6 Jy, respectively, as given in Table 4.1 of the same document. The final [3.6] and [4.5] photometry is listed in columns 9-10 of Table~\ref{tab:Photometry}.

Prior studies have shown that the amplitude of [3.6] and/or [4.5] variability in T0--T8 dwarfs can can reach the 10\% level, with some objects varying more in one band than the other (\citealt{metchev2015}). This amplitude increases at later spectral types. In fact, one Y dwarf, WISE 1405+5534, has already been observed to vary at levels as high as 3.5\% in [3.6] and [4.5] based on a limited data set ({\citealt{cushing2016}). Another Y dwarf observed for variability, WISE 1738+2732, showed peak-to-peak variability of $\sim$ 3 \%, at [4.5] with potentially up to 30\% variability in the near-infrared \citep{leggett2016}. Therefore, our tabulated values list photometry only for the one AOR having concurrent  [3.6] and [4.5] observations so that the resulting [3.6]$-$[4.5] value represents a physical snapshot of the color at a specific time rather than a possibly non-physical color created from disparate epochs of [3.6] and [4.5] observations. The AORs from which the [3.6] and [4.5] photometry is measured are listed in column 4 of Table~\ref{tab:SpitzerObs}.

The centroid locations determined by the MOPEX routine on each of the epochal coadds (average positions across multiple dithers) were then used as the fundamental source of our astrometric measurements. Our resulting inputs to our astrometric fitting routine at each epoch were the source location, time of observation of the middle frame, and geometric cooordinates of $Spitzer$ during the observations.

\section{Astrometric Analysis}
\added{Using the astrometric measurements described in the previous section, we then re-registered each frame onto a common reference frame, determined positional uncertainties, and then solved for the proper motion and parallax of each source, as described below. Target coordinates at each epoch are recorded in Table~\ref{tab:coords} and our final astrometric solutions are detailed in Table~\ref{tab:astrometric_solutions}.}

\subsection{Coordinate Re-Registration}
Prior to fitting our astrometric solutions for each target, we re-registered the coordinates of our targets in each epoch onto a single reference frame. We chose to align our coordinates to those provided by the Gaia Mission in Data Release 1 (DR1; \citealt{gaia2016a, gaia2016b}). These positions are the best that are currently available across the whole sky. 2MASS positions, which provide the basis for the WCS coordinates given by MOPEX/APEX, have positional uncertainties on the order of $\sim$ 70 mas \citep{mccallon2007}, while Gaia DR1 positions for the brightest, unsaturated reference stars have positional uncertainties on the order of $\lesssim$ 1 mas. 

\begin{figure*}
\begin{center}
\includegraphics[scale=.7, angle=0]{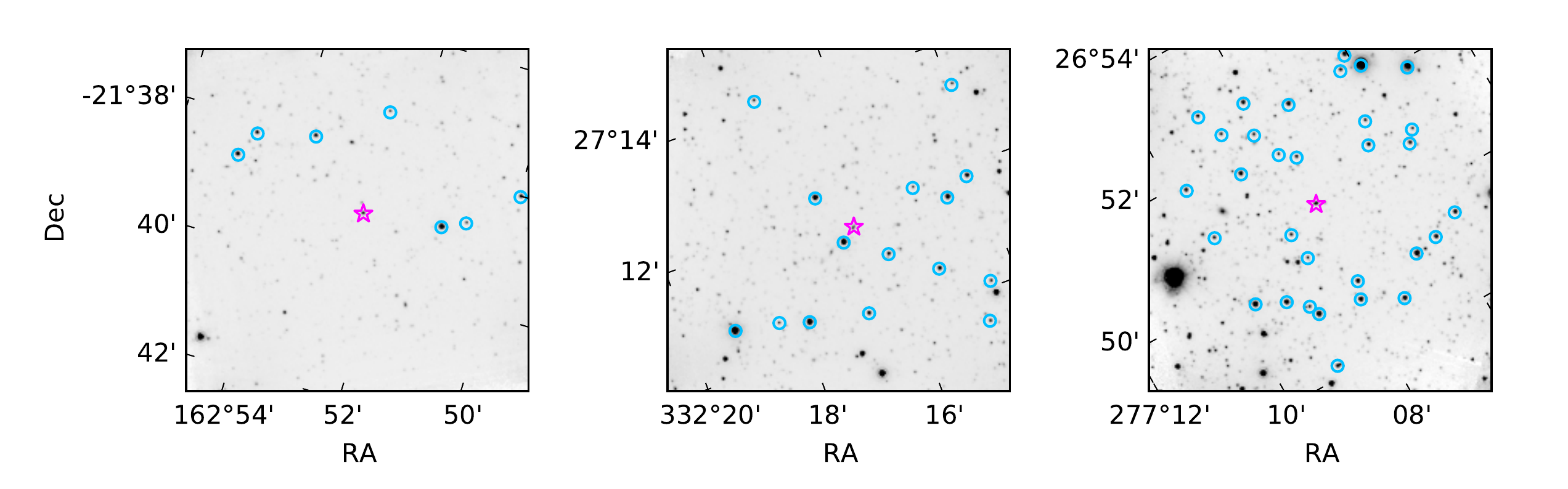}
\caption{Supermosaic frames of WISE 1051$-$2138, WISE 2209+2711, and WISE 1828+2650, from left to right. Reference stars in each of the fields are circled in blue. Targets are marked by a magenta star. These frames show examples of different target fields, ranging from few reference stars to many.
\label{fig:ref_stars}}
\end{center}
\end{figure*}

We selected reference background sources for the re-registration process by requiring that the sources be detected in all epochal co-adds, have SNR>100, and have positional uncertainties within 2$\sigma$ of the median positional uncertainty of the field. Requiring a detection in every co-added frame cut sources on the extreme edge of the field, while the positional uncertainty cut removed any sources with any significant proper motion. We then evaluated each reference target by-eye to discard any non point-like sources. We obtained Gaia coordinates for each reference star, where available, and excluded any with exceptionally high uncertainties ($\gtrsim$ 1 mas) in the Gaia DR1, as well as reference sources that were lacking Gaia coordinates. The resulting set of reference stars varied from 7 to 96, depending on the stellar density in the field. Thumbnail images of three example fields with the target and reference sources highlighted can be found in Figure~\ref{fig:ref_stars}, showcasing fields with low, moderate, and high numbers of reference stars.

\subsection{Positional Uncertainties}
\begin{figure}
\begin{center}
\includegraphics[scale=0.85, angle=0]{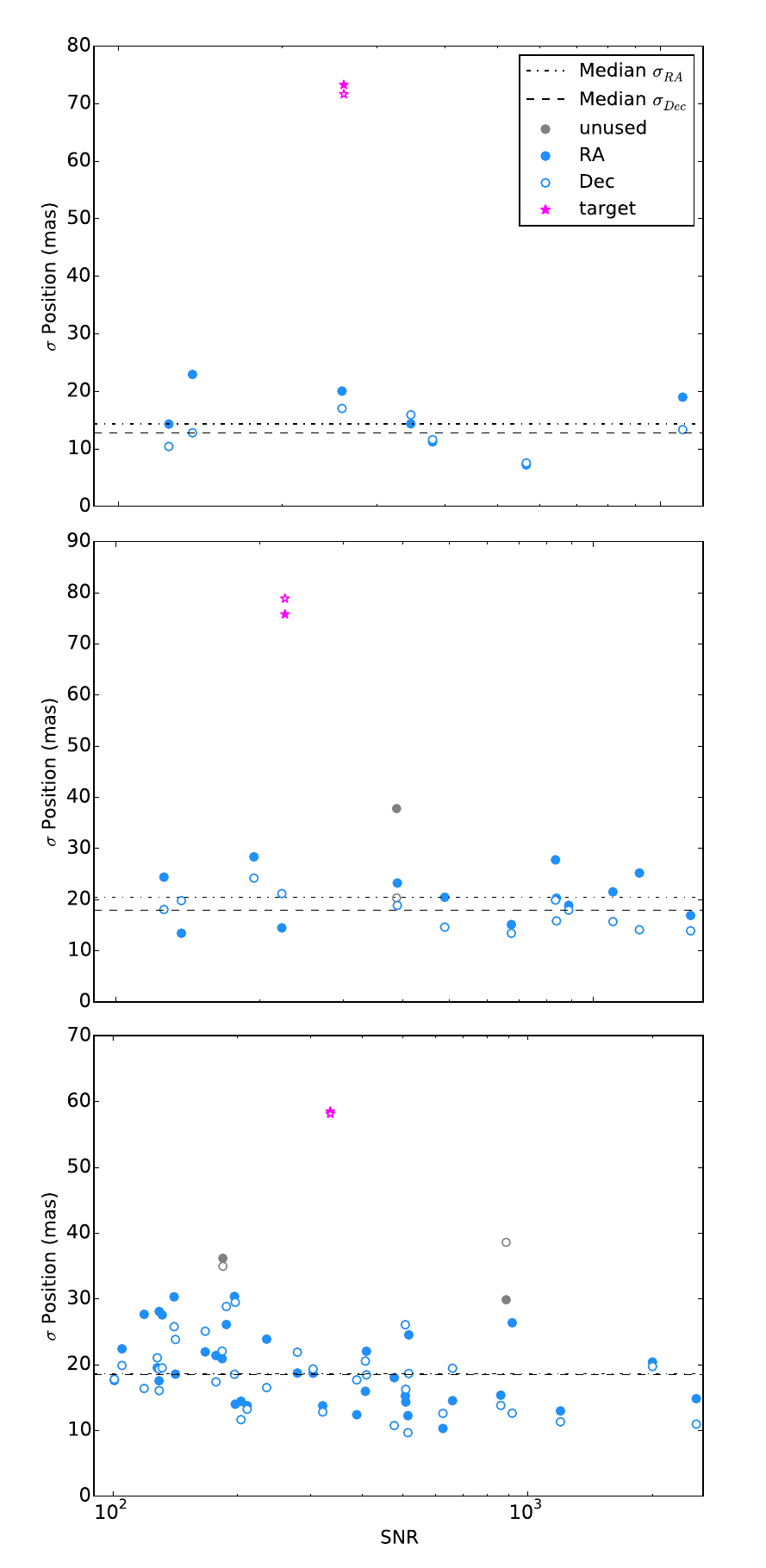}
\caption{Positional uncertainties vs. SNR for stars in the fields of WISE 1051$-$2138, WISE 2209+2711, and WISE 1828+2650, from left to right. Reference star $\sigma_{\rm RA}$ are in red and $\sigma_{\rm Dec}$ are in blue. Uncertainties were calculated by taking the standard deviation of the centroid location across all epochs, post re-registration. Positional uncertainty drops with increasing SNR, until reaching a systematic floor. We measure the median positional uncertainty for reference stars with SNR $>$ 100 after performing a 1$\sigma$ clipping to remove outliers. The median value (horizontal lines) is used as the target positional uncertainty, in lieu of the MOPEX-given $\sigma_{\rm RA}$ and $\sigma_{\rm Dec}$ (stars), which significantly overestimates the positional uncertainties.
\label{fig:0410pos_unc}}
\end{center}
\end{figure}
We found that the positional uncertainties output by the MOPEX/APEX centroid extractions were overestimated by a factor of $\gtrsim$ 2 compared to the uncertainties on background stars with similar SNR. Instead of using these inflated uncertainties, we determined empirical positional uncertainties for each target by comparison to the positional uncertainties of the presumably non-moving field reference stars. For each field, we re-registered the locations of all stars in the field using the correction determined by the reference field stars as detailed above. We then calculated the positional uncertainty of every source in both RA and Dec as the standard deviation of the centroid location across all epochs, post-re-registration. As expected, positional uncertainty drops with increasing SNR until it reaches a systematic floor of $\sim$15--40 mas, depending on the field. Figure~\ref{fig:0410pos_unc} shows three examples of positional uncertainty vs. source SNR, given a low, medium, or high number of reference stars. Our target SNRs are typically high enough that their positional uncertainties can be determined from the asymptotic portion of the graph. We measure the median positional uncertainty above a cutoff SNR $>$ 100 after performing a 2$\sigma$ clipping to remove significant outliers. These outliers could be non-point-like sources, e.g. galaxies, or they could have significant proper motion. The median value rounded to the nearest 5 mas is then the positional uncertainty that we use in each epoch to determine the astrometric fits for each of our targets, with a floor of 15 mas. Positional uncertainties for each target are listed in column 10 of Table~\ref{tab:astrometric_solutions}.

After determining our target uncertainties, the selected reference stars that likewise met the sigma clipping requirement were then used to perform a final re-registration. We performed a least-squares affine transformation to adjust each frame onto the Gaia reference frame. To do this, we projected both the Gaia and MOPEX coordinates onto a tangent plane ($\xi, \eta$) and then solved for the best-fit generalized 6-term solution, allowing for offsets, rotation, and scaling between the two planes.

\begin{center}
\begin{deluxetable*}{lccccc}
\tabletypesize{\footnotesize}
\tablecaption{Target coordinates at each epoch\label{tab:coords}}
\tablehead{
\colhead{Name} &  
\colhead{RA} &
\colhead{Dec} &
\colhead{$\Delta$RA} &
\colhead{$\Delta$Dec} &
\colhead{Obs. Date}\\ 
\colhead{---} &  
\colhead{(deg)} &
\colhead{(deg)} &
\colhead{(arcsec)} &
\colhead{(arcsec)} &
\colhead{MJD}\\ 
\colhead{(1)} &                          
\colhead{(2)} &  
\colhead{(3)} &
\colhead{(4)} &
\colhead{(5)} &
\colhead{(6)} 
}
\startdata
WISE0146+4234&26.7360341575 & 42.5694215003 &0.02 & 0.02 & 55656.090572  \\
             &26.7358573158 &42.5694154093  &0.02  &0.02  &55993.0453281 \\
             &26.7358123414 &42.569430009   &0.02  &0.02  &56215.0761909 \\
             &26.7355032955 &42.5693971935  &0.02  &0.02  &56768.0673261 \\
             &26.7355150149 &42.569396338   &0.02  &0.02  &56758.3544525 \\
             &26.7355269728 &42.5694010066  &0.02  &0.02  &56750.4605196 \\
             &26.7355337547 &42.5694111367  &0.02  &0.02  &56737.3844441 \\
             &26.7356177374 &42.5694257248  &0.02  &0.02  &56616.0666252 \\
             &26.735610704  &42.5694074714  &0.02  &0.02  &56602.4785276 \\
             &26.7356193002 &42.5694181532  &0.02  &0.02  &56592.4703411 \\
             &26.7356217406 &42.5694100106  &0.02  &0.02  &56579.1881263 \\
             &26.7356996309 &42.5694108774  &0.02  &0.02  &56393.1344496 \\
             &26.7356927974 &42.5694167189  &0.02  &0.02  &56388.816662  \\
             &26.7356863574 &42.5694080177  &0.02  &0.02  &56372.3134272 \\
             &26.7356981636 &42.5694173971  &0.02  &0.02  &56364.2577053 \\
\enddata
\tablecomments{Table~\ref{tab:coords} is published in its entirety in the machine-readable format. A portion is shown here for guidance regarding its form and content.}
\end{deluxetable*}
\end{center}

\subsection{Astrometric Solutions}
After re-registering each source onto a common reference frame, we then solved simultaneously for 5 parameters: trigonometric parallax $\pi$, proper motion in both RA and dec ($\mu_{\alpha}$, $\mu_{\delta}$), and initial position ($\alpha_0$, $\delta_0$) at a fiducial time of $T_0$=2014.0, which falls roughly in the middle of our time baseline for each object. We used the standard astrometric equations \citep{smart1977, green1985}, inputing epochal coordinates, time of observation, and rectangular observatory coordinates obtained from the image headers. We then used Pythons' Scipy least squares minimization module\footnote{https://docs.scipy.org/doc/scipy-0.19.0/reference/generated/scipy.optimize.leastsq.html} to solve for the best fit. Our best-fit astrometric solutions are listed in Table~\ref{tab:astrometric_solutions} and plotted in Figure~\ref{fig:astrometric_fits}. \added{Three examples are printed here. The complete figure set (22 objects) is available in the online journal}. We show both the overall astrometric fit, as well as the parallactic ellipse, after removing the best-fit proper motion component. The best-fit model shown makes use of the $Spitzer$ ephemerides from JPL's Horizons\footnote{https://ssd.jpl.nasa.gov/?horizons} to calculate the heliocentric rectangular coordinates of $Spitzer$ over a longer time baseline and with higher cadence than our observations. These measurements are for relative parallaxes, not absolute. We estimate that the correction for the systematic offset of the average parallax of the background stars is $\sim$ 1 mas, well within the random errors of our solutions. 

One caveat for the targets at high declination ($\lvert \delta \rvert \gtrsim 70 \degr $) is that an unidentifiable problem in the MOPEX mosaicking code leads to much more uncertain astrometry. This is reflected in the larger uncertainties we adopt for their epochal positions and the generally larger reduced chi-squared ($\chi_{\nu}^2$) values we measure. This issue will be further discussed in our forthcoming paper presenting parallaxes for all of the T6 and later brown dwarfs in our parallax program (Kirkpatrick et al., in prep).

\clearpage
\movetabledown=5.5 in
\begin{rotatetable}
\begin{deluxetable}{lcccccccccccc}
\tabletypesize{\scriptsize}
\tablecaption{Best-fit Astrometric Solutions \label{tab:astrometric_solutions}}
\tablehead{
\colhead{Object} &  
\colhead{$\alpha_{0, 2014}$} &
\colhead{$\delta_{0, 2014}$} &
\colhead{$\sigma_{\alpha_0}$} & 
\colhead{$\sigma_{\delta_0}$} &
\colhead{$\mu_{\alpha}$} &
\colhead{$\mu_{\delta}$} & 
\colhead{$\pi_{ \rm trig}$} &
\colhead{Distance} &
\colhead{$\sigma_{\rm pos}$} &
\colhead{$n_{\rm epochs}$} & 
\colhead{$n_{\rm ref}$} &
\colhead{$\chi^2$ / dof = $\chi^2_{\nu}$} \\ 
\colhead{Name} &
\colhead{(Deg, J2000)} &
\colhead{(Deg, J2000)} &
\colhead{(mas)} &
\colhead{(mas)} &
\colhead{(mas/yr)} &
\colhead{(mas/yr)} &
\colhead{(mas, relative)} &
\colhead{(pc)} &
\colhead{(mas)} &
\colhead{} &
\colhead{} &
\colhead{} \\
\colhead{(1)} &                          
\colhead{(2)} &  
\colhead{(3)} &     
\colhead{(4)} &
\colhead{(5)} &                          
\colhead{(6)} &
\colhead{(7)} &
\colhead{(8)} &
\colhead{(9)} &
\colhead{(10)} &
\colhead{(11)} &
\colhead{(12)} &
\colhead{(13)}
}
\startdata
WISE 0146+4234   & 26.735579  &  42.569408  & 8.69 & 6.24 & $-$450.67 $\pm$ 6.29 & $-$27.90$\pm$6.34   &  45.575 $\pm$ 5.74 & 21.94$\substack{+3.16 \\ -2.45}$ & 20 & 15 & 18 & $34.0/25=1.36$ \\[5pt] 
WISE 0336$-$0143 & 54.020862  &  $-$1.732170  & 6.65 & 6.48 & $-$247.35 $\pm$ 6.05 & $-$1213.46$\pm$6.03 & 100.90 $\pm$ 5.86  & 9.91$\substack{+0.61 \\ -0.54}$ & 20 & 14 & 13 & $29.44/23=1.28$ \\[5pt] 
WISE 0350$-$5658 & 57.500996  &  $-$56.975638 & 17.80 & 9.66 & $-$206.94 $\pm$ 6.52 & $-$577.67$\pm$6.68  &  168.84 $\pm$ 8.53 & 5.92$\substack{+0.32 \\ -0.28}$ & 30 &14 & 8 & $23.69/23=1.03$ \\[5pt] 
WISE 0359$-$5401 & 59.891827  &  $-$54.032517 & 11.96 & 6.90 & $-$152.70 $\pm$ 4.83 &  $-$783.66$\pm$4.93 &  75.36 $\pm$ 6.62  & 13.27$\substack{+1.28 \\ -1.07}$ & 25 &16 &  8 & $31.32/27=1.16$ \\[5pt] 
WISE 0410+1502   & 62.595853  &  15.044417  & 5.00 & 4.71 &  959.86 $\pm$ 3.57 & $-$2218.64$\pm$3.46 &  153.42 $\pm$ 4.05 & 6.52$\substack{+0.18 \\ -0.17}$  & 15 &16 & 12 & $31.32/27=1.16$ \\[5pt] 
WISE 0535$-$7500 & 83.819477  &  $-$75.006740 & 39.31 & 10.08 & $-$113.23 $\pm$ 7.71 &  23.72$\pm$7.52   &  79.51 $\pm$ 8.79  & 12.58$\substack{+1.56 \\ -1.25}$ & 30 &12 & 28 & $32.3/19=1.70$\\[5pt] 
WISE 0647$-$6232 & 101.846784 &  $-$62.542832 & 14.92 & 6.74 & 1.015 $\pm$ 5.08  &  390.97$\pm$4.61  &  83.73 $\pm$ 5.68  & 11.94$\substack{+0.87 \\ -0.76}$ & 20 &13 & 21 & $24.43/21=1.16$ \\[5pt] 
WISE 0713$-$2917 & 108.344414 &  $-$29.298188 & 5.51 & 4.72 & 341.10 $\pm$ 6.57  & $-$411.13$\pm$6.00  &  100.73 $\pm$ 4.74  & 9.93$\substack{+0.49 \\ -0.45}$ & 15 &13 & 68 & $9.87/21=0.47$ \\[5pt] 
WISE 0734$-$7157 & 113.681539 &  $-$71.962325 & 32.07 & 9.94 & $-$566.22 $\pm$ 8.85 &  $-$77.54$\pm$8.82  &  67.63 $\pm$ 8.68  & 14.79$\substack{+2.18 \\ -1.68}$ & 30 &12 & 26 & $21.59/19=1.14$ \\[5pt] 
WISE 0825+2805   & 126.280554 &  28.096545  & 5.40 & 4.72 & $-$64.35 $\pm$ 5.56  &  $-$234.73$\pm$5.36 &  139.02 $\pm$ 4.33 & 7.19$\substack{+0.23 \\ -0.22}$ & 15 &14 & 13 & $24.25/23=1.05$ \\[5pt] 
WISE 1051$-$2138 & 162.875233 &  $-$21.650040 & 6.74 & 6.28 & 145.57 $\pm$ 6.84  &  $-$160.68$\pm$6.60 &  49.27 $\pm$ 6.47  & 20.3$\substack{+3.1 \\ -2.4}$ & 20 &12 &  7 & $17.79/19=0.94$ \\[5pt] 
WISE 1055$-$1652 & 163.972546 &  $-$16.870930 & 6.98 & 6.48 & $-$1001.7 $\pm$ 9.2 &  432.16$\pm$9.17 &  71.21 $\pm$ 6.82 & 14.04$\substack{+1.5 \\ -1.2}$ & 20 &10 &  12 & $22.8/15=1.52$ \\[5pt] 
WISE 1206+8401   & 181.512553 &  84.019282  & 93.16 & 9.09 & $-$557.69 $\pm$ 6.54 &  $-$241.31$\pm$6.51 &  85.12 $\pm$ 9.27  & 11.75$\substack{+1.44 \\ -1.15}$ & 30 &14 &   7 & $19.46/14=1.39$ \\[5pt] 
WISE 1318$-$1758 & 199.641070 &  $-$17.974002 & 7.37 & 6.97 & $-$514.59 $\pm$ 7.20 &  3.70$\pm$6.86   &  48.06 $\pm$ 7.33  & 20.81$\substack{+3.74 \\ -2.75}$ & 25 &15 &   7 & $30.79/15=1.23$ \\[5pt] 
WISE 1405+5534   & 211.322480 &  55.572793  & 14.60 & 8.54 & $-$2336.04$\pm$ 6.91 &  238.02$\pm$7.40  &  144.35 $\pm$ 8.60 & 6.93 $\substack{+0.44 \\ -0.39}$ & 25 &13 &   7 & $32.84/21=1.56$ \\[5pt] 
WISE 1541$-$2250 & 235.464061 &  $-$22.840554 & 5.03 & 4.51 & $-$895.05 $\pm$ 4.68 &  $-$94.73$\pm$4.66  &  167.05 $\pm$ 4.19 & 5.99 $\substack{+0.154 \\ -0.147}$ & 15 &15 &  26 & $24.38/25=0.98$ \\[5pt] 
WISE 1639$-$6847 & 249.921736 &  $-$68.797280 & 22.37 & 7.68 &  579.09 $\pm$ 12.52 & $-$3104.54$\pm$12.25 &  228.05 $\pm$ 8.93 & 4.39$\substack{+0.18 \\ -0.17}$ & 25 &12 & 96 & $23.94/19=1.26$ \\[5pt] 
WISE 1738+2732   & 264.648443 &  27.549315  & 5.32 & 4.59 &  343.27 $\pm$ 3.45 & $-$340.63$\pm$3.35  &  136.26 $\pm$ 4.27 & 7.34$\substack{+0.24 \\ -0.22}$ & 15 &16 &  13 & $28.95/27=1.07$ \\[5pt] 
WISE 1828+2650   & 277.130717 &  26.844012  & 5.04 & 4.58 & 1020.99 $\pm$ 3.20 & 175.55$\pm$3.09   &  100.21 $\pm$ 4.23  & 9.98$\substack{+0.44 \\ -0.40}$ & 15 &16 &  31 & $30.58/27=1.13$ \\[5pt] 
WISE 2056+1459   & 314.121287 &  14.998666  & 4.54 & 4.39 &  822.99 $\pm$ 3.37 & 535.72$\pm$3.36   &  138.32 $\pm$ 3.86 & 7.23$\substack{+0.21 \\ -0.20}$ & 15 &16 &  38 & $18.19/27=0.67$ \\[5pt] 
WISE 2209+2711   & 332.275281 &  27.194171  & 6.58 & 5.95 & 1199.55$\pm$ 4.94  & $-$1359.00$\pm$4.76 &  154.41 $\pm$ 5.67 & 6.48$\substack{+0.25 \\ -0.23}$ & 20 &15 &  15 &$23.41/25=0.94$ \\[5pt] 
WISE 2220$-$3628 & 335.230875 &  $-$36.471639 & 7.61 & 6.00 &  292.91 $\pm$ 7.43 &  $-$61.46$\pm$7.04  &  84.10 $\pm$ 5.90  & 11.89$\substack{+0.90 \\ -0.78}$ & 20 &14 &  13 &$22.71/23=0.99$ \\[5pt] 
\enddata
\end{deluxetable}
\end{rotatetable}
\clearpage

\figsetstart
\figsetnum{5}
\figsettitle{Astrometric Fits}

\figsetgrpstart
\figsetgrpnum{5.1}
\figsetgrptitle{}
\figsetplot{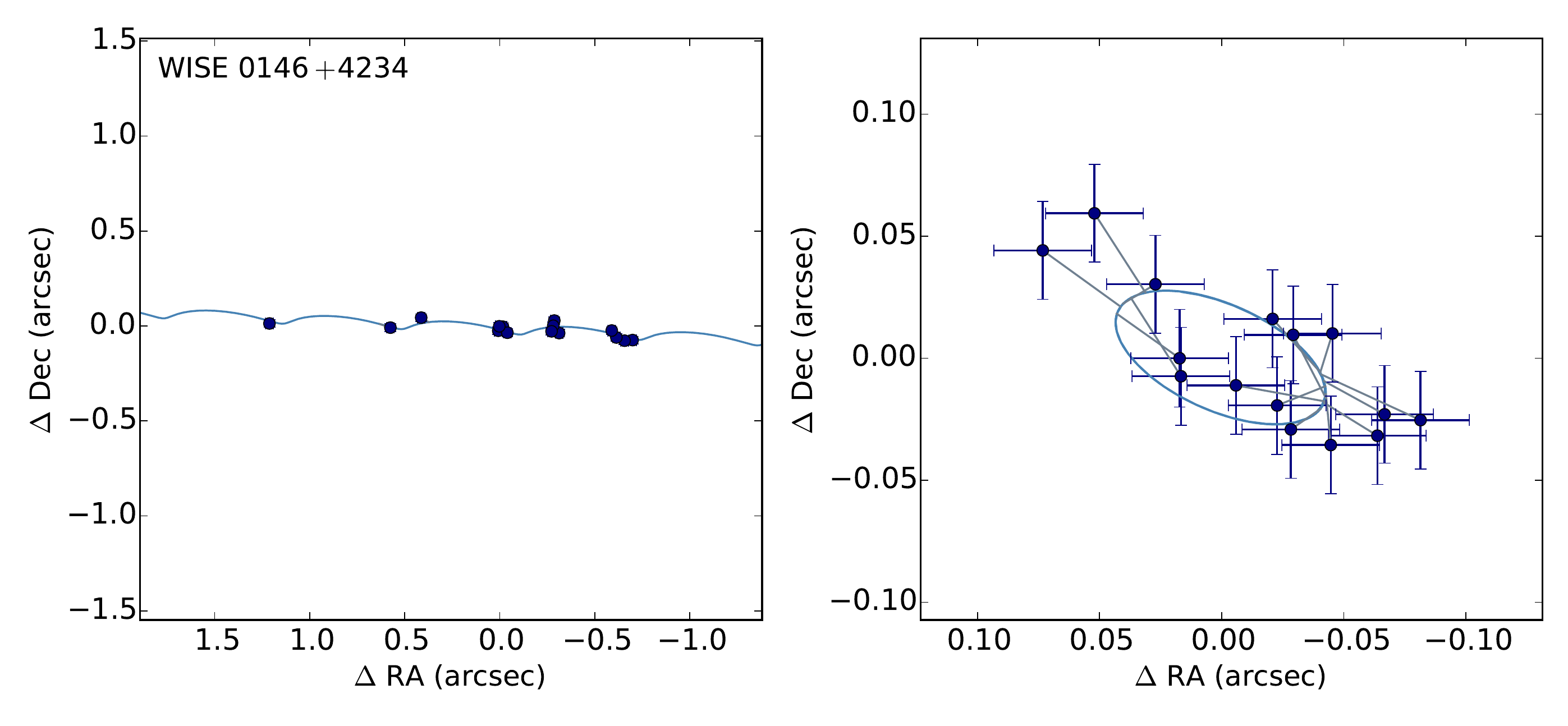}
\figsetgrpnote{Astrometric fits for each of our targets. We maintained a square scaling for the $\Delta$ Declination and $\Delta$ RA. Our observations are plotted in navy and the best-fit astrometric model is plotted in light blue. The left plots include proper motion and parallax and the right plots have proper-motion removed. Note the differing scales between the left and right plots. WISE 0146+4234 is an un-resolved binary, which produces systematic offsets of our astrometry and causes the parallactic ellipse to appear smaller than it is.}
\figsetgrpend

\figsetgrpstart
\figsetgrpnum{5.2}
\figsetgrptitle{}
\figsetplot{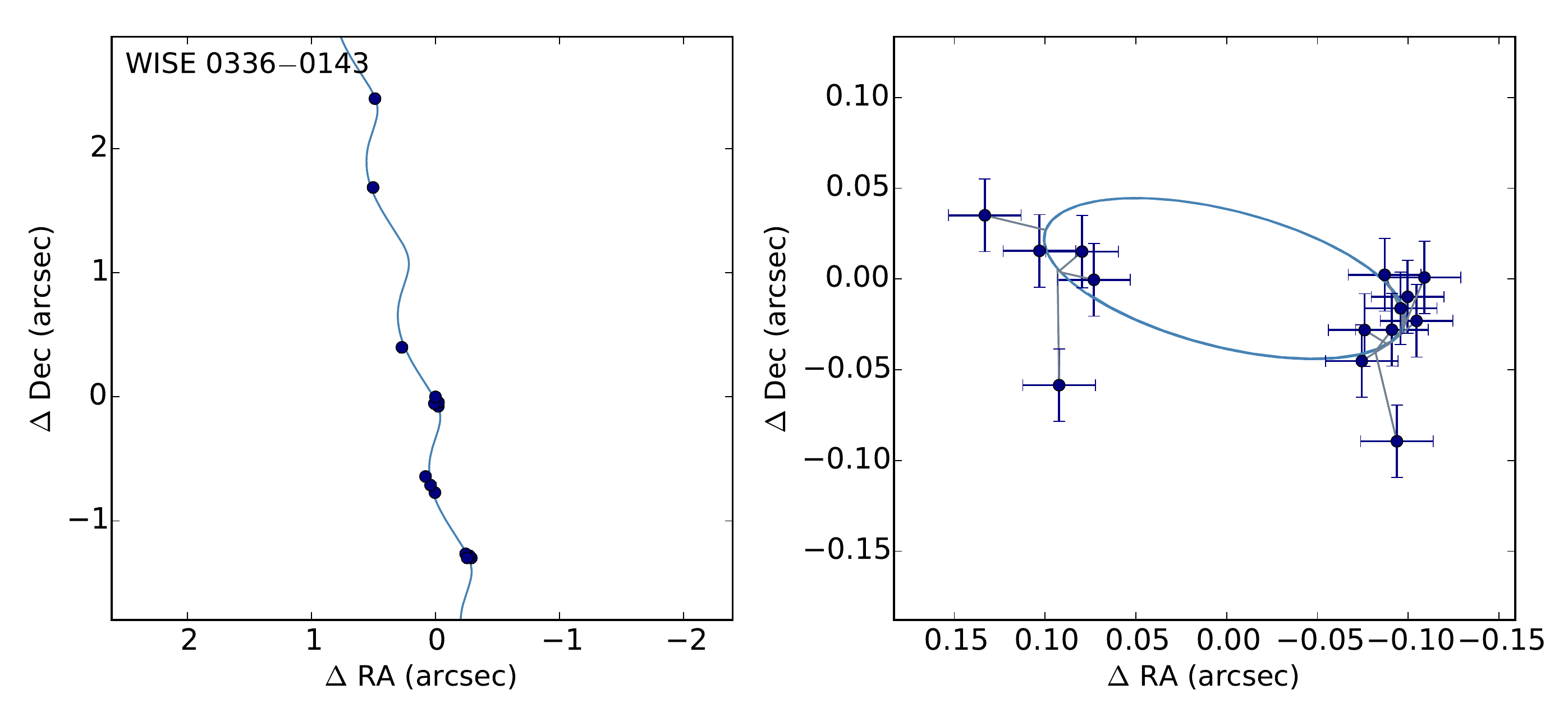}
\figsetgrpnote{Astrometric fits for each of our targets. We maintained a square scaling for the $\Delta$ Declination and $\Delta$ RA. Our observations are plotted in navy and the best-fit astrometric model is plotted in light blue. The left plots include proper motion and parallax and the right plots have proper-motion removed. Note the differing scales between the left and right plots.}
\figsetgrpend

\figsetgrpstart
\figsetgrpnum{5.3}
\figsetgrptitle{}
\figsetplot{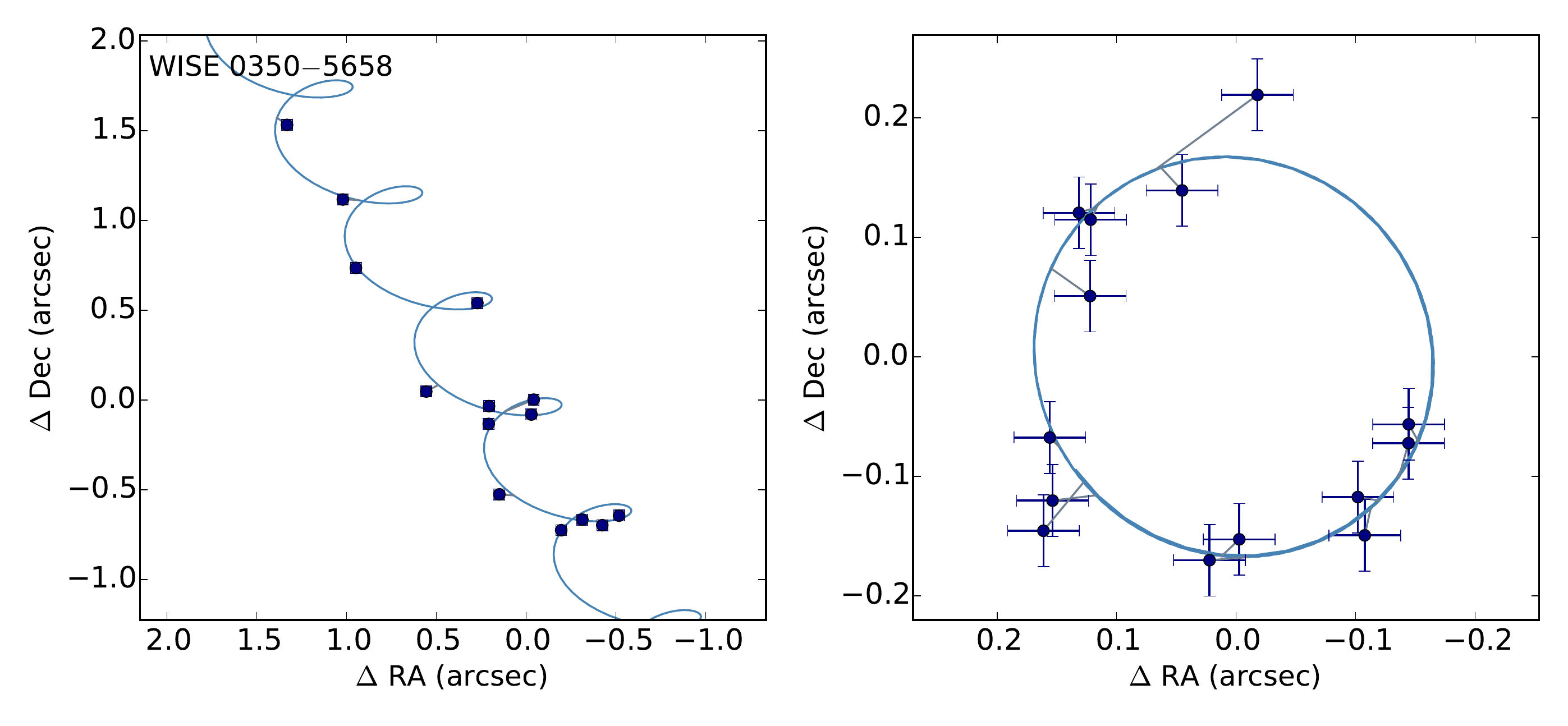}
\figsetgrpnote{Astrometric fits for each of our targets. We maintained a square scaling for the $\Delta$ Declination and $\Delta$ RA. Our observations are plotted in navy and the best-fit astrometric model is plotted in light blue. The left plots include proper motion and parallax and the right plots have proper-motion removed. Note the differing scales between the left and right plots.}
\figsetgrpend

\figsetgrpstart
\figsetgrpnum{5.4}
\figsetgrptitle{}
\figsetplot{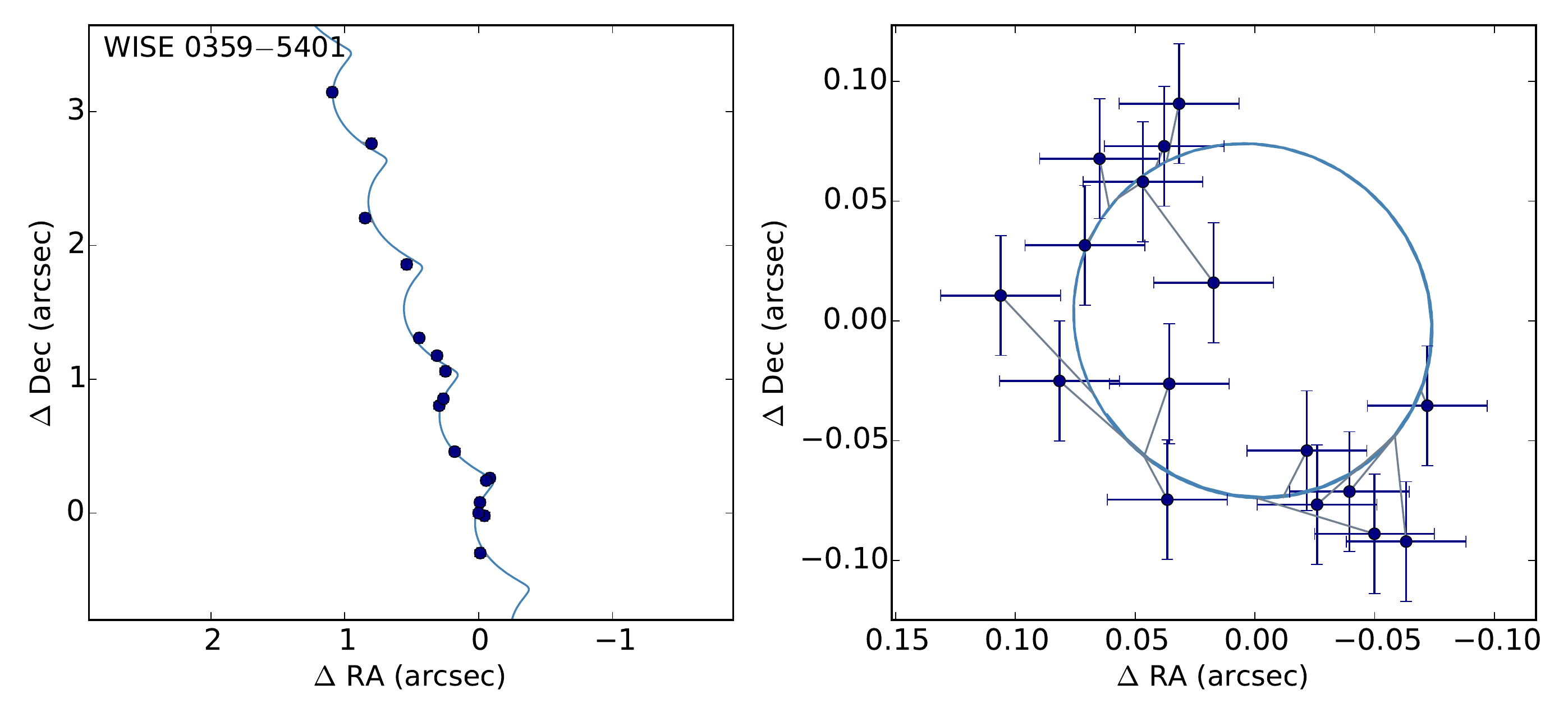}
\figsetgrpnote{Astrometric fits for each of our targets. We maintained a square scaling for the $\Delta$ Declination and $\Delta$ RA. Our observations are plotted in navy and the best-fit astrometric model is plotted in light blue. The left plots include proper motion and parallax and the right plots have proper-motion removed. Note the differing scales between the left and right plots.}
\figsetgrpend

\figsetgrpstart
\figsetgrpnum{5.5}
\figsetgrptitle{}
\figsetplot{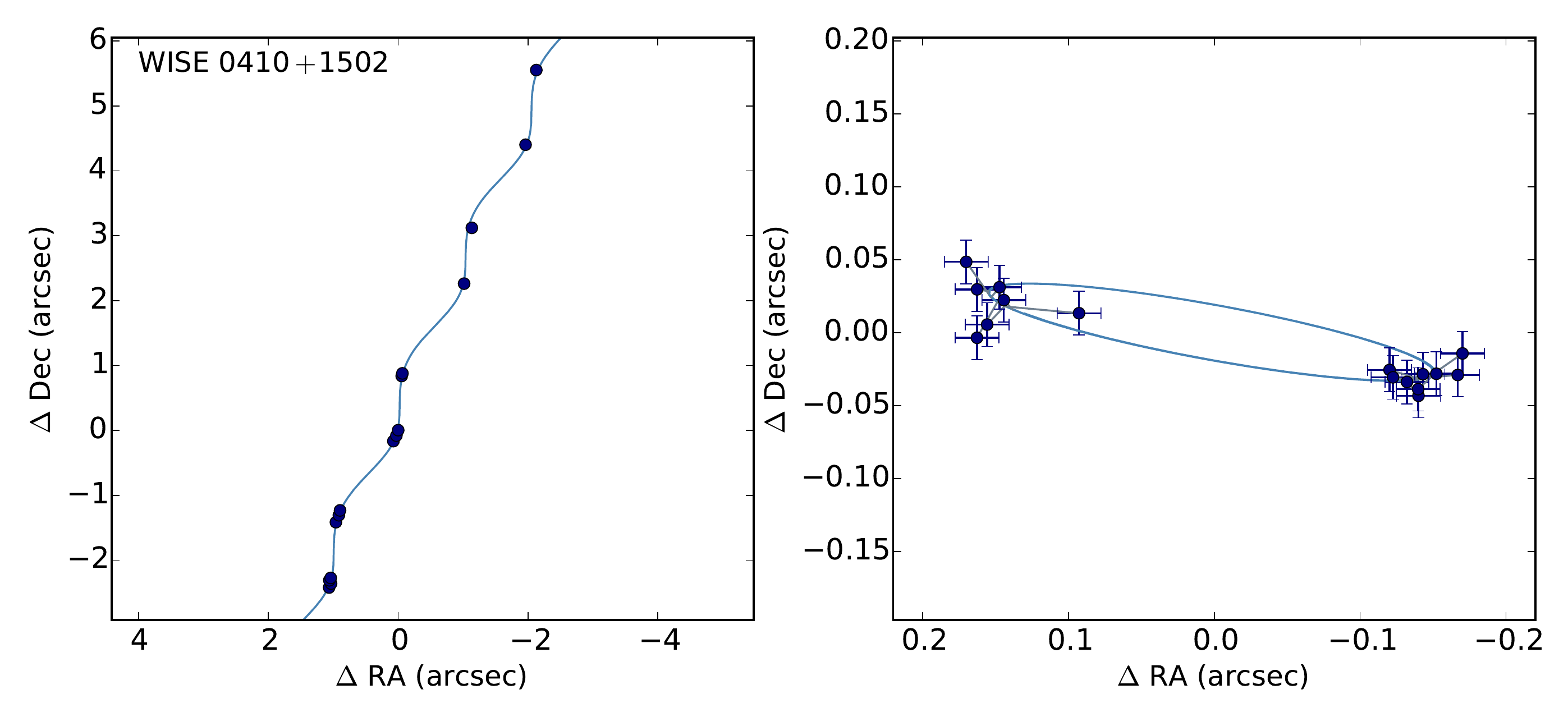}
\figsetgrpnote{Astrometric fits for each of our targets. We maintained a square scaling for the $\Delta$ Declination and $\Delta$ RA. Our observations are plotted in navy and the best-fit astrometric model is plotted in light blue. The left plots include proper motion and parallax and the right plots have proper-motion removed. Note the differing scales between the left and right plots.}
\figsetgrpend

\figsetgrpstart
\figsetgrpnum{5.6}
\figsetgrptitle{}
\figsetplot{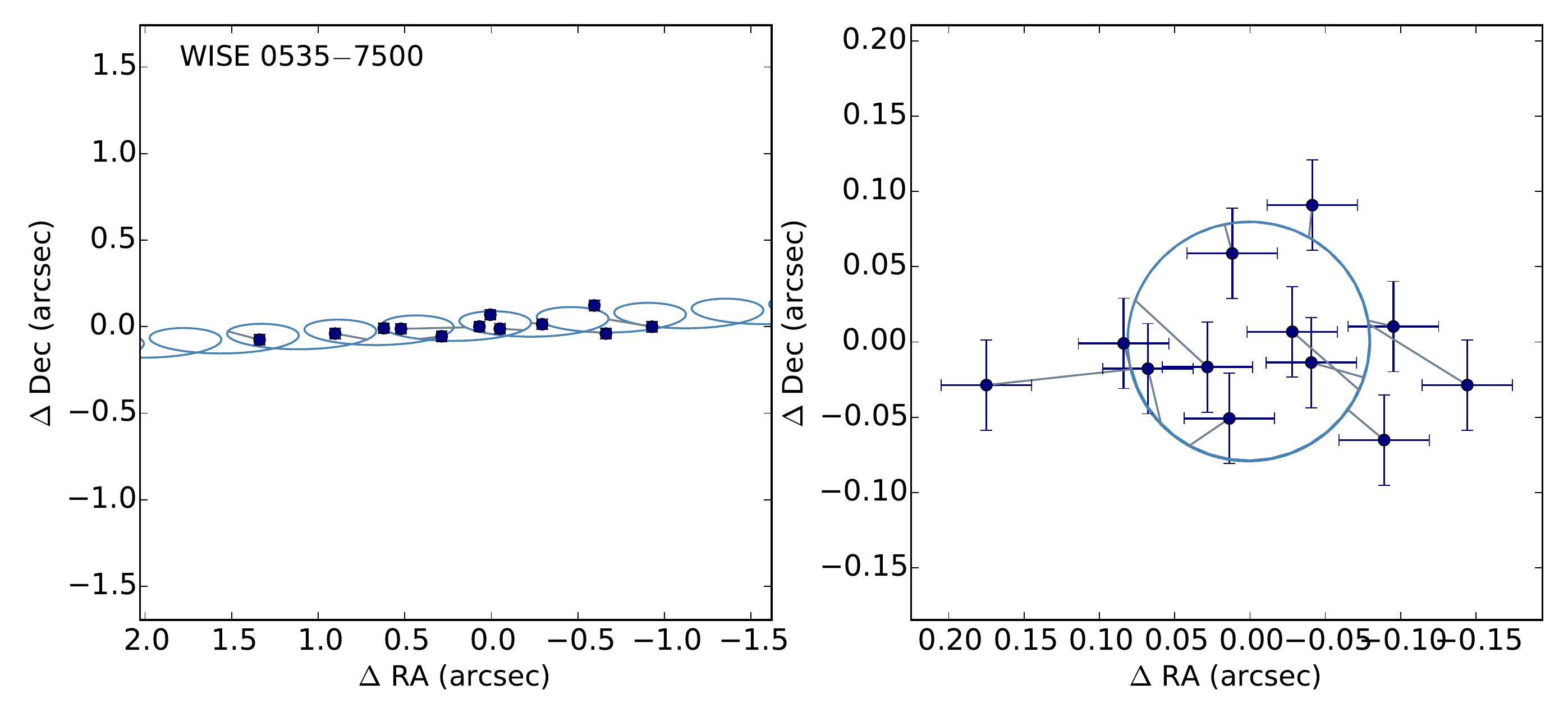}
\figsetgrpnote{Astrometric fits for each of our targets. We maintained a square scaling for the $\Delta$ Declination and $\Delta$ RA. Our observations are plotted in navy and the best-fit astrometric model is plotted in light blue. The left plots include proper motion and parallax and the right plots have proper-motion removed. Note the differing scales between the left and right plots.}
\figsetgrpend

\figsetgrpstart
\figsetgrpnum{5.7}
\figsetgrptitle{}
\figsetplot{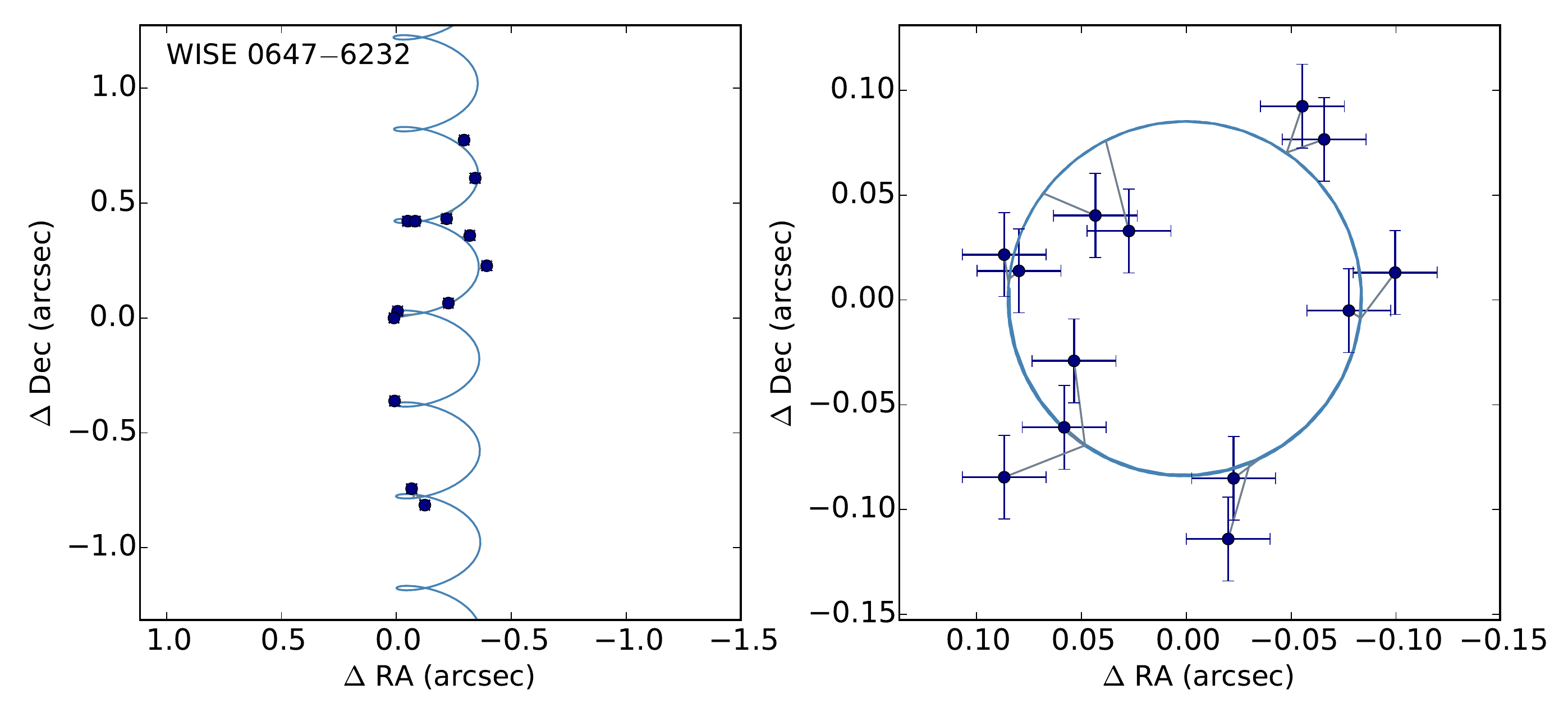}
\figsetgrpnote{Astrometric fits for each of our targets. We maintained a square scaling for the $\Delta$ Declination and $\Delta$ RA. Our observations are plotted in navy and the best-fit astrometric model is plotted in light blue. The left plots include proper motion and parallax and the right plots have proper-motion removed. Note the differing scales between the left and right plots.}
\figsetgrpend

\figsetgrpstart
\figsetgrpnum{5.8}
\figsetgrptitle{}
\figsetplot{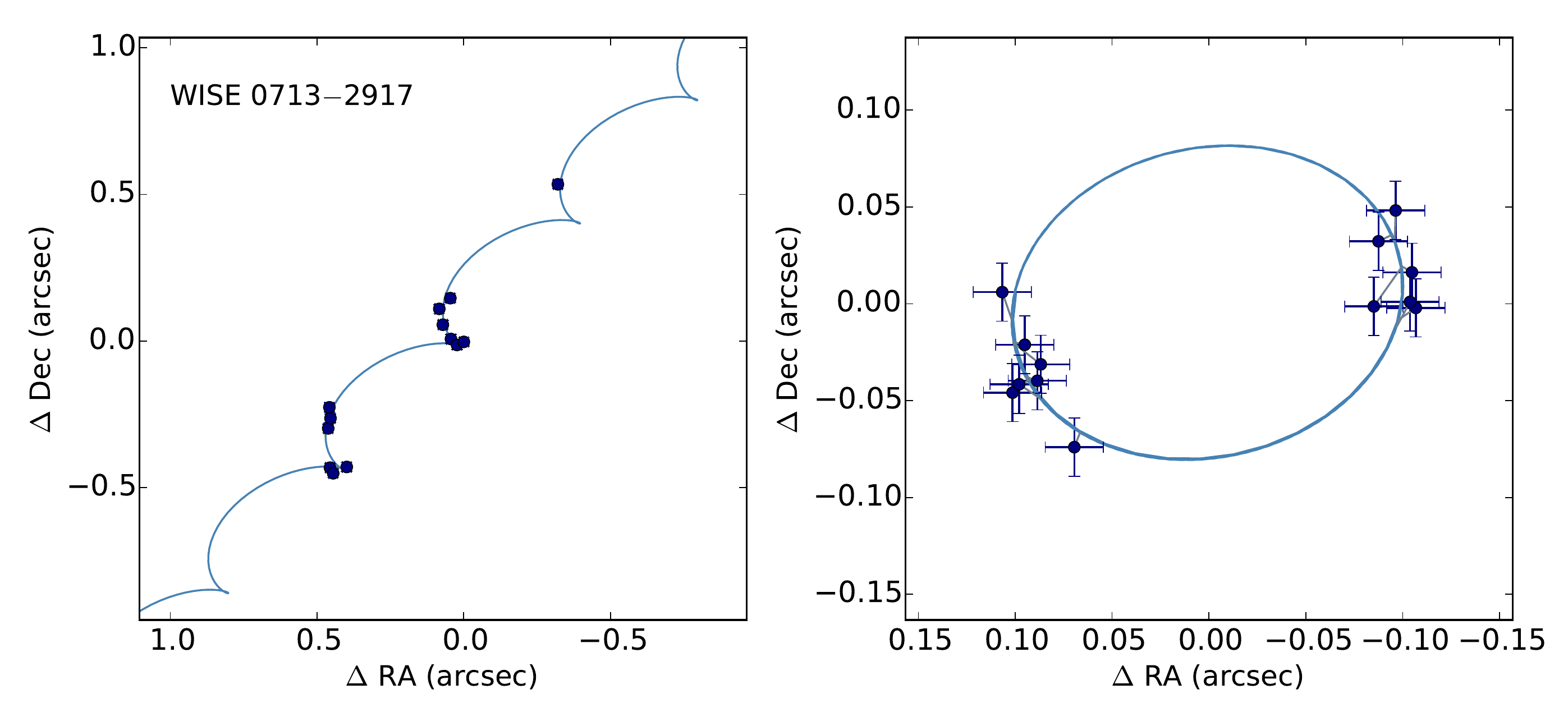}
\figsetgrpnote{Astrometric fits for each of our targets. We maintained a square scaling for the $\Delta$ Declination and $\Delta$ RA. Our observations are plotted in navy and the best-fit astrometric model is plotted in light blue. The left plots include proper motion and parallax and the right plots have proper-motion removed. Note the differing scales between the left and right plots.}
\figsetgrpend

\figsetgrpstart
\figsetgrpnum{5.9}
\figsetgrptitle{}
\figsetplot{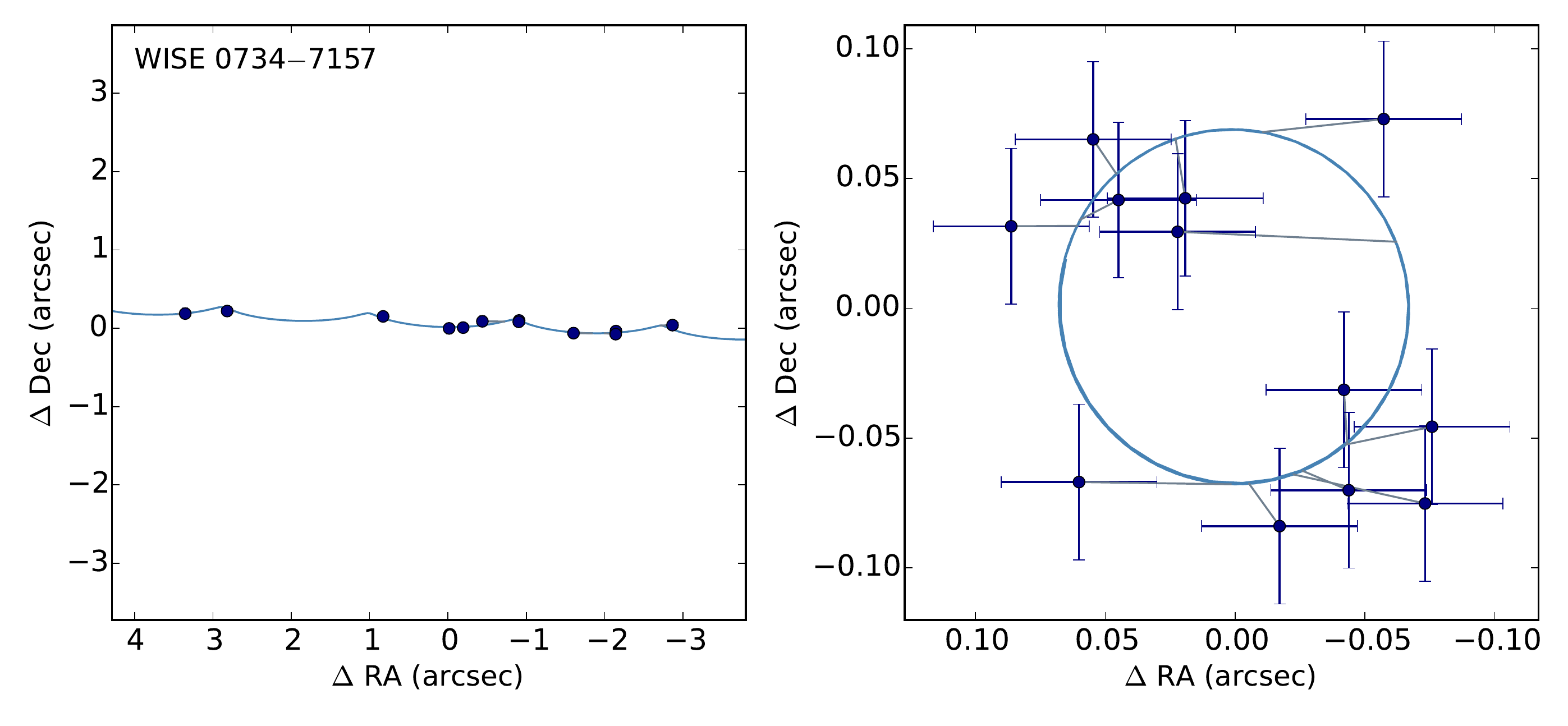}
\figsetgrpnote{Astrometric fits for each of our targets. We maintained a square scaling for the $\Delta$ Declination and $\Delta$ RA. Our observations are plotted in navy and the best-fit astrometric model is plotted in light blue. The left plots include proper motion and parallax and the right plots have proper-motion removed. Note the differing scales between the left and right plots.}
\figsetgrpend

\figsetgrpstart
\figsetgrpnum{5.10}
\figsetgrptitle{}
\figsetplot{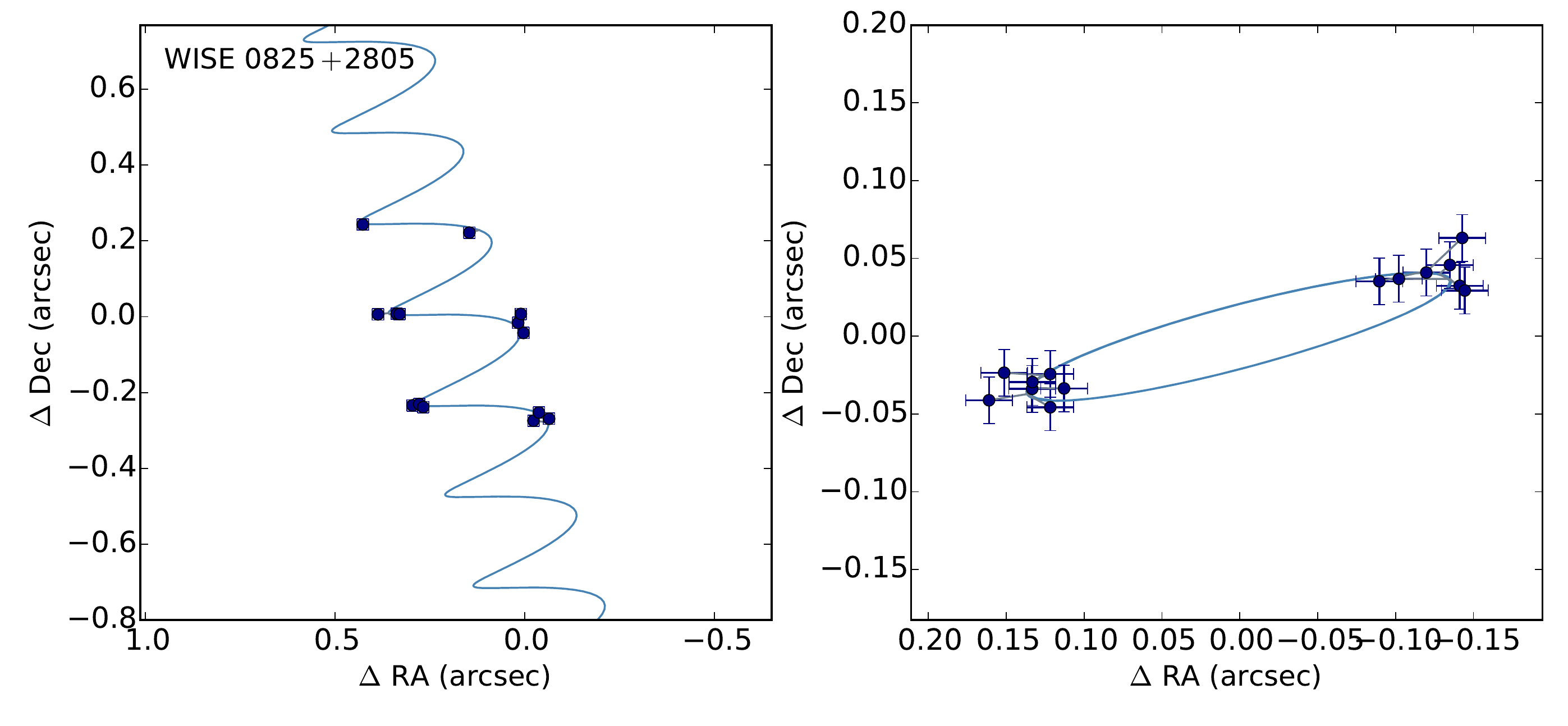}
\figsetgrpnote{Astrometric fits for each of our targets. We maintained a square scaling for the $\Delta$ Declination and $\Delta$ RA. Our observations are plotted in navy and the best-fit astrometric model is plotted in light blue. The left plots include proper motion and parallax and the right plots have proper-motion removed. Note the differing scales between the left and right plots.}
\figsetgrpend

\figsetgrpstart
\figsetgrpnum{5.11}
\figsetgrptitle{}
\figsetplot{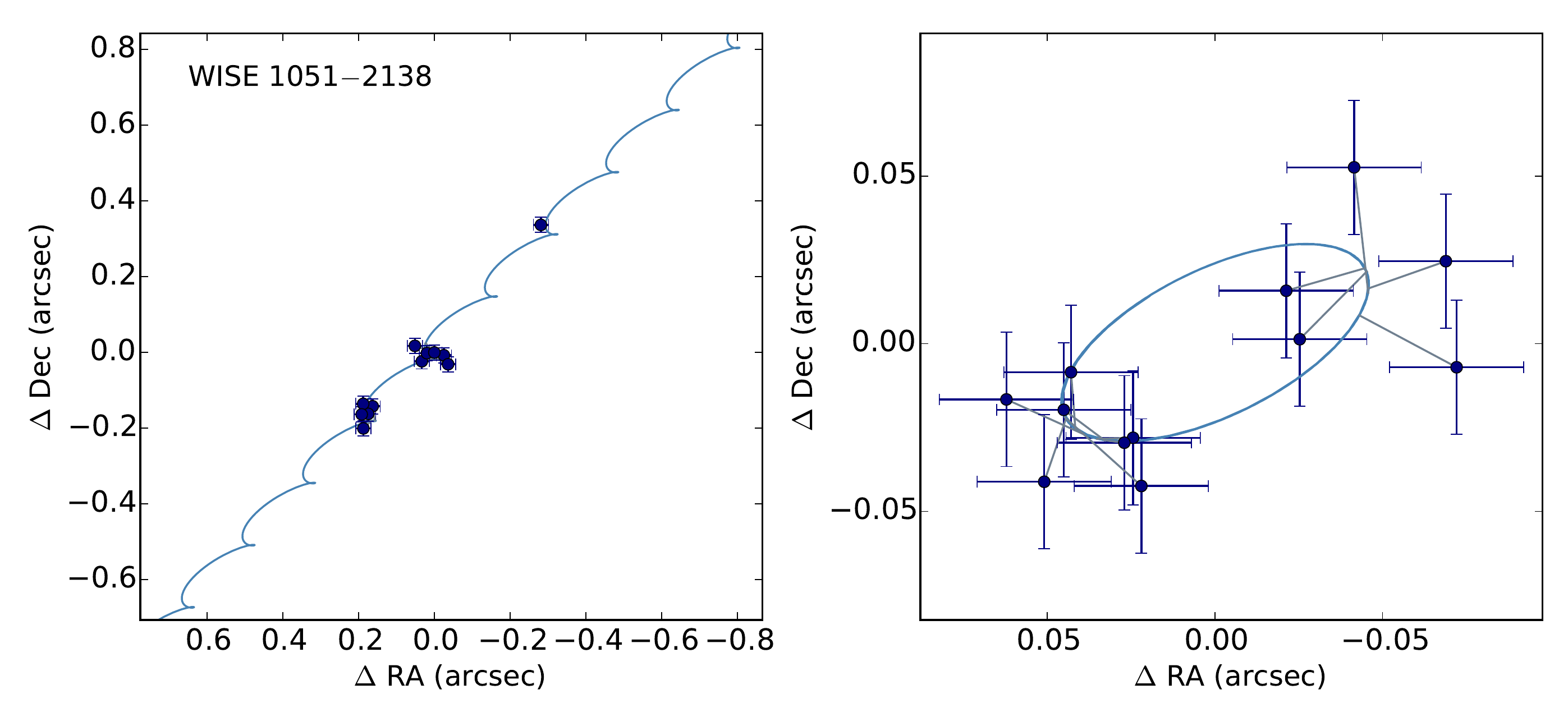}
\figsetgrpnote{Astrometric fits for each of our targets. We maintained a square scaling for the $\Delta$ Declination and $\Delta$ RA. Our observations are plotted in navy and the best-fit astrometric model is plotted in light blue. The left plots include proper motion and parallax and the right plots have proper-motion removed. Note the differing scales between the left and right plots.}
\figsetgrpend

\figsetgrpstart
\figsetgrpnum{5.12}
\figsetgrptitle{}
\figsetplot{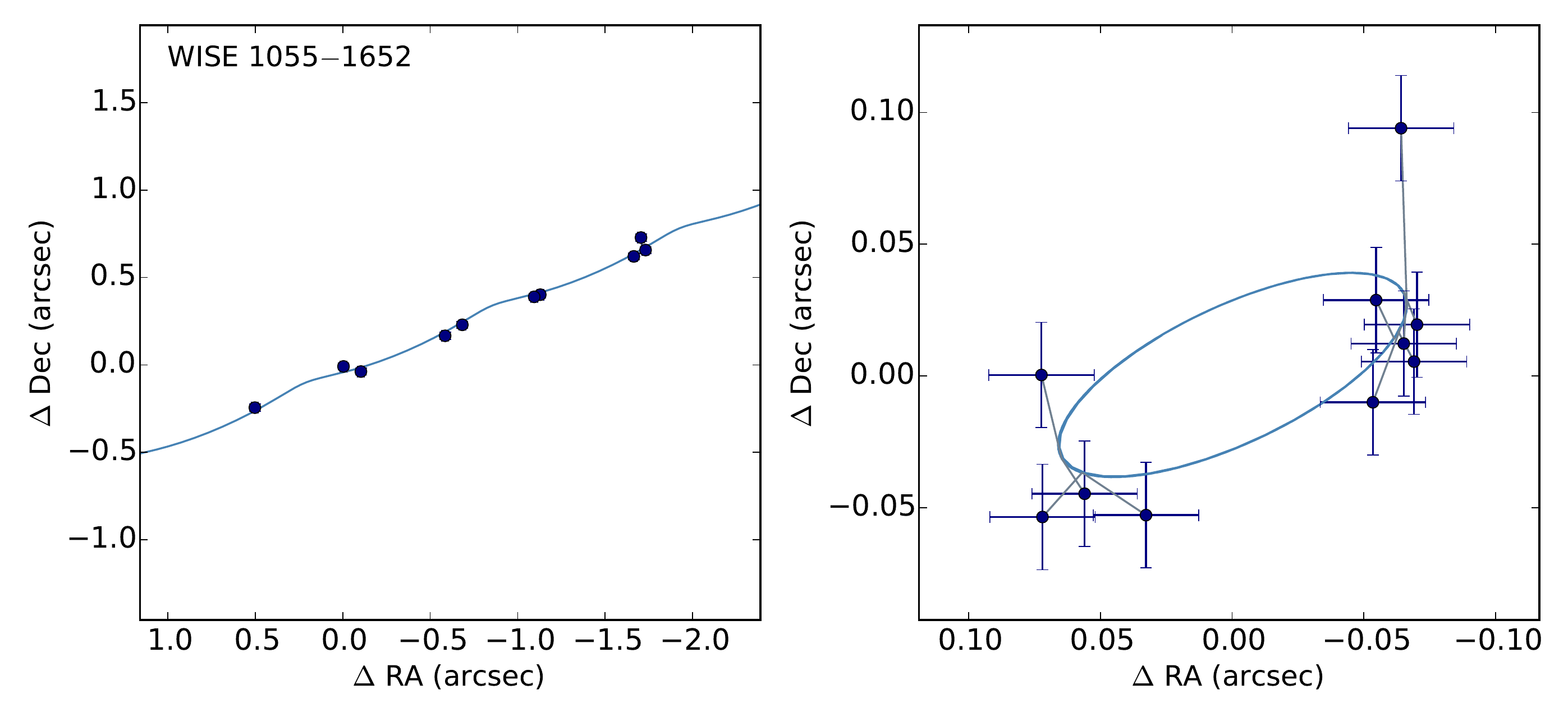}
\figsetgrpnote{Astrometric fits for each of our targets. We maintained a square scaling for the $\Delta$ Declination and $\Delta$ RA. Our observations are plotted in navy and the best-fit astrometric model is plotted in light blue. The left plots include proper motion and parallax and the right plots have proper-motion removed. Note the differing scales between the left and right plots.}
\figsetgrpend

\figsetgrpstart
\figsetgrpnum{5.13}
\figsetgrptitle{}
\figsetplot{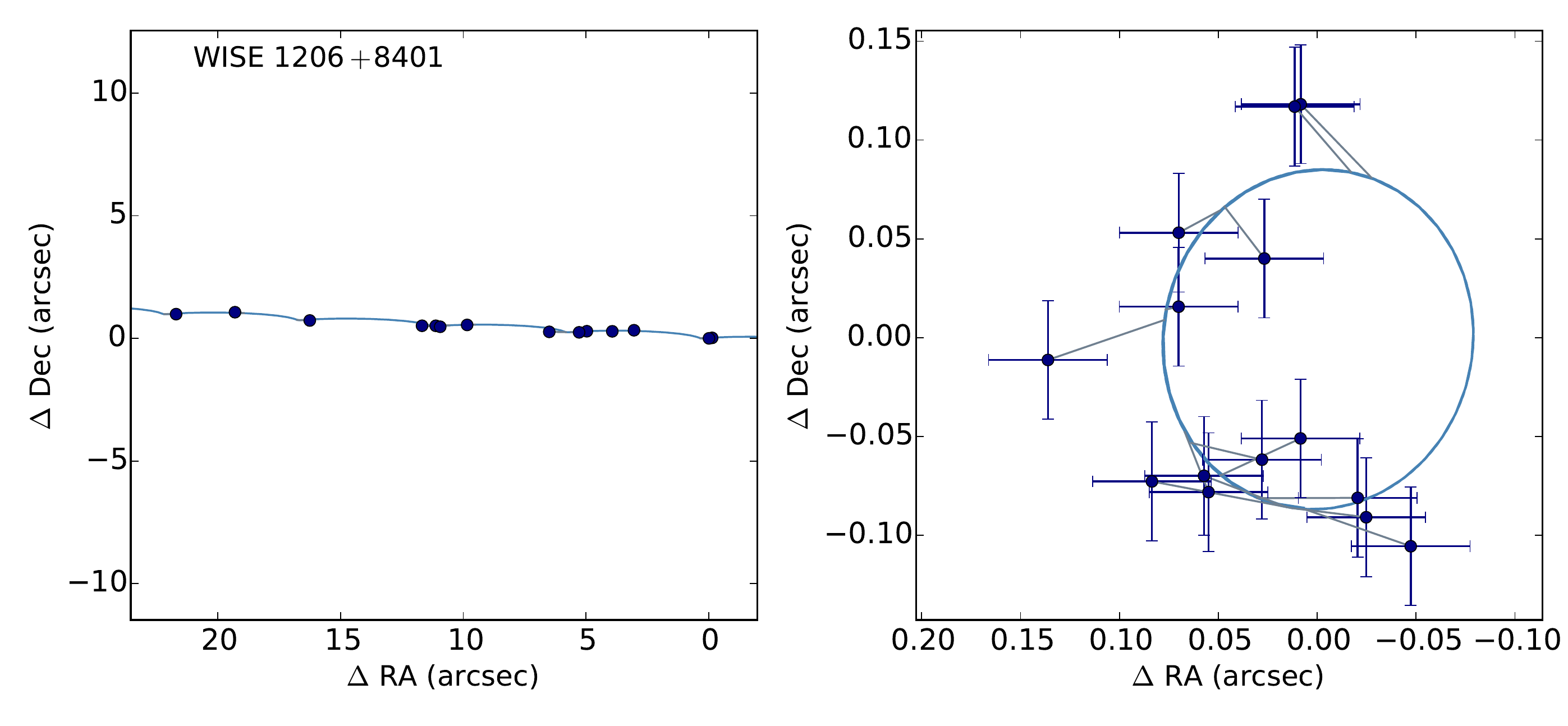}
\figsetgrpnote{Astrometric fits for each of our targets. We maintained a square scaling for the $\Delta$ Declination and $\Delta$ RA. Our observations are plotted in navy and the best-fit astrometric model is plotted in light blue. The left plots include proper motion and parallax and the right plots have proper-motion removed. Note the differing scales between the left and right plots.}
\figsetgrpend

\figsetgrpstart
\figsetgrpnum{5.14}
\figsetgrptitle{}
\figsetplot{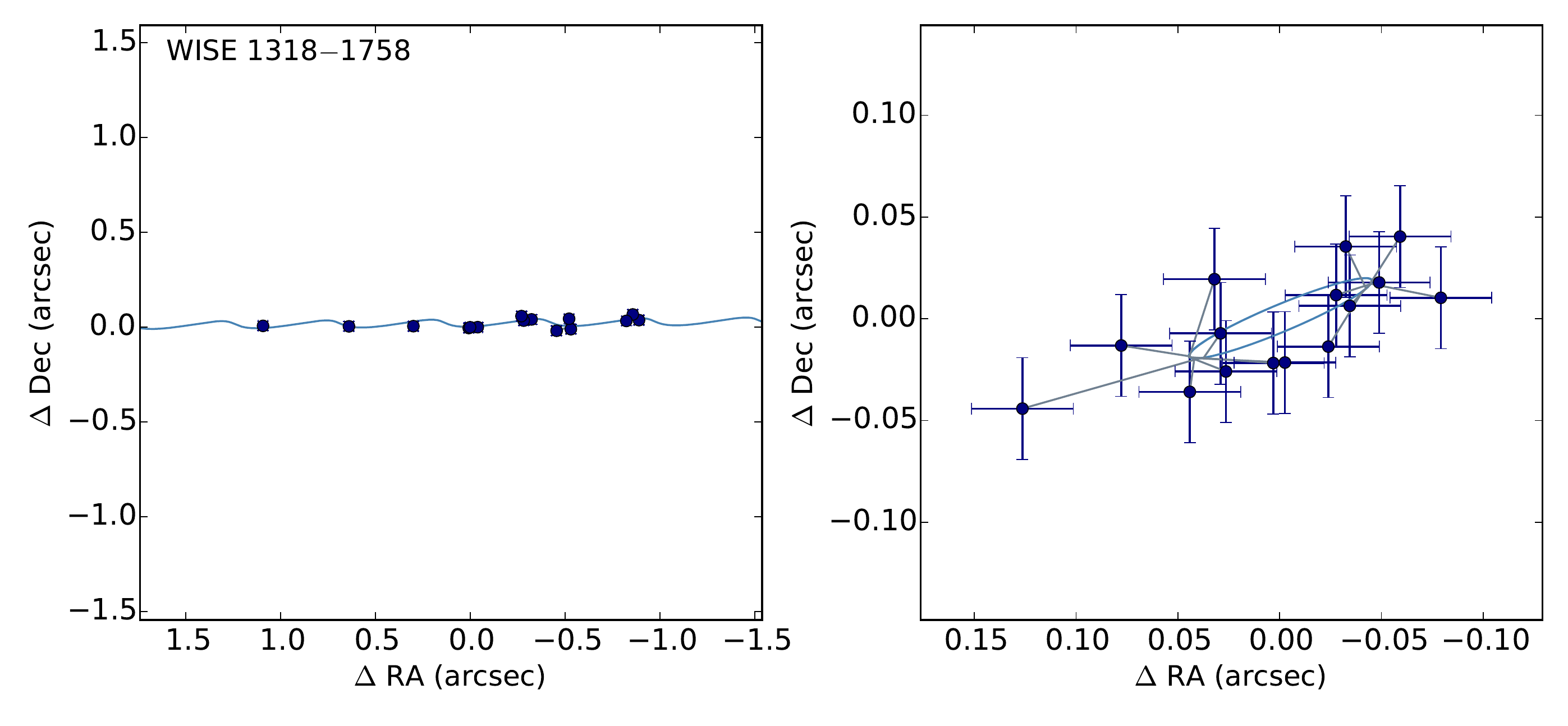}
\figsetgrpnote{Astrometric fits for each of our targets. We maintained a square scaling for the $\Delta$ Declination and $\Delta$ RA. Our observations are plotted in navy and the best-fit astrometric model is plotted in light blue. The left plots include proper motion and parallax and the right plots have proper-motion removed. Note the differing scales between the left and right plots. }
\figsetgrpend

\figsetgrpstart
\figsetgrpnum{5.15}
\figsetgrptitle{}
\figsetplot{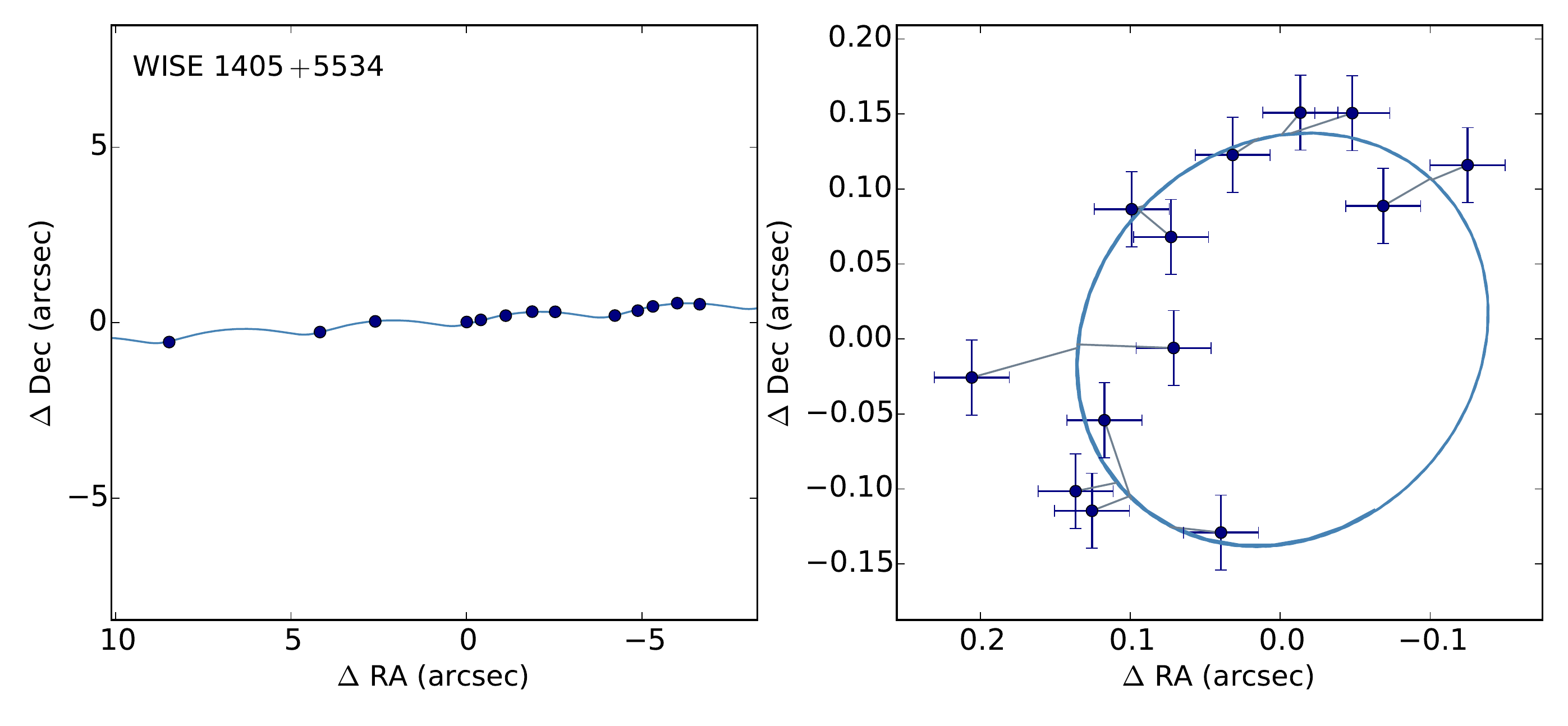}
\figsetgrpnote{Astrometric fits for each of our targets. We maintained a square scaling for the $\Delta$ Declination and $\Delta$ RA. Our observations are plotted in navy and the best-fit astrometric model is plotted in light blue. The left plots include proper motion and parallax and the right plots have proper-motion removed. Note the differing scales between the left and right plots. }
\figsetgrpend

\figsetgrpstart
\figsetgrpnum{5.16}
\figsetgrptitle{}
\figsetplot{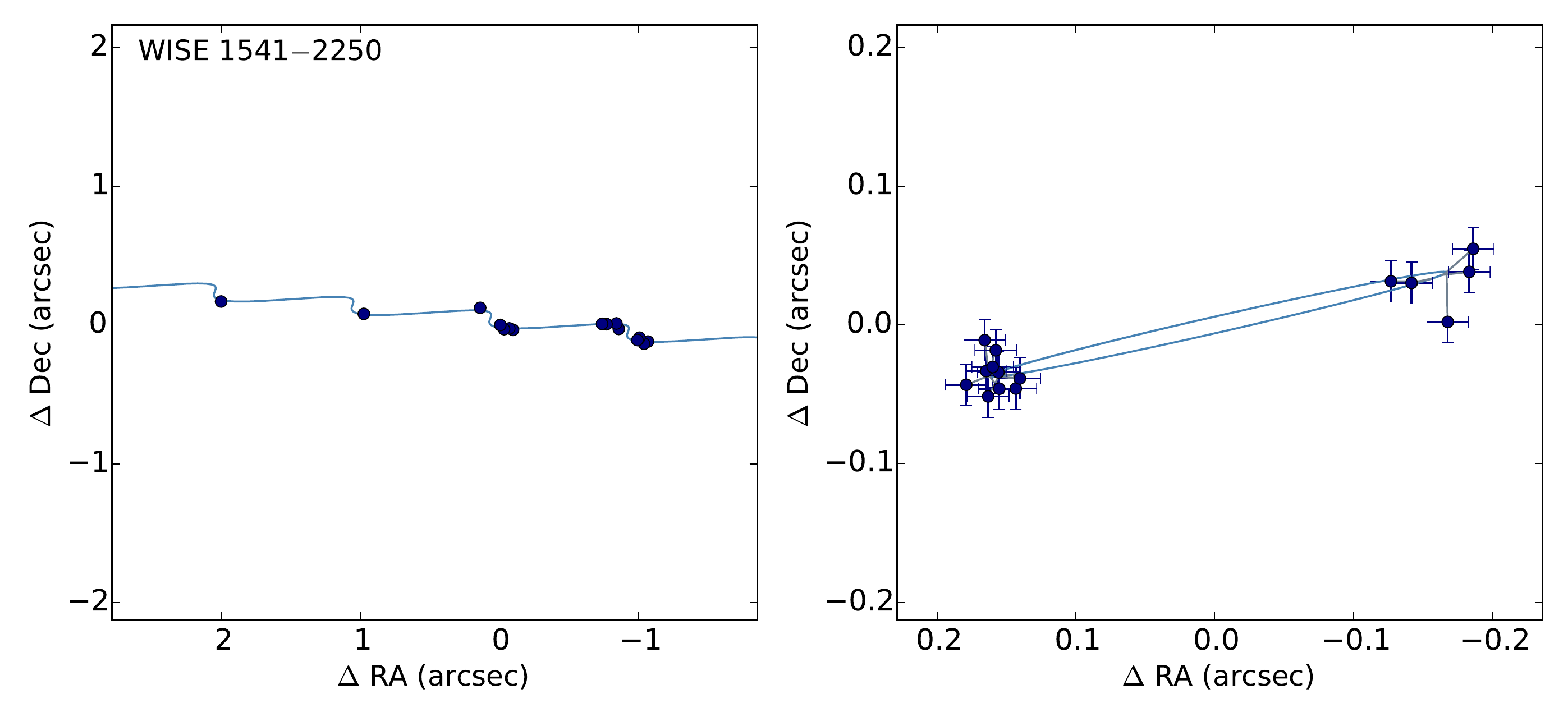}
\figsetgrpnote{Astrometric fits for each of our targets. We maintained a square scaling for the $\Delta$ Declination and $\Delta$ RA. Our observations are plotted in navy and the best-fit astrometric model is plotted in light blue. The left plots include proper motion and parallax and the right plots have proper-motion removed. Note the differing scales between the left and right plots. }
\figsetgrpend

\figsetgrpstart
\figsetgrpnum{5.17}
\figsetgrptitle{Title 17}
\figsetplot{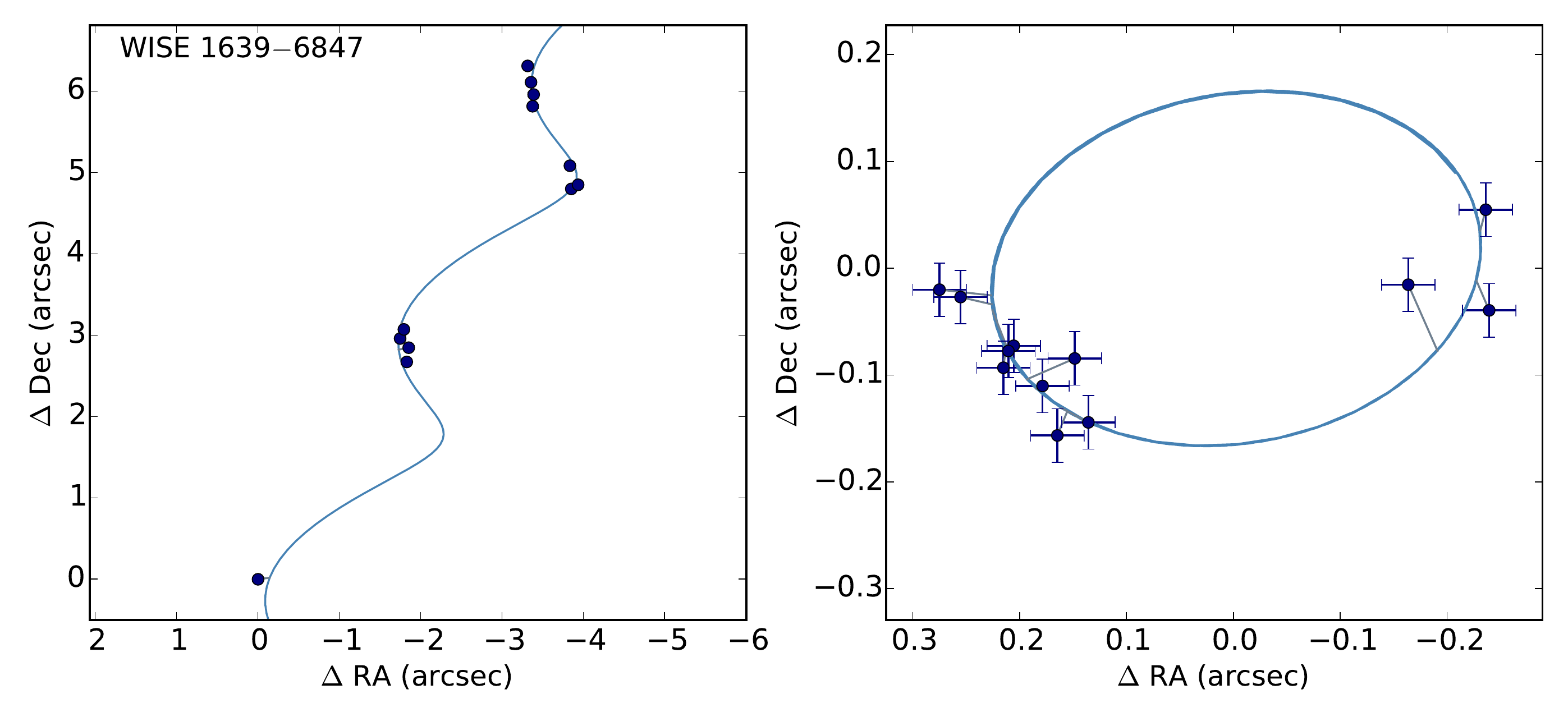}
\figsetgrpnote{Astrometric fits for each of our targets. We maintained a square scaling for the $\Delta$ Declination and $\Delta$ RA. Our observations are plotted in navy and the best-fit astrometric model is plotted in light blue. The left plots include proper motion and parallax and the right plots have proper-motion removed. Note the differing scales between the left and right plots. WISE 0146+4234 is an un-resolved binary, which produces systematic offsets of our astrometry and causes the parallactic ellipse to appear smaller than it is.}
\figsetgrpend

\figsetgrpstart
\figsetgrpnum{5.18}
\figsetgrptitle{}
\figsetplot{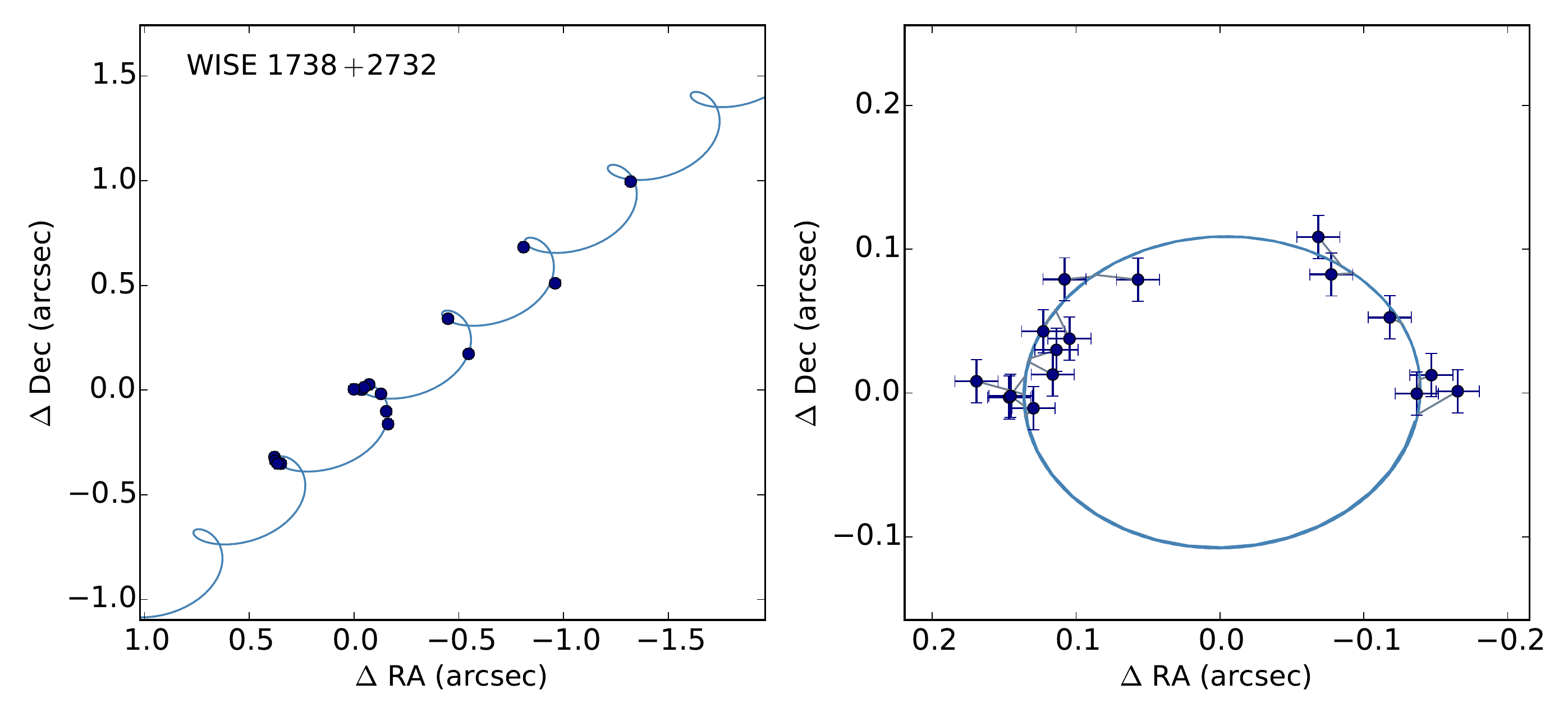}
\figsetgrpnote{Astrometric fits for each of our targets. We maintained a square scaling for the $\Delta$ Declination and $\Delta$ RA. Our observations are plotted in navy and the best-fit astrometric model is plotted in light blue. The left plots include proper motion and parallax and the right plots have proper-motion removed. Note the differing scales between the left and right plots. }
\figsetgrpend

\figsetgrpstart
\figsetgrpnum{5.19}
\figsetgrptitle{}
\figsetplot{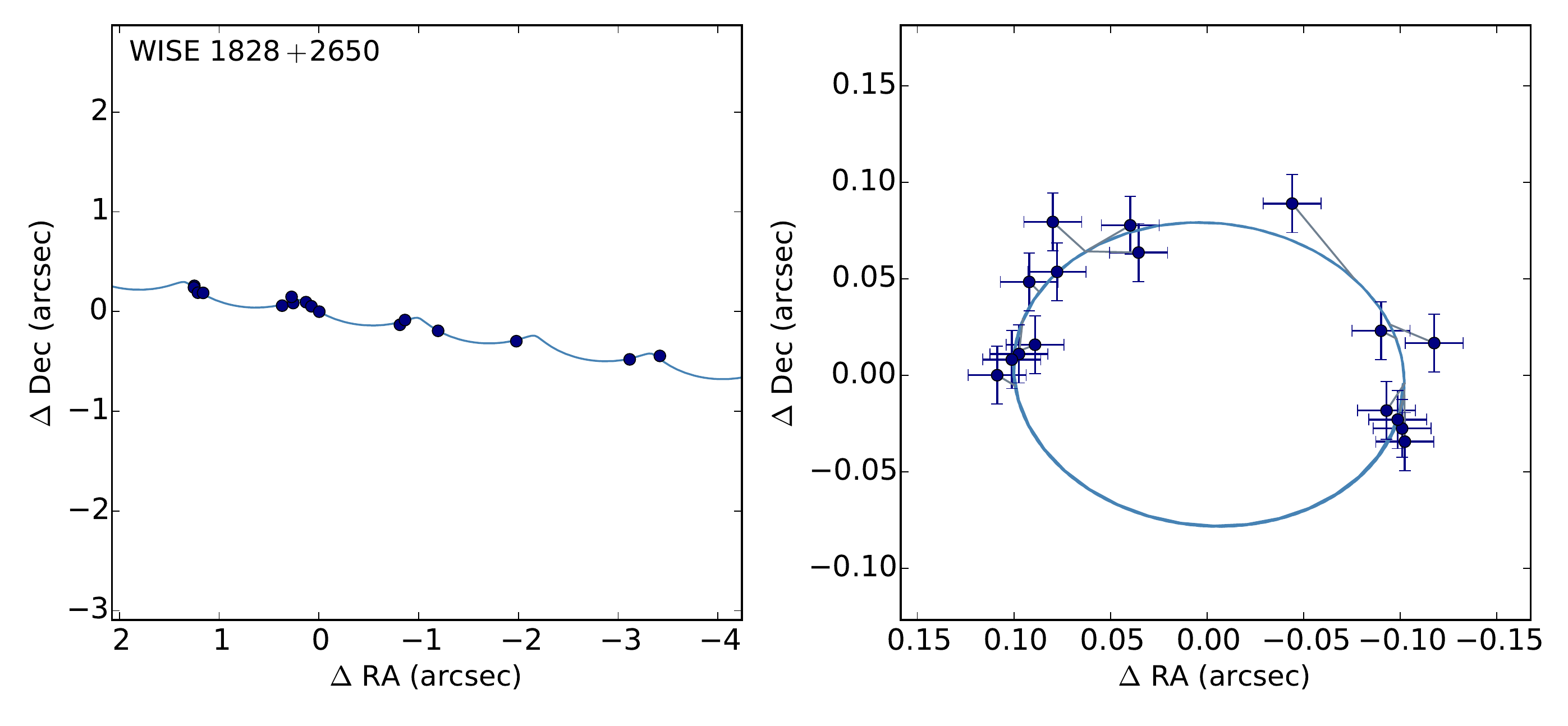}
\figsetgrpnote{Astrometric fits for each of our targets. We maintained a square scaling for the $\Delta$ Declination and $\Delta$ RA. Our observations are plotted in navy and the best-fit astrometric model is plotted in light blue. The left plots include proper motion and parallax and the right plots have proper-motion removed. Note the differing scales between the left and right plots. }
\figsetgrpend

\figsetgrpstart
\figsetgrpnum{5.20}
\figsetgrptitle{}
\figsetplot{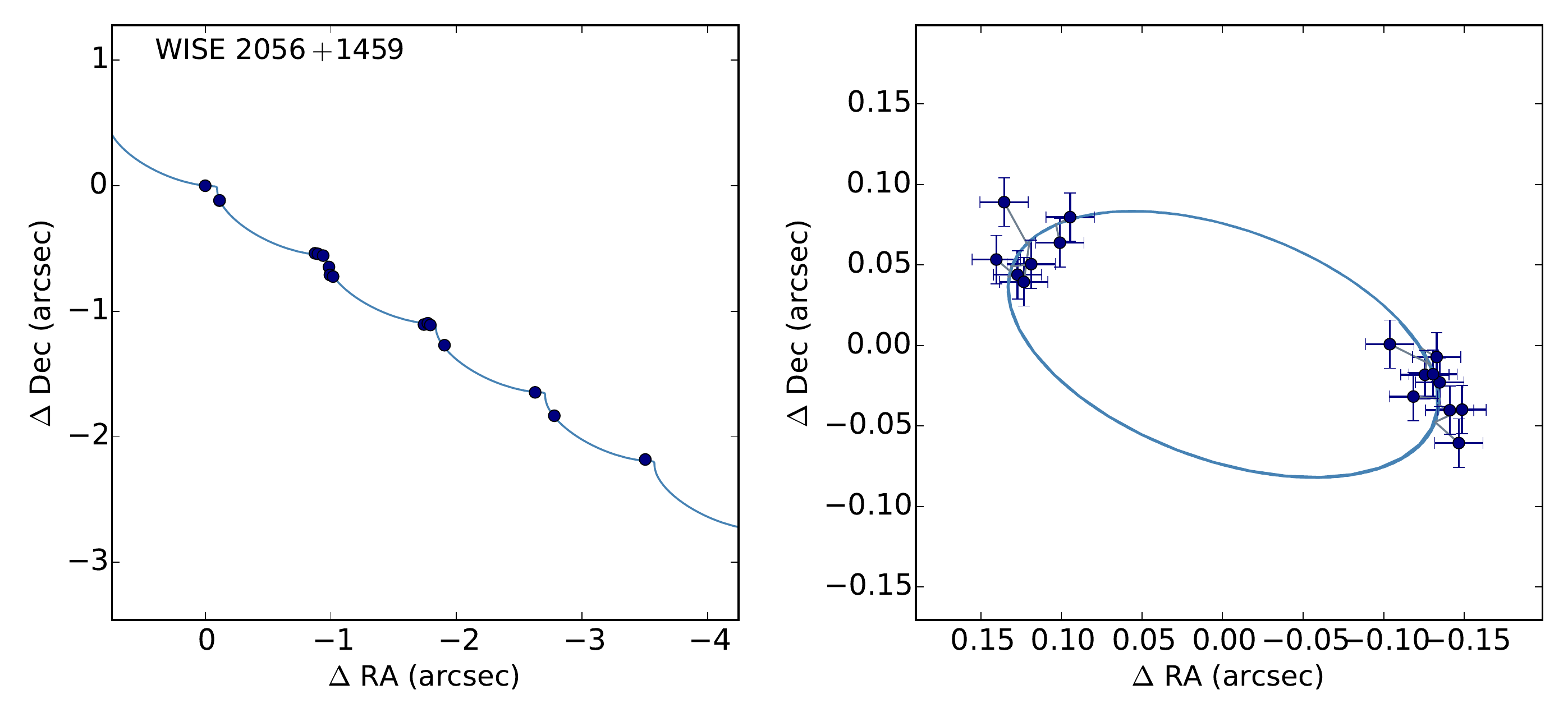}
\figsetgrpnote{Astrometric fits for each of our targets. We maintained a square scaling for the $\Delta$ Declination and $\Delta$ RA. Our observations are plotted in navy and the best-fit astrometric model is plotted in light blue. The left plots include proper motion and parallax and the right plots have proper-motion removed. Note the differing scales between the left and right plots.}
\figsetgrpend

\figsetgrpstart
\figsetgrpnum{5.21}
\figsetgrptitle{}
\figsetplot{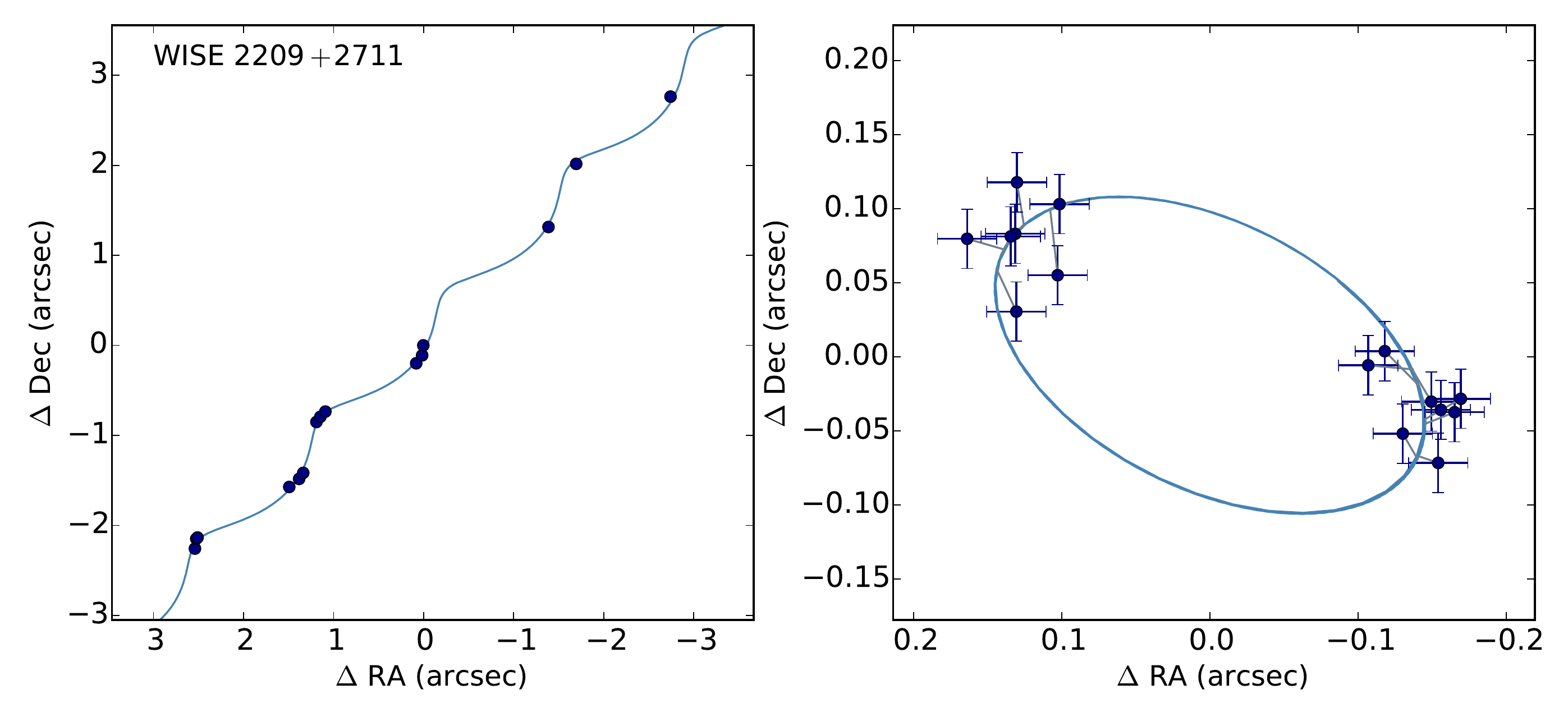}
\figsetgrpnote{Astrometric fits for each of our targets. We maintained a square scaling for the $\Delta$ Declination and $\Delta$ RA. Our observations are plotted in navy and the best-fit astrometric model is plotted in light blue. The left plots include proper motion and parallax and the right plots have proper-motion removed. Note the differing scales between the left and right plots.}
\figsetgrpend

\figsetgrpstart
\figsetgrpnum{5.22}
\figsetgrptitle{}
\figsetplot{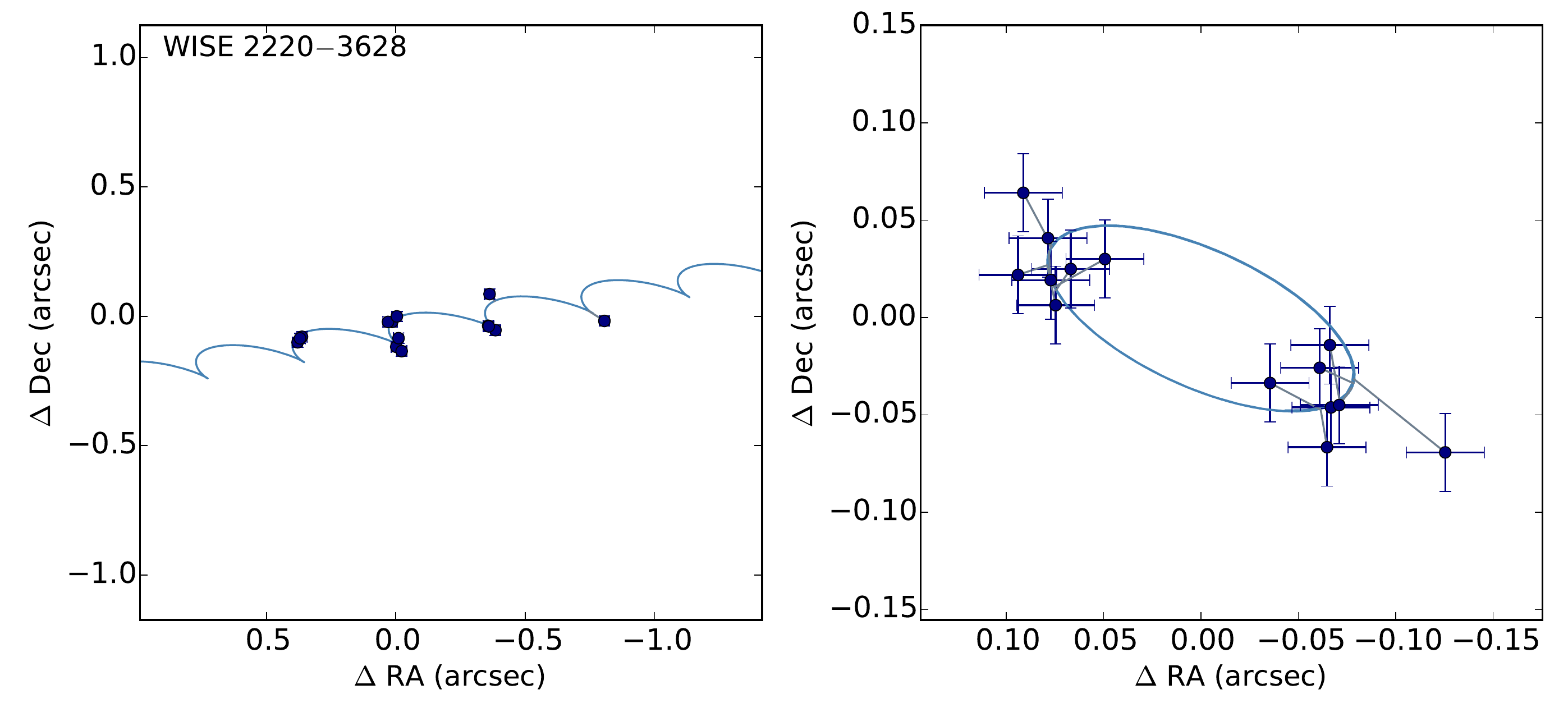}
\figsetgrpnote{Astrometric fits for each of our targets. We maintained a square scaling for the $\Delta$ Declination and $\Delta$ RA. Our observations are plotted in navy and the best-fit astrometric model is plotted in light blue. The left plots include proper motion and parallax and the right plots have proper-motion removed. Note the differing scales between the left and right plots.}
\figsetgrpend

\figsetend

\begin{figure*}
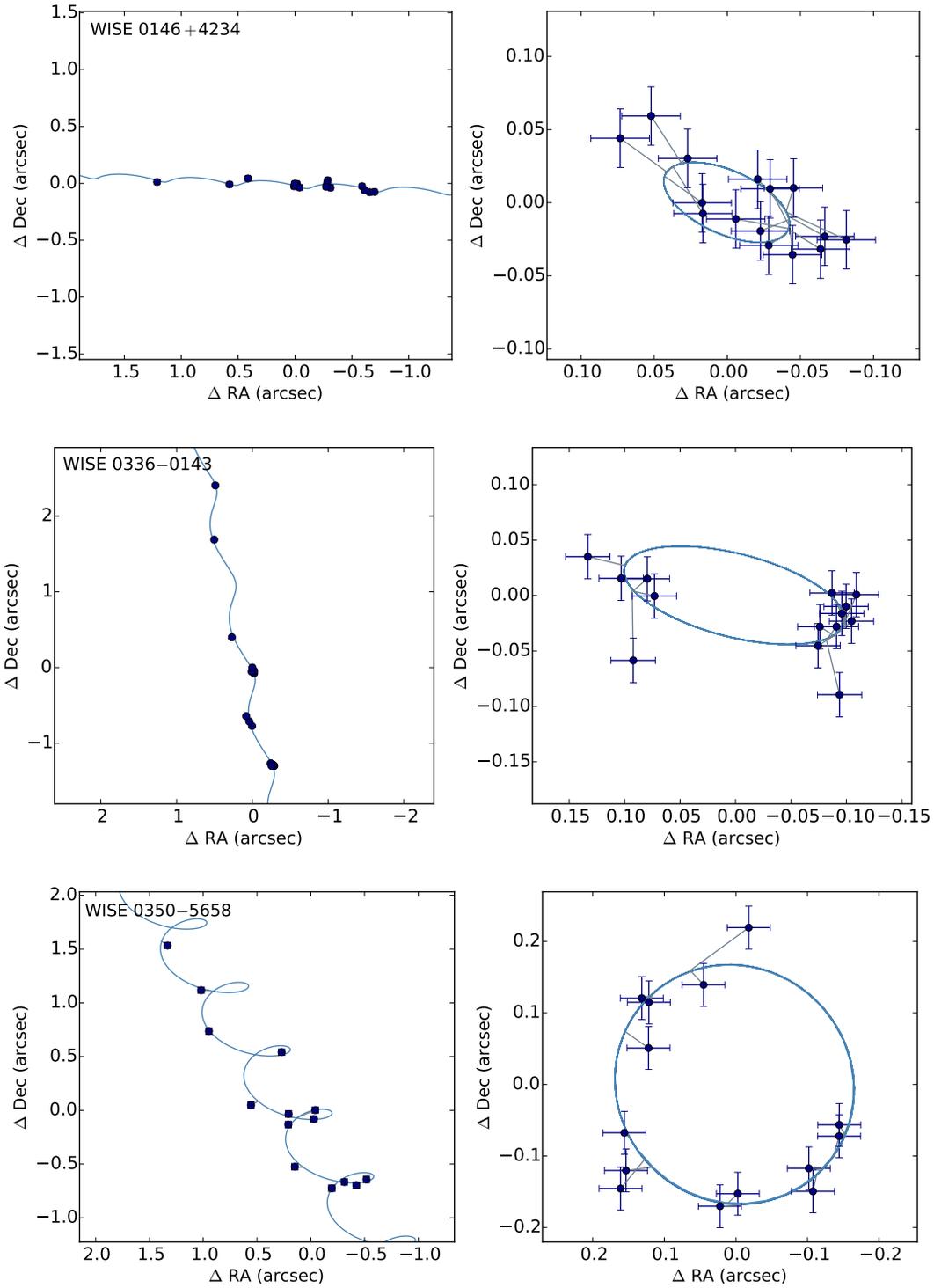

\begin{center}
\includegraphics[scale=0.5]{fig5_1.pdf}
\includegraphics[scale=0.51]{fig5_2.pdf}
\includegraphics[scale=0.5]{fig5_3.pdf}
\caption{Astrometric fits for three of our targets. \added{The complete figure set (22 objects) is available in the online journal}. We maintained a square scaling for the $\Delta$ Declination and $\Delta$ RA. Our observations are plotted in navy and the best-fit astrometric model is plotted in light blue. The left plots include proper motion and parallax and the right plots have proper-motion removed. Note the differing scales between the left and right plots. WISE 0146+4234 is an un-resolved binary, which produces systematic offsets of our astrometry and causes the parallactic ellipse to appear smaller than it is.  \label{fig:astrometric_fits}}
\end{center}
\end{figure*}

\section{Results}
\label{sec:results}

\subsection{Color Magnitude Diagrams \label{sec:cmd}}
We \replaced{list}{used our measured distances to calculate} the absolute magnitudes for all objects in $J_{MKO}$, $H_{MKO}$, [3.6], [4.5], $W1$, and $W2$, when available, \added{listed} in Table~\ref{tab:abs_mag}. In Figure~\ref{fig:cmd}, we plot four different color-magnitude diagrams (CMD)s, showing $M_J$ vs. $J-W2$, $M_H$ vs. $H-W2$, $M_{W2}$ vs. $J-W2$, and $M_{[4.5]}$ vs. [3.6]$-$[4.5]. Data from this paper are plotted as filled circles and data from the literature are open symbols (\citealt{tinney2014}: circles; \citealt{dupuy2013}: diamonds). Every object is colored according to its spectral type, as shown in the legend. The CMDs show a tight trend, particularly in $M_J$ vs. $J-W2$ and $M_H$ vs. $H-W2$, in which the trends previously seen for earlier spectral types are continued, showing decreasing absolute magnitudes in the near-infrared as $J-W2$ colors redden. 

We determined a weighted linear fit to both $M_J$ vs. $J-W2$ and $M_H$ vs. $H-W2$ and tabulate the coefficients in Table~\ref{tab:cmd_distance}. Although these relations require two photometric observations to obtain a photometric distance estimate, we find that this relationship is much tighter than if we were to determine fits to the absolute magnitude vs. Spectral Type. 

$M_{W2}$ vs. $J-W2$ shows more scatter than the near-infrared color magnitude diagrams. Interestingly, $M_{W2}$ vs. $J-W2$ appears to plateau in $M_{W2}$ across the T/Y transition. It is unclear if this feature is real, or due to a bias (systematic or otherwise). \added{It is possible that this represents a T/Y transition, perhaps due to the rainout of an opacity source or the appearance of the salt/sulfide clouds \citep{morley2012}.}

The M$_{[4.5]}$ vs. [3.6]$-$[4.5] plot shows significantly more cosmic scatter than the other panels in Figure~\ref{fig:cmd}. This is likely due to [3.6] being a non-ideal band for observing objects with significant \methane \ absorption. The blue tail of the 4.5 \microm \ bandpass falls into the [3.6] filter transmission, giving late-T and Y an overall very red slope in [3.6]. It's possible that variations in gravity and/or metallicity cause this slope to shift, producing the observed scatter. It is likely that the $W2$ vs. $W1-W2$  CMD would show a much tighter correlation, because the $W1$ and $W2$ bandpasses were designed specifically for cold brown dwarfs; however, many targets only have limits on their $W1$ magnitudes. 

\subsection{Absolute Magnitude vs SpT}
\added{Figure~\ref{fig:magspt} shows absolute magnitude in various near and mid-infrared bands as a function of spectral type, for this sample as well as other values taken from the literature.} Studying the relationship of the absolute magnitude emitted at each bandpass as a function of spectral type provides us with insight on the evolution of the brown dwarf spectral energy distribution as it cools over time. Earlier-type brown dwarfs tend to follow a narrow trend in absolute magnitude, with flux decreasing in each of the bands monotonically as a function of spectral type. Because spectral typing historically sorts objects by effective temperatures, we expected the Y dwarf sample to continue this trend. However, instead of a tight correlation between spectral type and absolute magnitude, we see a large amount of scatter, spanning as much as $\sim$ 5 magnitudes within the Y0 spectral class alone. \deleted{Figure~\ref{fig:magspt} shows absolute magnitude in various near and mid-infrared bands as a function of spectral type, for this sample as well as other values taken from the literature.}

Such a large spread in absolute properties cannot be explained by typical levels of variability (\citealt{cushing2016, leggett2016}) and must be indicative of a different physical mechanism. In Figure~\ref{fig:magspt}, each of the objects is colored according to $J-W2$ color cutoffs, as detailed in the legend. Here, we are using $J-W2$ as a proxy for temperature, based on Figure 18 from \citet{schneider2015}, which in turn utilizes the atmospheric models of \citet{saumon2012}, \citet{morley2012}, and \citet{morley2014}. Regardless of the type of clouds used in the atmospheric models, they all show a monotonic reddening of $J-W2$ as temperature decreases. When we separate objects by their $J-W2$ color, new trends appear in Figure~\ref{fig:magspt}. In particular, the Y0 dwarfs appear to cover a very broad range in effective temperatures, likely accounting for the $\sim$ 5 orders of absolute magnitudes observed in the $J$ band.

\begin{figure*}
\begin{center}
\includegraphics[scale=.9, angle=0]{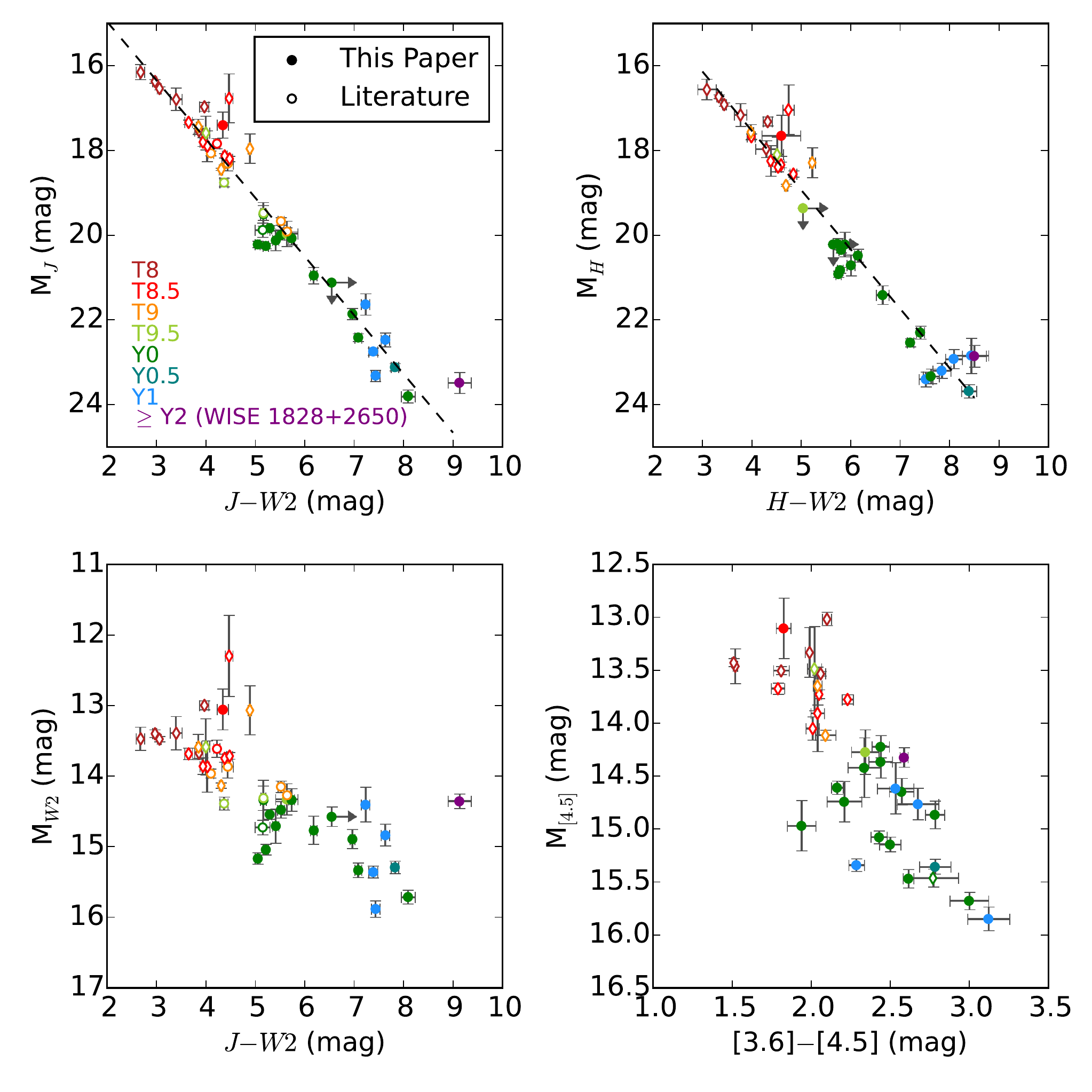}
\caption{Color Magnitude diagrams for M$_J$ vs. $J-W2$, M$_H$ vs. $H-W2$, M$_{W2}$ vs. $J-W2$, and M$_{[4.5]}$ vs. [3.6]-[4.5]. Open circles are from \citet{tinney2014} and open diamonds are from \citet{dupuy2013}. Filled circles are from this paper. Objects are shaded according to the spectral types listed in the legend. Weighted linear fits to M$_J$ vs. $J-W2$ and M$_H$ vs. $H-W2$ are plotted in dashed black lines. 
\label{fig:cmd}}
\end{center}
\end{figure*}

\begin{center}
\begin{deluxetable*}{lcccccccc}
\tabletypesize{\footnotesize}
\tablecaption{Absolute Magnitudes\label{tab:abs_mag}}
\tablehead{
\colhead{Object} &  
\colhead{$\pi_{\rm trig}$} &
\colhead{Distance} &
\colhead{$M_{J}$} & 
\colhead{$M_{H}$} &
\colhead{$M_{[3.6]}$} &
\colhead{$M_{[4.5]}$} &
\colhead{$M_{W1}$} &
\colhead{$M_{W2}$} \\ 
\colhead{Name} &
\colhead{(mas)} &
\colhead{(pc)} &
\colhead{(mag)} &
\colhead{(mag)} &
\colhead{(mag)} &
\colhead{(mag)} &
\colhead{(mag)} &
\colhead{(mag)}  \\
\colhead{(1)} &                          
\colhead{(2)} &  
\colhead{(3)} &     
\colhead{(4)} &
\colhead{(5)} &                          
\colhead{(6)} &
\colhead{(7)} &
\colhead{(8)} &
\colhead{(9)}
}
\startdata
WISE 0146+4234   & 45.575 $\pm$ 5.74 & 21.94$\substack{+3.16 \\ -2.45}$ & 17.69 $\pm$ 0.37  &  17.00 $\pm$ 0.36  & 15.65 $\pm$ 0.29  & 13.36 $\pm$ 0.27 & $>$17.43 & 13.38 $\pm$ 0.28   \\[5pt] 
WISE 0336$-$0143 & 100.90$\pm$5.86  & 9.91$\substack{+0.61 \\ -0.54}$ & $>$21.12 &  $>$20.22  & 17.22 $\pm$0.15 & 14.65 $\pm$ 0.13 & 18.47 $\pm$ 0.49 & 14.58 $\pm$ 0.14  \\[5pt] 
WISE 0350$-$5658 & 168.84 $\pm$ 8.53 & 5.92$\substack{+0.32 \\ -0.28}$ & 23.32 $\pm$ 0.13  &  23.40 $\pm$ 0.17 & 18.97 $\pm$ 0.17 & 15.85 $\pm$ 0.11 & $>$ 19.84 & 15.88 $\pm$ 0.12   \\[5pt] 
WISE 0359$-$5401 & 75.36 $\pm$ 6.62  & 13.27$\substack{+1.28 \\ -1.07}$ & 20.95 $\pm$ 0.20 &  21.41 $\pm$ 0.22 & 16.95 $\pm$ 0.22 & 14.74 $\pm$ 0.19 & $>$ 18.42 &  14.77 $\pm$ 0.20 \\[5pt] 
WISE 0410+1502   & 153.42 $\pm$ 4.05 & 6.52$\substack{+0.18 \\ -0.17}$  & 20.25 $\pm$ 0.06  &  20.83 $\pm$ 0.07  & 17.51 $\pm$ 0.07 & 15.08 $\pm$ 0.06 &  $>$ 19.10 & 15.04 $\pm$ 0.07  \\[5pt] 
WISE 0535$-$7500 & 79.51 $\pm$ 8.79  & 12.58$\substack{+1.56 \\ -1.25}$ & 21.63 $\pm$ 0.25  & 22.84 $\pm$ 0.42 & 17.15 $\pm$ 0.26 & 14.62 $\pm$ 0.24 & 17.44 $\pm$ 0.28 &  14.41 $\pm$ 0.24   \\[5pt] 
WISE 0647$-$6232 & 83.73 $\pm$ 5.68  & 11.94$\substack{+0.87 \\ -0.76}$ & 22.47 $\pm$ 0.16 &  22.92 $\pm$ 0.22 & 17.44 $\pm$ 0.20 & 14.77 $\pm$ 0.15 & $>$ 19.15  &  14.84 $\pm$ 0.16   \\[5pt] 
WISE 0713$-$2917 & 100.73 $\pm$ 4.74  & 9.93$\substack{+0.49 \\ -0.45}$ & 20.00 $\pm$ 0.11 &  20.21 $\pm$ 0.13 & 16.66 $\pm$ 0.11 & 14.22 $\pm$ 0.10 & $>$ 18.79  & 14.48 $\pm$ 0.11 \\[5pt] 
WISE 0734$-$7157 & 67.63 $\pm$ 8.68  & 14.79$\substack{+2.18 \\ -1.68}$ & 19.50 $\pm$ 0.28 &  20.22 $\pm$ 0.29 & 16.76 $\pm$ 0.30 & 14.42 $\pm$ 0.28 & 17.90 $\pm$ 0.40 &  14.34 $\pm$ 0.28    \\[5pt] 
WISE 0825+2805   & 139.02 $\pm$ 4.33 & 7.19$\substack{+0.23 \\ -0.22}$ & 23.12 $\pm$ 0.08 &  23.68 $\pm$ 0.15  & 18.14 $\pm$ 0.12 & 15.36 $\pm$ 0.07 & $>$ 19.16  &  15.29 $\pm$ 0.09  \\[5pt] 
WISE 1051$-$2138 & 49.27 $\pm$ 6.47  & 20.3$\substack{+3.1 \\ -2.4}$ & 17.40 $\pm$ 0.30 &  17.65 $\pm$ 0.48 & 14.93 $\pm$ 0.29 & 13.10 $\pm$ 0.29 & 15.76 $\pm$ 6.84  & 13.06 $\pm$ 0.29  \\[5pt] 
WISE 1055$-$1652 & 71.21 $\pm$ 6.82 & 14.04$\substack{+1.5 \\ -1.2}$ & 19.97 $\pm$ 0.30 &  $>$ 19.36 & 16.61 $\pm$ 0.22 & 14.27 $\pm$ 0.21 & $>$ 17.37 &  14.33 $\pm$ 0.22  \\[5pt] 
WISE 1206+8401   & 85.12 $\pm$ 9.27  & 11.75$\substack{+1.44 \\ -1.15}$ & 20.12 $\pm$ 0.24 &  20.71 $\pm$ 0.24  & 16.91 $\pm$ 0.25 & 14.97 $\pm$ 0.24 & $>$ 18.38 &  14.71 $\pm$ 0.24 \\[5pt] 
WISE 1318$-$1758 & 48.06 $\pm$ 7.33  & 20.81$\substack{+3.74 \\ -2.75}$ & 16.84 $\pm$ 0.38 &  16.12 $\pm$ 0.40 & 15.20 $\pm$ 0.34 & 13.12 $\pm$ 0.33 & 15.92 $\pm$ 0.37 &  13.07 $\pm$ 0.34    \\[5pt] 
WISE 1405+5534   & 144.35 $\pm$ 8.60 & 6.93 $\substack{+0.44 \\ -0.39}$ &  21.86 $\pm$ 0.13 &  22.30 $\pm$ 0.15  & 17.65 $\pm$ 0.14 & 14.87 $\pm$ 0.13 & 19.56 $\pm$ 0.42 &  14.89 $\pm$ 0.13   \\[5pt] 
WISE 1541$-$2250 & 167.05 $\pm$ 4.19 & 5.99 $\substack{+0.154 \\ -0.147}$ & 22.75 $\pm$ 0.08 &  23.20 $\pm$ 0.18 & 17.63 $\pm$ 0.07 & 15.34 $\pm$ 0.06 & 17.85 $\pm$ 0.17 &  15.36 $\pm$ 0.08   \\[5pt] 
WISE 1639$-$6847 & 228.05 $\pm$ 8.93 & 4.39$\substack{+0.18 \\ -0.17}$ &  22.42 $\pm$ 0.09 &  22.54 $\pm$ 0.09 & 18.08 $\pm$ 0.09 & 15.47 $\pm$ 0.09 &  19.06 $\pm$ 0.21 & 15.33 $\pm$0.10 \\[5pt] 
WISE 1738+2732   & 136.26 $\pm$ 4.27 & 7.34$\substack{+0.24 \\ -0.22}$ & 20.22 $\pm$ 0.07 &  20.92 $\pm$ 0.07  & 17.64 $\pm$ 0.09 & 15.15 $\pm$ 0.07 &  18.38 $\pm$ 0.17 & 15.17 $\pm$ 0.08  \\[5pt] 
WISE 1828+2650   & 100.21 $\pm$ 4.23  & 9.98$\substack{+0.44 \\ -0.40}$ & 23.48 $\pm$ 0.25 &  22.85 $\pm$ 0.26  & 16.91 $\pm$ 0.09 & 14.33 $\pm$ 0.09 & $>$ 18.25 & 14.36 $\pm$ 0.10   \\[5pt] 
WISE 2056+1459   & 138.32 $\pm$ 3.86 & 7.23$\substack{+0.21 \\ -0.20}$ & 19.83 $\pm$ 0.06 &  20.35 $\pm$ 0.07  & 16.77 $\pm$ 0.07 & 14.61 $\pm$ 0.06 &  17.18 $\pm$ 0.10 & 14.54 $\pm$ 0.07   \\[5pt] 
WISE 2209+2711   & 154.41 $\pm$ 5.67 & 6.48$\substack{+0.25 \\ -0.23}$ & 23.80 $\pm$ 0.15 &  23.33 $\pm$ 0.17  & 18.68 $\pm$ 0.14 & 15.68 $\pm$ 0.08 & $>$ 19.77 & 15.71 $\pm$ 0.10 \\[5pt] 
WISE 2220$-$3628 & 84.10 $\pm$ 5.90  & 11.89$\substack{+0.90 \\ -0.78}$ & 20.07 $\pm$ 0.15 &  20.48 $\pm$ 0.16 & 16.80 $\pm$ 0.17 & 14.37 $\pm$ 0.15 &  $>$ 18.40 &  14.34 $\pm$ 0.16  \\[5pt] 
\enddata
\end{deluxetable*}
\end{center}

\begin{center}
\begin{deluxetable}{lccc}
\tabletypesize{\footnotesize}
\tablecaption{Coefficients for linear fits to Color-Magnitude Relations\label{tab:cmd_distance}}
\tablehead{
\colhead{Color} &  
\colhead{c0} &
\colhead{c1} &
\colhead{rms}\\ 
\colhead{(1)} &                          
\colhead{(2)} &  
\colhead{(3)} &
\colhead{(4)}
}
\startdata
$M_J$ vs. $J-W2$ & 12.186 & 1.386 & 0.475  \\
$M_H$ vs. $H-W2$ & 11.935 & 1.401 & 0.544 \\ 
\enddata
\tablecomments{These coefficients fit a line such that $M_X = c0 + c1\times(M_X - W2)$ , where X is the $J$ or $H$ photometry on the MKO system.}
\end{deluxetable}
\end{center}

\begin{figure*}
\begin{center}
\includegraphics[scale=.74, angle=0]{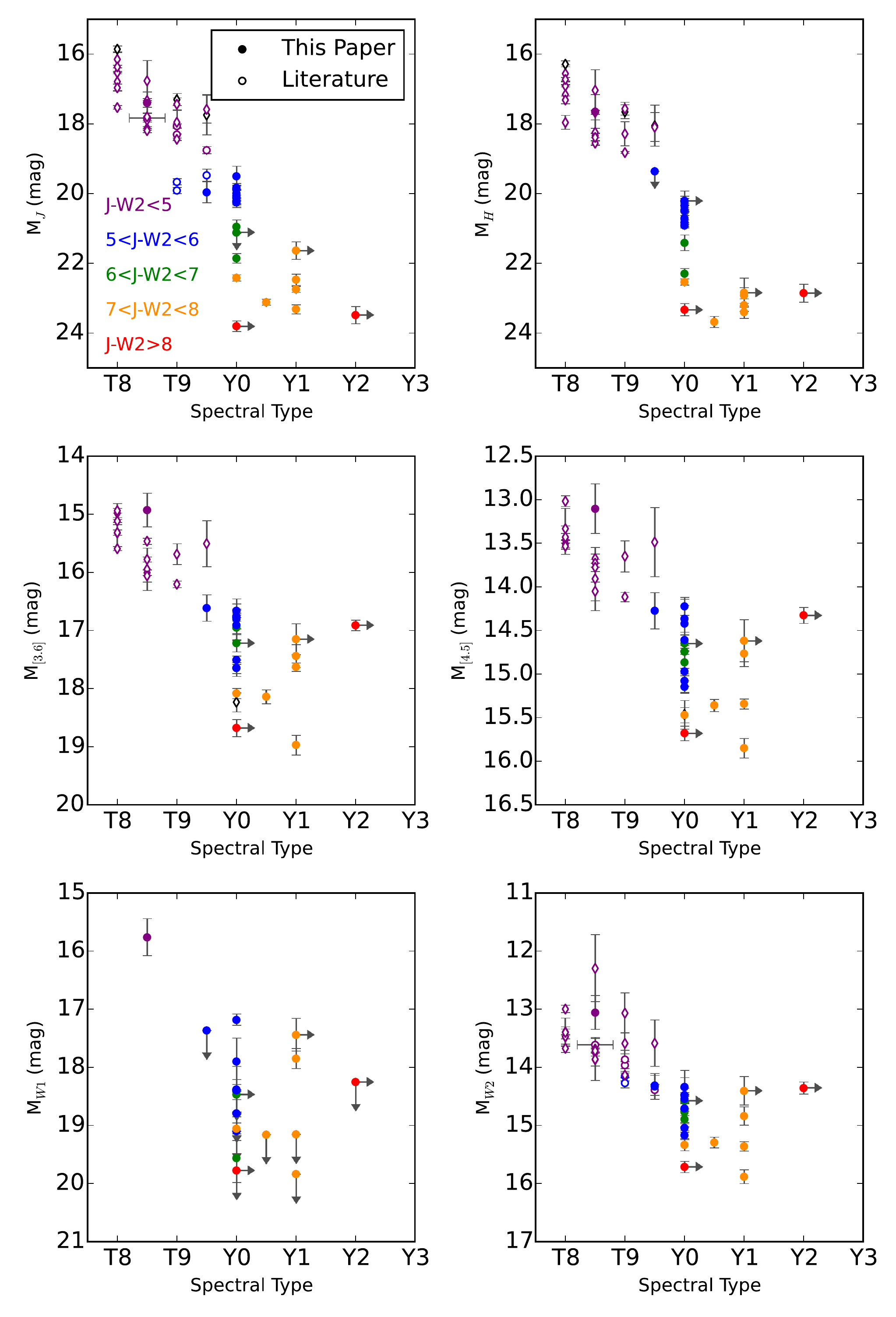}
\caption{Absolute magnitude vs. Spectral Type. Open circles are from \citet{tinney2014} and open diamonds are from \citet{dupuy2013}. Shaded objects are from this paper. Objects are shaded according to J-W2 color, as shown in the legend. 
\label{fig:magspt}}
\end{center}
\end{figure*}

\subsection{Spectrophotometric and Photometric Distances for New Discoveries}

In \S~\ref{sec:cmd} we determine photometric distance relationships based on linear fits to $M_J$ vs. $J-W2$ and $M_H$ vs. $H-W2$. These fits are valid for objects with \added{2 $<$ $J-W2$ $<$ 9} \deleted{colors ranging from 2--9} and \added{3 $<$ $H-W2$ $<$ 9} \deleted{from 3--9}. Below, we use this photometric distance relationship to estimate distances to the new \added{$\geq$} T8 objects presented here. For objects $<$ T8, we use the spectrophotometric distance relations from \citet{filippazzo2015}.

\textit{WISE 0550$-$1950}: We do not have an adequate baseline to measure the parallax of this new T6.5 dwarf, but using the spectrophotometric distance estimates from \citet{filippazzo2015}, we estimate a distance of 32.9 pc to this object.

\textit{WISE 0615+1526}: We estimate a photometric distance of 22.3 pc for this object. 

\textit{WISE 0642+0423}: We estimate its photometric distance to be 29.6 pc.

\textit{WISE 1220+5407} Our photometric distance estimate puts it at 22.5 pc.

\textit{WISE 2203+4619} is estimated to be 18.9 pc away, based on our photometric distance relationships in \S~\ref{sec:cmd}.

\subsection{Comparison to Literature}

In Table~\ref{tab:lit_compare} and Figure~\ref{fig:delta_pi} we compare our results to previously published astrometric fits \added{for} all of our targets with previous parallax measurements. We find that our results are mostly consistent with previously published values in the literature, with a few notable exceptions. 

\added{\subsubsection{Comparison to \citet{tinney2014} and \citet{beichman2014}}}
\added{Our results are consistent with \citet{tinney2014} and \citet{beichman2014} for most objects, with the exception of WISE 2220$-$3628. For this object, we find a consistent astrometric fit to the \citet{tinney2014} dataset, but our results are discrepant from those of \citet{beichman2014}.} Upon further review of the \citet{beichman2014} dataset, we noticed that their measurements only cover one side of the parallactic ellipse, leaving the other side unconstrained and biasing the measurement. This is likely the cause of their discrepant fit.

\added{\subsubsection{Comparison to \citet{dupuy2013}}}
We also measure significant offsets in parallax values from \citet{dupuy2013}. We find systematically larger parallax values (closer distances) than they do for 5 out of the 6 targets we have in common. Each object has at least one other measurement in the literature and we find that we are consistently in agreement with the other reference. \citet{tinney2014} and \citet{smart2017} also note systematic offsets between their parallax measurements and those of \citet{dupuy2013}, concluding that these are likely due to the smaller number of measurements and thus a degeneracy between the parallax and proper motion parameters. For the extreme case of 1541$-$2250 (not plotted in comparison figures), we note, as did \citet{beichman2014} and \citet{tinney2014} that this object has several epochs skewed by a blend with a background star that throw off the fit in the [3.6] data, which explains the $>$ 3-$\sigma$ difference between the \citet{dupuy2013} results and others in the literature. 

We explored several hypotheses to explain the discrepancies between the \citet{dupuy2013} measurements and those presented here. Similar to our parallax measurements, \citet{dupuy2013} uses the IRAC instrument on $Spitzer$ to measure the positions of each target. However, they observed in [3.6], whereas the measurements presented here were made using [4.5] data. 

Our first hypothesis is that the use of [3.6] data causes a chromatic distortion on the image plane that is different for the target than the background stars, and which would cause a systematic offset in the positions of the targets in the \citet{dupuy2013} dataset. The IRAC instrument design utilizes beam splitters in each of its two fields of view to refract shorter wavelength light ([3.6] and [4.5]) to separate focal planes from the longer wavelength light (ch3 and ch4). Both [3.6] and [4.5] have similar background characteristics during the warm mission (\citealt{carey2010}). The brown dwarf targets are significantly fainter in [3.6] compared to [4.5], requiring longer integration times (thus providing more background stars in each field). Late-T and Y dwarfs exhibit extreme methane absorption near the methane fundamental bandhead at 3.3 \microm , which produces a dramatic upward slope in the spectral energy distribution within the [3.6] bandpass. Thus the targets have significantly redder effective central wavelengths compared to the relatively flat spectral energy distributions of the background stars. This reddening effect in [3.6] would lead to a slightly different average angle of refraction, compared to the background stars' average angles of refraction. The target spectral energy distributions in the [4.5] bandpass peak much closer to the center of the bandpass and should not have significantly different effective wavelengths from the background stars, so they should be immune to this effect. We thus expect that this effect would be evident by comparing the offset parallax measurement to the [3.6]--[4.5] color (Figure~\ref{fig:ch1ch2trent}). We see a slight correlation between [3.6]$-$[4.5] color and parallax offset, but there is not enough data to draw a firm conclusion.

\begin{center}
\begin{deluxetable*}{lccccccc}
\tabletypesize{\scriptsize}
\tablecaption{Comparison to Published Parallaxes and Proper Motions\label{tab:lit_compare}}
\tablehead{
\colhead{Object} &
\colhead{Measurement} &
\colhead{This paper} &
\colhead{Smart et al. (2017)} &
\colhead{Beichman et al. (2014)} &
\colhead{Tinney et al. (2014)} &
\colhead{Dupuy \& Kraus (2013)} &
\colhead{Leggett et al. (2017)\tablenotemark{a}}\\
\colhead{(1)} &                          
\colhead{(2)} &  
\colhead{(3)} &     
\colhead{(4)} &
\colhead{(5)} &                          
\colhead{(6)} &
\colhead{(7)} &
\colhead{(8)}
}
\startdata
& $\pi_{\rm trig}$ (mas) & 45.6$\pm$5.7 & \nodata & 94$\pm$14 & \nodata & \nodata &54$\pm$5 \\
WISE 0146+4234 & $\mu_{\alpha}$ (mas/yr) &  $-$450.67$\pm$6.3 & \nodata & $-$441$\pm$13 & \nodata & \nodata &$-$455$\pm$4 \\
&  $\mu_{\delta}$ (mas/yr) & $-$27.9$\pm$6.3 & \nodata & $-$26$\pm$16 & \nodata & \nodata &$-$24$\pm$4 \\
\tableline
& $\pi_{\rm trig}$ (mas) & 168.8$\pm$8.5 & \nodata & \nodata & \nodata & \nodata &184$\pm$10 \\
WISE 0350$-$5658 & $\mu_{\alpha}$ (mas/yr) &  $-$206.9$\pm$6.5 & \nodata & \nodata & \nodata & \nodata &$-$206$\pm$7 \\
&  $\mu_{\delta}$ (mas/yr) & $-$577.7$\pm$6.7 & \nodata & \nodata & \nodata & \nodata &$-$578$\pm$8 \\
\tableline
& $\pi_{\rm trig}$ (mas) & 75.4$\pm$6.62 & \nodata & \nodata & 63.2$\pm$6.0 & \nodata &\nodata \\
WISE 0359$-$5401 & $\mu_{\alpha}$ (mas/yr) &  $-$152.7$\pm$4.8 & \nodata & \nodata & $-$176.0$\pm$10.8 & \nodata &\nodata \\
&  $\mu_{\delta}$ (mas/yr) & $-$783.7$\pm$4.9 & \nodata & \nodata & $-$744.5$\pm$11.9 & \nodata &\nodata \\
\tableline
& $\pi_{\rm trig}$ (mas) & 153.4$\pm$4.0 & 144.3$\pm$9.9 & 160$\pm$9  & \nodata & 132$\pm$15 &\nodata \\
WISE 0410+1502 & $\mu_{\alpha}$ (mas/yr) &  959.9$\pm$3.6  & 956.8$\pm$5.6 & 966$\pm$13 & \nodata & 958$\pm$37 &\nodata \\
&  $\mu_{\delta}$ (mas/yr) & $-$2218.6$\pm$3.5 & $-$2221.2$\pm$5.5 & $-$2218$\pm$13 & \nodata & $-$2229$\pm$29 &\nodata \\
\tableline
& $\pi_{\rm trig}$ (mas) & 79.5$\pm$8.8 & \nodata & \nodata & 74$\pm$14 & \nodata &70$\pm$5 \\
WISE 0535$-$7500 & $\mu_{\alpha}$ (mas/yr) &  $-$113.2$\pm$7.7 & \nodata & \nodata & $-$113.4$\pm$15.4 & \nodata &$-$127$\pm$4 \\
&  $\mu_{\delta}$ (mas/yr) & 23.7$\pm$7.5 & \nodata & \nodata & 36.2$\pm$8.8 & \nodata &13$\pm$4 \\
\tableline
& $\pi_{\rm trig}$ (mas) & 83.7$\pm$5.7 & \nodata & \nodata & 93$\pm$13  & \nodata &\nodata \\
WISE 0647$-$6232 & $\mu_{\alpha}$ (mas/yr) &  1.0$\pm$5.1 & \nodata & \nodata & 0.6$\pm$16.1 & \nodata &\nodata \\
&  $\mu_{\delta}$ (mas/yr) & 391.0$\pm$4.6 & \nodata & \nodata & 368.0$\pm$18.0 & \nodata &\nodata \\
\tableline
& $\pi_{\rm trig}$ (mas) & 100.7$\pm$4.7 & \nodata & 106$\pm$13 & 08.7$\pm$4.0 & \nodata &\nodata \\
WISE 0713$-$2917 & $\mu_{\alpha}$ (mas/yr) &  341.1$\pm$6.6 & \nodata & 388$\pm$20 & 350.1$\pm$4.8 & \nodata &\nodata \\
&  $\mu_{\delta}$ (mas/yr) & $-$411.1$\pm$6.0 & \nodata & $-$419$\pm$22 & $-$411.4$\pm$5.6 & \nodata &\nodata \\
\tableline
& $\pi_{\rm trig}$ (mas) & 67.6$\pm$8.7 & \nodata & \nodata & 73.7$\pm$6.6 & \nodata &\nodata \\
WISE 0734$-$7157 & $\mu_{\alpha}$ (mas/yr) &  $-$566.2$\pm$8.8  & \nodata & \nodata & $-$565.8$\pm$7.7 & \nodata &\nodata \\
&  $\mu_{\delta}$ (mas/yr) & $-$77.5$\pm$8.8 & \nodata & \nodata & $-$81.5$\pm$8.0 & \nodata &\nodata \\
\tableline
& $\pi_{\rm trig}$ (mas) & 139.0$\pm$4.3 & \nodata & \nodata & \nodata & \nodata &158$\pm$7 \\
WISE 0825+2805 & $\mu_{\alpha}$ (mas/yr) &  $-$64.4$\pm$5.6 & \nodata & \nodata & \nodata & \nodata &$-$66$\pm$8 \\
&  $\mu_{\delta}$ (mas/yr) & $-$234.7$\pm$5.4 & \nodata & \nodata & \nodata & \nodata &$-$247$\pm$10 \\
\tableline
& $\pi_{\rm trig}$ (mas) & 85.1$\pm$9.3 & \nodata & \nodata & \nodata & \nodata & 85$\pm$7 \\
WISE 1206+8401 & $\mu_{\alpha}$ (mas/yr) &  $-$557.7$\pm$6.5 & \nodata & \nodata & \nodata & \nodata &$-$585$\pm$4 \\
&  $\mu_{\delta}$ (mas/yr) & $-$241.3$\pm$6.5 & \nodata & \nodata & \nodata & \nodata &$-$253$\pm$5 \\
\tableline
& $\pi_{\rm trig}$ (mas) & 144.3$\pm$8.6 & \nodata & \nodata & \nodata & 129$\pm$19 &155$\pm$6 \\
WISE 1405+5534 & $\mu_{\alpha}$ (mas/yr) &  $-$2336.0$\pm$6.9  & \nodata & \nodata & \nodata & $-$2263$\pm$47 &$-$2334$\pm$5 \\
&  $\mu_{\delta}$ (mas/yr) & 238.0$\pm$7.40 & \nodata & \nodata & \nodata & 288$\pm$41 &232$\pm$5 \\
\tableline
& $\pi_{\rm trig}$ (mas) & 167.1$\pm$4.2 & \nodata & 176$\pm$9 & 175.1$\pm$4.4 & 74$\pm$31 &\nodata \\
WISE 1541$-$2250 & $\mu_{\alpha}$ (mas/yr) &  $-$895.0$\pm$4.7 & \nodata & $-$857$\pm$12 & $-$894.7$\pm$4.2 & $-$870$\pm$130 &\nodata \\
&  $\mu_{\delta}$ (mas/yr) & $-$94.7$\pm$4.7 & \nodata & $-$87$\pm$13 & $-$87.7$\pm$4.7 & $-$13$\pm$58 &\nodata \\
\tableline
& $\pi_{\rm trig}$ (mas) & 228.1$\pm$8.9 & \nodata & \nodata & 202.3$\pm$3.1 & \nodata &\nodata \\
WISE 1639$-$6847 & $\mu_{\alpha}$ (mas/yr) &  579.1$\pm$12.5 & \nodata & \nodata & 586.0$\pm$5.5 & \nodata &\nodata \\
&  $\mu_{\delta}$ (mas/yr) & $-$3104.5$\pm$12.2 & \nodata & \nodata & $-$3101.1$\pm$3.6 & \nodata &\nodata \\
\tableline
& $\pi_{\rm trig}$ (mas) & 136.3$\pm$4.3 & 128.5$\pm$6.3 & 128$\pm$10 & \nodata & 102$\pm$18 &\nodata \\
WISE 1738+2732 & $\mu_{\alpha}$ (mas/yr) &  343.3$\pm$3.5 & 345.0$\pm$5.7 & 317$\pm$9 & \nodata & 292$\pm$63 &\nodata \\
&  $\mu_{\delta}$ (mas/yr) & $-$340.6$\pm$3.4 & $-$340.1$\pm$5.1 & $-$321$\pm$11 & \nodata & $-$396$\pm$22 &\nodata \\
\tableline
& $\pi_{\rm trig}$ (mas) & 100.2$\pm$4.2 & \nodata & 106$\pm$7 & \nodata & 70$\pm$14 &\nodata \\
WISE 1828+2650 & $\mu_{\alpha}$ (mas/yr) &  1021.0$\pm$3.2 & \nodata & 1024$\pm$7 & \nodata & 1020$\pm$15 &\nodata \\
&  $\mu_{\delta}$ (mas/yr) & 175.6$\pm$3.1 & \nodata & \nodata & \nodata & 173$\pm$16 &\nodata \\
\tableline
& $\pi_{\rm trig}$ (mas) & 138.3$\pm$3.9  & 148.9$\pm$8.2 & 140$\pm$9  & \nodata & 144$\pm$23 &\nodata \\
WISE 2056+1459 & $\mu_{\alpha}$ (mas/yr) &  823.0$\pm$3.3 & 826.4$\pm$5.5 & 812$\pm$9 & \nodata & 761$\pm$46 &\nodata \\
&  $\mu_{\delta}$ (mas/yr) & 535.7$\pm$3.4 & 530.7$\pm$8.5 & 34$\pm$8 & \nodata & 500$\pm$21 &\nodata \\
\tableline
& $\pi_{\rm trig}$ (mas) & 154.4$\pm$5.7 & \nodata & 147$\pm$11 & \nodata & \nodata &\nodata \\
WISE 2209+2711 & $\mu_{\alpha}$ (mas/yr) & 1199.6$\pm$4.9 & \nodata & c1217$\pm$13 & \nodata & \nodata &\nodata \\
&  $\mu_{\delta}$ (mas/yr) & $-$1359.0$\pm$4.8 & \nodata & $-$1372$\pm$15 & \nodata & \nodata &\nodata \\
\tableline
& $\pi_{\rm trig}$ (mas) & 84.1$\pm$5.9 & \nodata & 136$\pm$17 & 87.2$\pm$3.7 & \nodata &\nodata \\
WISE 2220$-$3628 & $\mu_{\alpha}$ (mas/yr) &  292.9$\pm$7.4 & \nodata & 283$\pm$13 & 282.7$\pm$5.0 & \nodata &\nodata \\
&  $\mu_{\delta}$ (mas/yr) & $-$61.5$\pm$7.0  & \nodata & $-$97$\pm$17 & $-$94.0$\pm$3.0 & \nodata &\nodata
\enddata
\tablenotetext{a}{The data presented in \citet{leggett2017} include astrometric data first published in \citet{luhman2016}.} 
\end{deluxetable*}
\end{center}

Our second hypothesis is that there is a fundamental difference between our fitting analysis and that of \citet{dupuy2013}. In Section 2.4 of \citet{dupuy2012}, they describe their methodology for determining astrometric fits: ``We fitted three parameters to
the combined ($\alpha,\delta$) data: proper motion in right ascension ($\mu_{\alpha}$), proper motion in declination ($\mu_{\delta}$ ), and parallax ($\pi$). This is notably different from one standard approach taken in the literature of fitting two separate values of the parallax in $\alpha$ and $\delta$ (...) MPFIT minimized the residuals in ($\alpha,\delta$) after subtracting the relative parallax and proper motion offsets (three parameters) and the mean ($\alpha,\delta$) position (effectively removing 2 additional degrees of freedom).'' 

We interpret this to meant that the subtraction of the average ($\alpha,\delta$) position requires the parallax solution to fit through one point located at the center of the parallactic ellipse. The effect would be averaged out over long time baselines, but we believe this method to be ineffectual for limited epochs. The sense of the bias that we see is in the expected direction; that is, their ellipse fits are artificially smaller because of their choice of data analysis method. To test this, we performed a reduction of the same [3.6] data used in their paper but employing our methodology described above. In this case, we used the [3.6] PRF appropriate for Warm $Spitzer$ data. The resulting astrometric fits are compared to our [4.5] parallax measurements in Figure~\ref{fig:dpitrent}. In Figure~\ref{fig:dpitrent}, the original measurements from \citet{dupuy2013} are shown in yellow, and the re-calculation using our fitting analysis and the \citet{dupuy2013} data are in blue. In most cases, our calculations measure parallax solutions that are closer to those measured with our [4.5] data, though consistent with the original \citet{dupuy2013} values within the uncertainties.

\begin{figure}
\begin{center}
\gridline{\fig{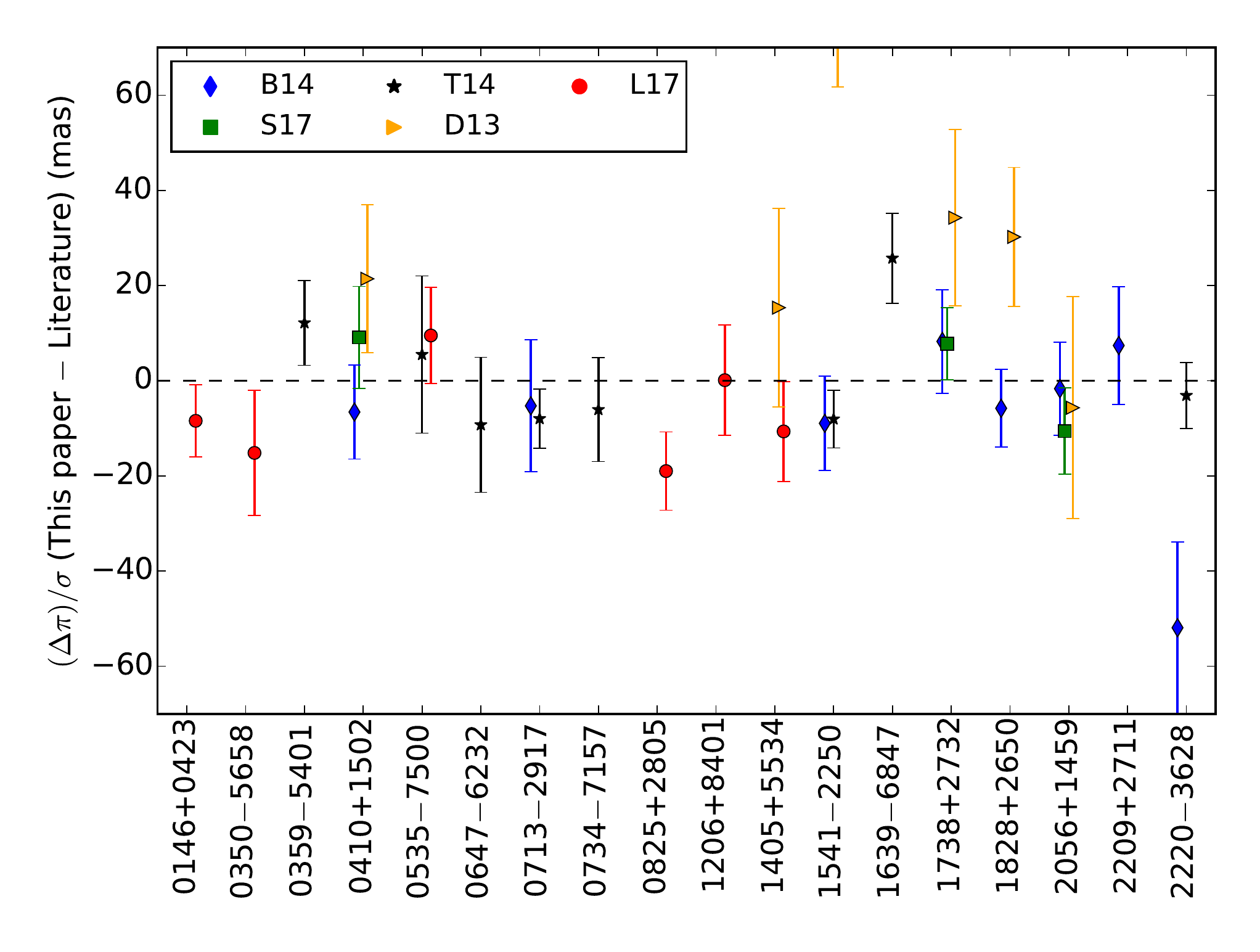}{3.5 in}{(a)}}
\gridline{\fig{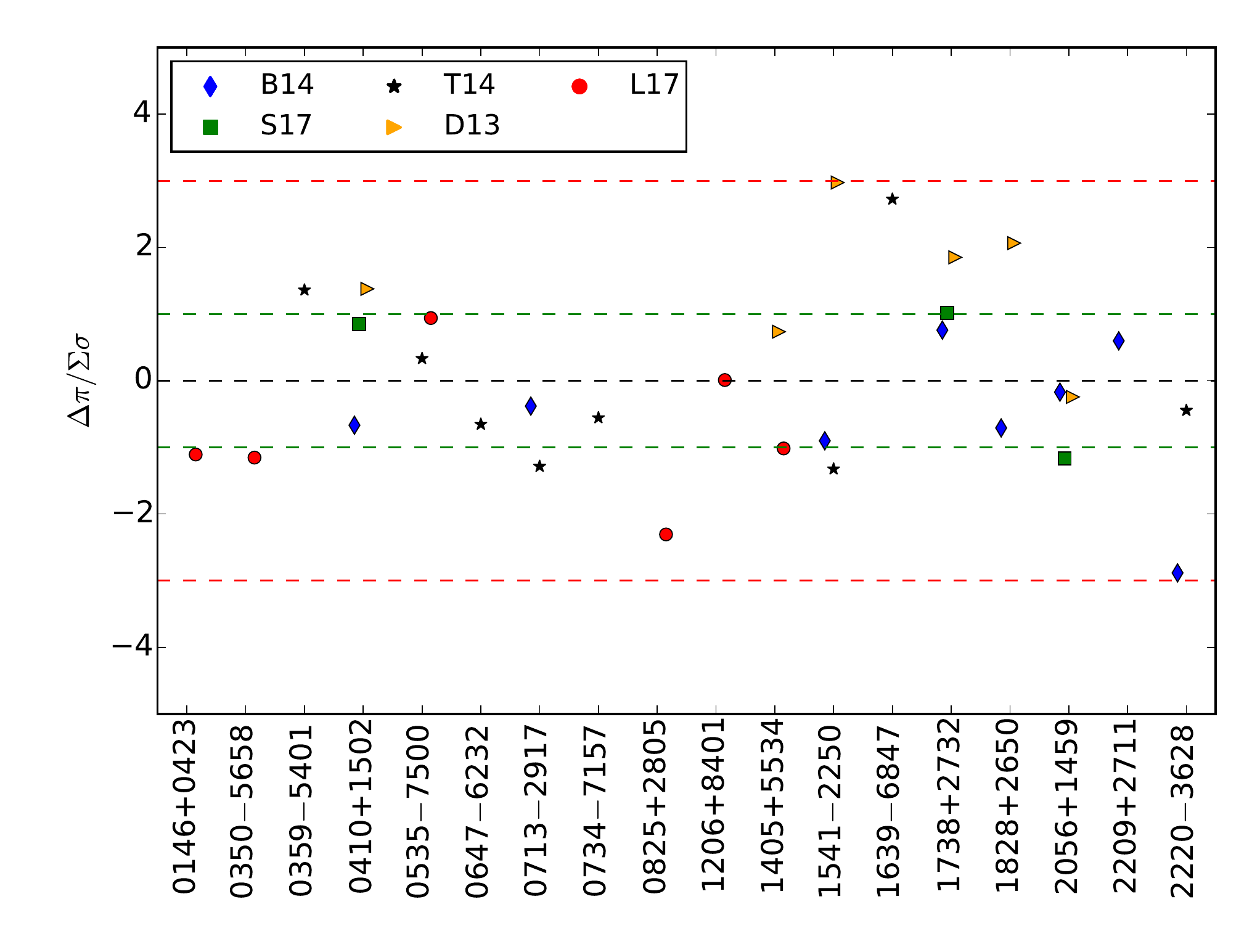}{3.5 in}{(b)}}
\caption{(a) Comparison of the difference in parallax values from this paper with the literature, vs. target name. Differences from \citet{beichman2014} are in blue diamonds, \citet{tinney2014} in black stars, \citet{leggett2017} in red circles, \citet{smart2017} in green squares, and \citet{dupuy2013} in yellow triangles. Note that the \citet{dupuy2013} value for 1541 is off the chart, their parallax being miscalculated due to a blend with a background star in their dataset.  \\
(b) Fractional $\sigma$ difference between this paper and the literature. Dashed green lines denote 1$\sigma$ offsets and dashed red lines denote 3$\sigma$ offsets. With a few exceptions, our measured parallaxes are consistent within 1$\sigma$ to previously published values. 
\label{fig:delta_pi}}
\end{center}
\end{figure}

\begin{figure}
\begin{center}
\includegraphics[scale=.41, angle=0]{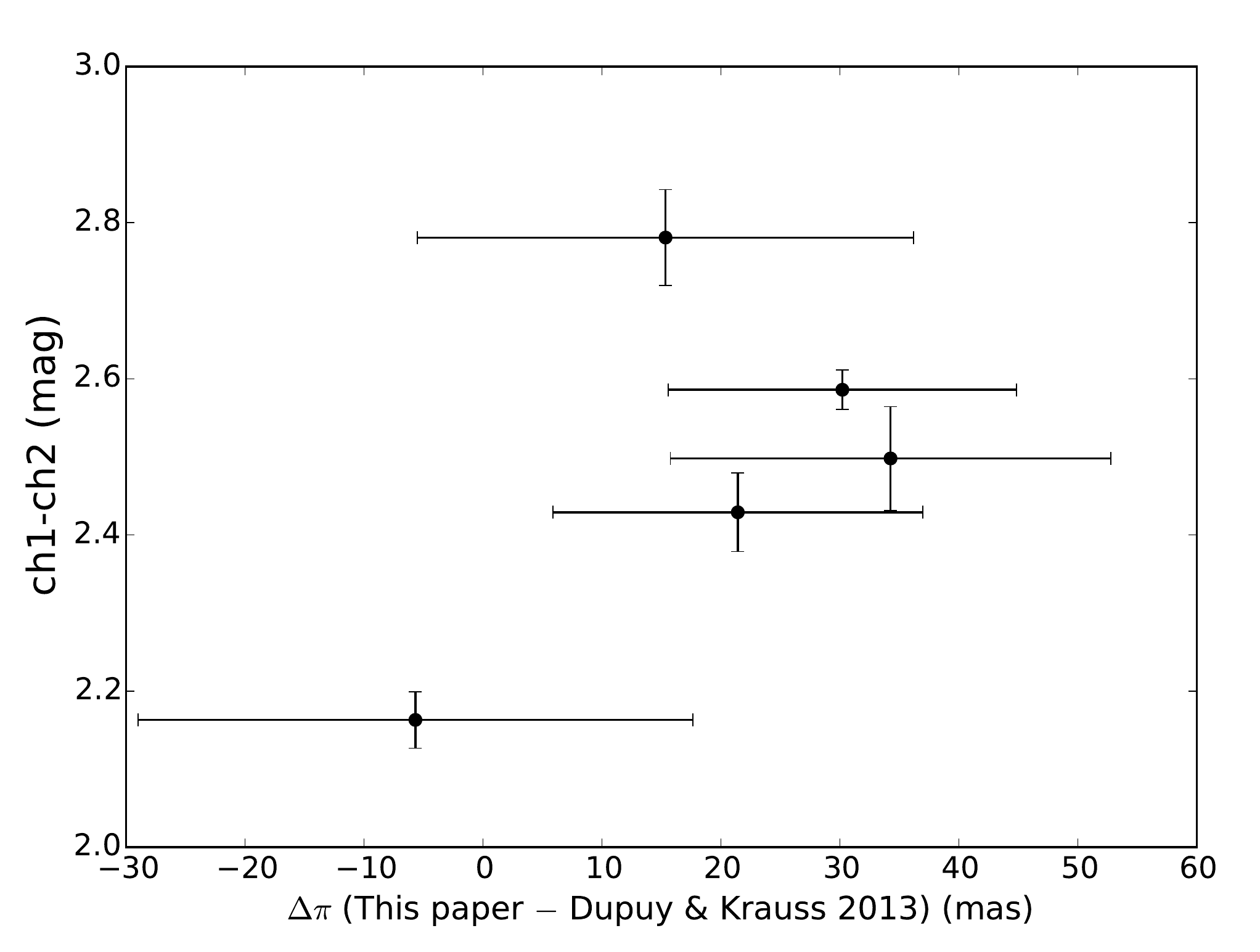}
\caption{Comparison of parallax offset between our values and those of \citet{dupuy2013} vs. [3.6]$-$[4.5] color. If the extremely red-sloped [3.6] bandpass were responsible for the offset, we would expect to see an increasing trend in offset vs. [3.6]$-$[4.5] color. A slight correlation is seen, though there is not enough data to draw a firm conclusion.
\label{fig:ch1ch2trent}}
\end{center}
\end{figure}

Our third hypothesis to explain the discrepant measurements is that the shorter time baseline of the \citet{dupuy2013} dataset made it difficult to disentangle the effects of proper motion when calculating the parallax. We explored this effect by reducing later epochs of [3.6] data, available on the $Spitzer$ archive. The addition of 9--10 epochs for each target cannot fully account for the earlier difference seen between the [3.6] and [4.5] parallax measurements. These differences are plotted in Figure~\ref{fig:dpitrent} in red. All targets except for 0410+1502 show an improved comparison, though the systematic offset remains. 

We believe that some combination of the three effects contributed to the systematically offset parallaxes published in \citet{dupuy2013}. After re-reducing the \citet{dupuy2013} data and adding additional epochs, we were unable to fully account for the discrepancy, but the offset as a function of color also appears to only have a slight trend.  

\begin{figure}
\begin{center}
\includegraphics[scale=.42, angle=0]{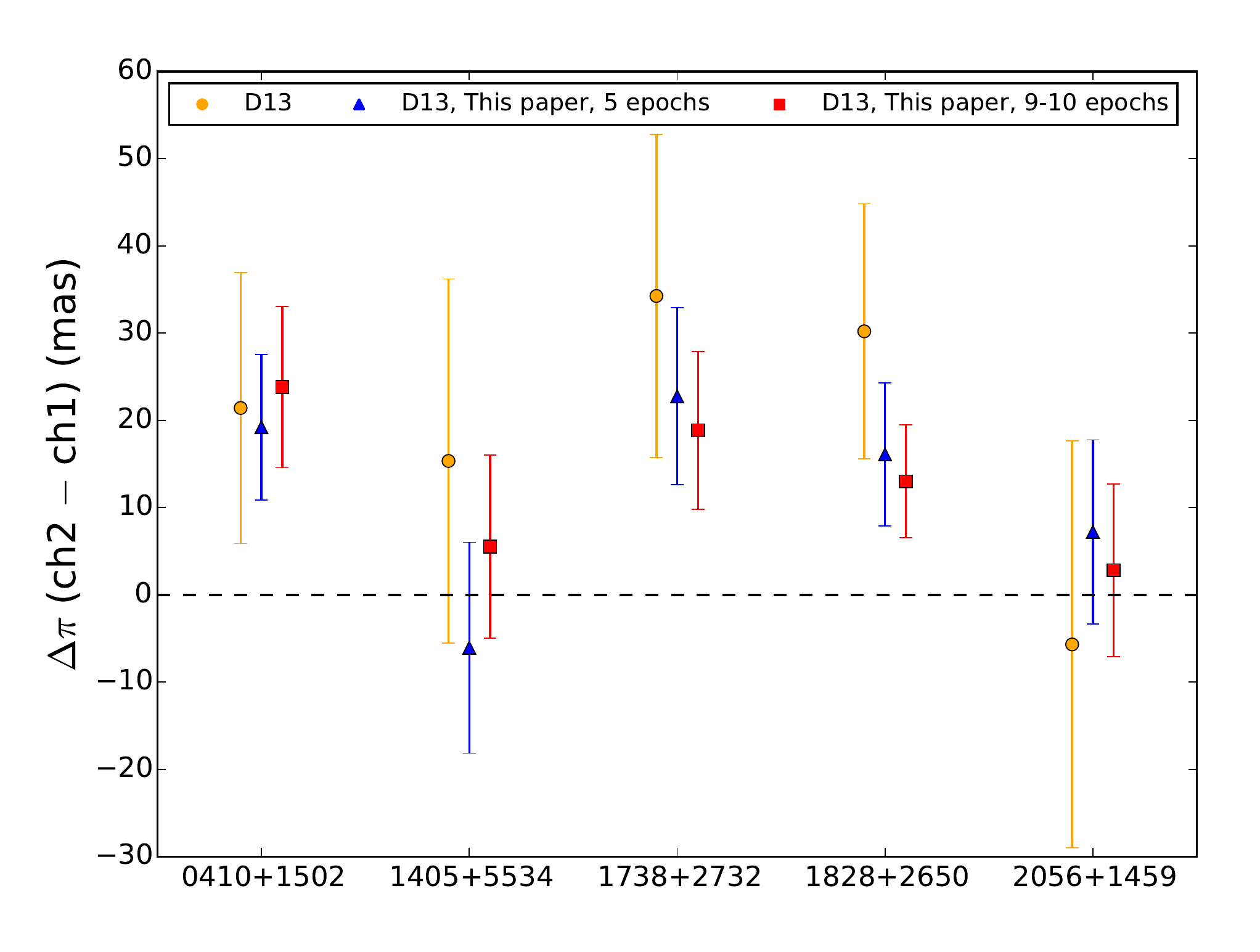}
\caption{Comparison of parallaxes measured with [4.5] and [3.6], for targets overlapping the \citet{dupuy2013} dataset. Parallax difference (mas) is plotted for each overlapping target. Data points have been offset to better show uncertainties. Yellow \replaced{points}{circles} are the original measurements from \citet{dupuy2013}. Blue \replaced{points}{triangles} were measured by re-reducing the \citet{dupuy2013} data and using our own fitting analysis. Red \replaced{points}{squares} were measured \replaced{similarly to the blue points}{using the \citet{dupuy2013} dataset} with 9-10 additional epochs of [3.6] data included from the {\it Spitzer} archive \added{and were analyzed using our fitting code}.  \deleted{1541 is not shown here, due to the blending with a background star and inability to determine a reasonable astrometric solution.}
\label{fig:dpitrent}}
\end{center}
\end{figure}

\added{\subsubsection{Comparison to Leggett et al. 2017}}
\citet{leggett2017} and \citet{luhman2016} \replaced{presented new astrometric measurements from the}{used} $Spitzer$ [4.5] data \replaced{in}{from} our parallax program to measure astrometry for several of the objects in this paper. We note a $\sim$ 1- to 2-$\sigma$ offset that is largely systematic between their measurements and our own. They find larger parallaxes than we do, by $\sim$ 10--20 mas for four out of six objects. It is unclear what is causing the difference between our parallax measurements.

\subsubsection{Note added during Proofs}
We also note that in \textit{Bedin, L. R. \& Fontanive, C.; submitted to MNRAS}, they measured a parallax for WISE J154151.65-225024.9 to be 169$\pm$2 mas, consistent with our own value of 167.1$\pm$4.2 mas. \\

\section{Discussion of Y dwarf Effective Temperatures}
\label{sec:discussion}
\subsection{Not all Y dwarfs are created equal}

Y0 dwarfs span several magnitudes in $M_J$, and nearly two in $W2$, based on the near-infrared classification of Y0 dwarfs. As previously mentioned, we used $J-W2$ as a proxy for temperature to separate populations in Figure~\ref{fig:magspt}. The color cuts show that the Y0 class spans $>$ 4 magnitudes in $J-W2$ and also overlaps the $J-W2$ color space occupied by the Y1 and later-typed objects. These findings indicate that the classical near-infrared spectral typing method of sorting M, L, and T dwarfs by their J band spectral morphologies does not efficiently separate Y dwarfs by their respective temperatures. Y dwarfs, with \teff $\lesssim$ 500 K, emit only a small fraction of their light in the near-infrared and would be best-characterized based on their mid-infrared spectra. This was noted in the Y dwarf discovery paper, \citet{cushing2011}; however, until the launch of JWST, observers have little hope of obtaining high SNR mid-infrared spectra of Y dwarfs- though some have tried (e.g. \citealt{skemer2016}). The peak emission of a $\lesssim$ 500K brown dwarf falls in the $\sim$ 3 -- 10 \microm \ range, causing the $J$ band to lie on the Wien tail of the blackbody spectrum. Considering the above, we recommend that mid-infrared spectra (i.e. from JWST) be used to more fully-characterize the physical properties of these extremely cold objects. Below we examine some of the more interesting targets in our sample.

\subsection{Notes on specific Y0 dwarfs}

\textit{WISE 0146$+$4234 AB} This object has discrepant near-infrared photometry in the literature due to its blended binary nature. For this reason, we have excluded it from our color-cuts and plot it in gray in Figures~\ref{fig:cmd} and \ref{fig:magspt}.

\textit{WISE 0336$-$0143} exhibits absolute magnitude and colors much more similar to Y dwarfs, than late-T dwarfs, as seen in Figures~\ref{fig:cmd} and \ref{fig:magspt}. We currently only have a limit on its near-infrared magnitudes, but our photometry agree with the later epoch of spectroscopy that this object is indeed a Y dwarf.

\textit{WISE 0734$-$7157} This particular dwarf is likely one of the warmest Y0's, based on its color and M$_J$. The best-fit temperature from \citet{schneider2015} is 450 K, and \citet{leggett2017} estimate its \teff \ to be 435--465 K.

\textit{WISE 1639$-$6847} is the second coldest Y0, based on $J-W2$ color. It's location in color-magnitude space is much more similar to the Y1 objects. \citet{leggett2017} estimate its \teff \ $\sim$ 360--390 K, coinciding with our findings.

\textit{WISE 2209$+$2711}
This is the faintest Y0 dwarf in every absolute magnitude band we measure. It is also the reddest in Y0 in $J-$W2. From \citet{schneider2015}, the best fit model gives \teff=500--550K, \logg=4--4.5, 0.2--1.5 Gyr old. \citet{leggett2017} estimate \teff=310--340 K, which agrees better with our estimates that this object is colder than most Y0's. Even if we re-classify this as a Y1, this would still be the faintest and reddest Y1. This object is also the reddest Y0 in $J-H$ and $H-W2$. It's mildly blue but not unusual in $Y-J$ \citep{schneider2015}. If we use the $J-W2$ vs. Temperature plot from \citet{schneider2015} to  determine an effective temperature, this object should be only $\sim$300K. At such cold temperatures, the near-infrared flux is solely coming from the Wien tail. Our observations are thus not able to fully sample the peak of the Planck function, and thus a small shift in \teff \ can cause a significant change in absolute magnitudes and colors. This particular target would be excellent for follow-up with JWST spectroscopy and imaging.

\subsection{Notes on specific Y0.5 dwarfs}
\textit{WISE 0825$+$2805}
This target is the third-reddest object in this sample in $J-$W2 after WISE 1828+2650 and WISE 2209+2711, likely indicating its extremely cold nature.

\subsection{Notes on specific >=Y1 dwarfs}
\textit{WISE 0350$-$5658} is the reddest in [3.6]$-$[4.5] in this sample, also the faintest in M$_{[3.6]}$ and M$_{[4.5]}$, and the faintest Y1 in M$_J$. It is likely extremely cold, probably matching the predicted $\sim$ 300 K from \citet{schneider2015} and the 310--340 K from \citet{leggett2017}.

\textit{WISE 0535$-$7500} is the brightest Y1-classified object and yet it was classified as $\geq$ Y1 in \citet{kirkpatrick2012}. WISE 0535 is located on the outskirts of the Large Magellanic Cloud and is in a highly crowded field that partially contaminated the $HST$ spectrum. This object would also benefit from follow-up observations in the mid-infrared.

\textit{WISE 1828$+$2650} is a known outlier that has thus far evaded a satisfactory explanation. \citet{leggett2017} propose that the peculiar near- and mid-infrared colors could be due to an unseen or equal-mass binary, however there are a couple of problems with the binarity hypothesis. First, extreme redness cannot be explained with binarity. Based on evolutionary models, extremely cold Y dwarfs effectively cannot be young, and so a protoplanetary or debris disk makes an unlikely culprit for the enhanced [3.6] and [4.5]. Second, the amount by which this object is over-luminous is at least one mag (depending on the band) and the maximum over-brightness observed from an equal-mass binary is 0.75 mag.

\subsection{Other Findings}
\deleted{The discovery of an additional Y dwarf, presented in this paper, brings the current total known Y dwarfs to 26. It has long been recognized that brown dwarfs cannot account for dark matter, and rather make up a fraction of the number of celestial objects compared to stars. However, it is likely that our sample of Y dwarfs within $\sim$10--20pc is incomplete. Their extremely cold nature makes them difficult to detect in proper-motion surveys. A dedicated 3--5 \microm \ all sky survey with a smaller pixel scale than $WISE$ would likely find a handful more.} 

\deleted{These objects are ideal for follow-up with JWST as we try to better understand star formation at the lowest masses and probe atmospheric conditions at the coldest temperatures. It may eventually be possible to spectroscopically differentiate between field brown dwarfs that have cooled to Y dwarf temperatures from the lowest-mass, Jupiter-sized exoplanets that have been ejected from their host system. Differing formation mechanisms predict different metallicity contents, but determining a metallicity will require 3--10 \microm \ spectroscopy with JWST.} 

\deleted{In the bottom-left panel of Figure~\ref{fig:cmd}, $M_{W2}$ seems to plateau between $J-$W2 4--6 mag, where the majority of the T9--Y0 objects lie. It is possible that this represents a T/Y transition, perhaps due to the rainout of an opacity source or the appearance of the salt/sulfide clouds \citep{morley2012}.}

\section{Summary and Final Remarks}
\label{sec:conclusions}
We present updated distance measurements for 22 late-T and Y dwarfs, measured from $Spitzer$/IRAC [4.5] data obtained over baselines of $\sim$ 2-7 years. We also present the discovery of one new Y dwarf and five new late-T dwarfs, based on spectra from Keck/NIRSPEC. With these distances, we probe the physical properties of Y dwarfs, and find that the Y dwarf spectral classifications are likely not ordering objects in a temperature-sensitive sequence. JWST mid-infrared spectra will probe the peak of the \replaced{blackbody functions}{spectral energy distributions} of these ultracool dwarfs and provide a better understanding of their physical characteristics.

\added{The discovery of an additional Y dwarf, presented in this paper, brings the current total known Y dwarfs to 26. It has long been recognized that brown dwarfs cannot account for dark matter, and rather make up a fraction of the number of celestial objects compared to stars. However, it is likely that our sample of Y dwarfs within $\sim$10--20pc is incomplete. Their extremely cold nature makes them difficult to detect in proper-motion surveys. A dedicated 3--5 \microm \ all sky survey with a smaller pixel scale than $WISE$ would likely find a handful more. CatWISE\footnote{\url{https://github.com/catwise}} is an upcoming survey that will use a re-processing of the AllWISE data to find fainter proper motion sources, including potential Y dwarfs. SPHEREx\footnote{\url{http://spherex.caltech.edu/index.html}}, a proposed NASA MIDEX mission, would conduct an all-sky spectral survey across 1--5 \microm \ and would also be likely to find more nearby T and Y dwarfs. }

\added{These objects are ideal for follow-up with JWST as we try to better understand star formation at the lowest masses and probe atmospheric conditions at the coldest temperatures. It may eventually be possible to spectroscopically differentiate between field brown dwarfs that have cooled to Y dwarf temperatures from the lowest-mass, Jupiter-sized exoplanets that have been ejected from their host system. Differing formation mechanisms predict different metallicity contents, but determining a metallicity will require 3--10 \microm \ spectroscopy with JWST.}

\acknowledgments
\label{sec:ack}
This work is based in part on observations made with the $Spitzer$ Space Telescope, which is operated by the Jet Propulsion Laboratory, California Institute of Technology under a contract with NASA. The authors wish to recognize and acknowledge the very significant cultural role and reverence that the summit of Mauna Kea has always had within the indigenous Hawaiian community.  We are most fortunate to have the opportunity to conduct observations from this mountain. This publication makes use of data products from the Wide-field Infrared Survey Explorer, which is a joint project of the University of California, Los Angeles, and the Jet Propulsion Laboratory/California Institute of Technology, funded by the National Aeronautics and Space Administration. This work has made use of data from the European Space Agency (ESA) mission {\it Gaia} (\url{http://www.cosmos.esa.int/gaia}), processed by the {\it Gaia} Data Processing and Analysis Consortium (DPAC, \url{http://www.cosmos.esa.int/web/gaia/dpac/consortium}). Funding
for the DPAC has been provided by national institutions, in particular the institutions participating in the {\it Gaia} Multilateral Agreement. This research has made use of the NASA/ IPAC Infrared Science Archive, which is operated by the Jet Propulsion Laboratory, California Institute of Technology, under contract with the National Aeronautics and Space Administration. This work is based in part on observations made with the Hale Telescope at Palomar Observatory, which is operated by the California Institute of Technology.

RLS's research was supported by the 2015  Henri Chr\'etien International Research Grant administered by the American Astronomical Society.

ECM thanks Dr. Gregory Mace for useful discussions, feedback, and mentoring.

\facilities{Spitzer(IRAC), Keck:II(NIRSPEC), Palomar(WIRC), IRSA}
\added{\software{REDSPEC (\url{http://www2.keck.hawaii.edu/inst/nirspec/redspec.html}), Astropy \citep{astropy1}, \citep{astropy2},(\url{https://doi.org/10.1051/0004-6361/201322068}); MOPEX (\url{http://irsa.ipac.caltech.edu/data/SPITZER/docs/dataanalysistools/tools/mopex/}); Scipy (\url{http://www.scipy.org/}); Numpy \citep{numpy}; Matplotlib \citep{matplotlib}, (\url{https://doi.org/DOI:10.1109/MCSE.2007.55} }}


\listofchanges 
\end{document}